\documentstyle[aps,twocolumn]{revtex}

\newcommand{\beq}{\begin{equation}}
\newcommand{\eeq}{\end{equation}}
\newcommand{\beqa}{\begin{eqnarray}}
\newcommand{\eeqa}{\end{eqnarray}}

\def\half{\frac{1}{2}}
\def\<{\langle}
\def\>{\rangle}

\def\up{\uparrow}
\def\down{\downarrow}
\def\lefta{\leftarrow}
\def\righta{\rightarrow}
\def\opone{\leavevmode\hbox{\small1\kern-3.8pt\normalsize1}}
\def\ml{\scriptstyle << \displaystyle}
\def\mr{\scriptstyle >> \displaystyle}
\def\F{{\cal F}}
\def\D{{\cal D}}
\def\H{{\cal H}}
\def\C{{\bf C}}

\newcommand{\infig}[2]{\begin{center}\mbox{\epsfxsize #2
\epsfbox{#1}}\end{center}}
\begin{document}
\input {epsf}
\title{Quantum cryptography}

\author
{Nicolas Gisin, Gr\'egoire Ribordy, Wolfgang Tittel and Hugo Zbinden \\
{\protect\small\em Group of Applied Physics, University of Geneva, 1211
Geneva 4, Switzerland}\\
}
\date{\today; submitted to Reviews of Modern Physics}

\maketitle

\begin{abstract}
Quantum cryptography could well be the first application of
quantum mechanics at the individual quanta level. The very fast
progress in both theory and experiments over the recent years are
reviewed, with emphasis on open questions and technological
issues.
\end{abstract}


\tableofcontents

\newpage
\section{Introduction}\label{Int}
Electrodynamics was discovered and formalized in the $19^{th}$ century. The $20^{th}$
century was then profoundly affected by its applications.
A similar adventure is possibly happening for quantum mechanics, discovered and
formalized during the last century. Indeed, although the laser and semiconductors are
already common, applications of the most radical predictions of quantum mechanics have
been thought of only recently and their full power remains a fresh gold mine for the
physicists and engineers of the $21^{st}$ century.

The most peculiar characteristics
of quantum mechanics are the existence of indivisible quanta and of entangled systems.
Both of these are at the root of Quantum Cryptography (QC) which could very well be the
first commercial application of quantum physics at the individual quantum level. In addition
to quantum
mechanics, the $20^{th}$ century has been marked by two other major scientific
revolutions: the theory of information and relativity. The status of the latter is
well recognized. It is less known that the concept of information, nowadays measured in bits,
and the formalization of probabilities is quite recent\footnote{The Russian mathematician A.N. Kolmogorow
(1956) is credited with being the first to have consistently formulated a mathematical theory
of probabilities in the 1940's.},
although they have a tremendous impact on our daily life. It is fascinating
to realize that QC lies at the intersection of quantum mechanics and
information theory and that, moreover, the tension between quantum mechanics and relativity
-- the famous EPR paradox (Einstein{\it et al.}1935) --
is closely connected to the security of QC. Let us add a further
point for the young physicists. Contrary to laser and semiconductor physics, which are
manifestations of quantum physics at the ensemble level and can thus be
described by semi-classical models, QC, and even much more quantum
computers, require a full quantum mechanical description (this may offer interesting
jobs for physicists well trained in the subtleties of their science).

This review article has several objectives. First we present the basic intuition
behind QC. Indeed the basic idea is so beautiful and simple that
every physicist and every student should be given the pleasure to enjoy it. The general
principle is then set in the broader context of modern cryptology (section \ref{TodayCrypto})
and made more precise (section \ref{BB84}).
Chapter \ref{TechChallenges} discusses the main technological challenges. Then,
chapters \ref{QCFaintPulses} and \ref{QCPhotonPair} present the most common implementation of
QC using weak laser pulses and photon pairs, respectively.
Finally, the important and difficult problems of eavesdropping
and of security proofs are discussed in chapter \ref{Eve}, where the emphasis is
more on the variety of questions than on technical issues. We tried to write the different
parts of this review in such a way that they can be read independently.

\section{A beautiful idea}\label{Idea}
The idea of QC was first proposed only in the 1970's by
Wiesner\footnote{Stephen Wiesner, then at Columbia University,
was the first one to propose ideas closely related to
QC, already in the 1970's. However, his revolutionary paper
appeared only a decade later. Since it is difficult to find, let us mention his
abstract: {\it The uncertainty principle imposes restrictions on the capacity of
certain types of communication channels. This paper will show that in compensation
for this ``quantum noise'', quantum mechanics allows us novel forms of coding without
analogue in communication channels adequately described by classical physics}.} (1983)
and by Charles H. Bennett from IBM and Gilles Brassard from Montr\'eal University
(1984, 1985)\footnote{Artur Ekert (1991) from Oxford University discovered QC
independently, though from a different perspective (see
paragraph \ref{EkertProtocol}).}. However, this idea is so simple that actually
every first year student since the infancy of quantum mechanics could have
discovered it!
Nevertheless, it is only nowadays that the matter is mature and
information security important enough, and -- interestingly --
only nowadays that physicists are ready to consider quantum mechanics, not only
as a strange theory good for paradoxes, but also as a tool for new engineering.
Apparently, information theory, classical cryptography,
quantum physics and quantum optics had first to develop into mature sciences.
It is certainly not a coincidence that QC and, more generally,
quantum information has been developed by a community including many computer
scientists and more mathematics oriented young physicists. A broader interest than
traditional physics was needed.

\subsection{The intuition}\label{Intuition}
Quantum Physics is well-known for being counter-intuitive, or even bizarre. We teach students
that Quantum Physics establishes a set of negative rules stating things that cannot be done.
For example:
\begin{enumerate}
\item{\it Every\ measurement perturbs the system}.
\item{\it One cannot determine simultaneously the position and the momentum of a particle
with arbitrary high accuracy.}
\item{\it One cannot measure the polarization of a photon in the vertical-horizontal basis
and simultaneously in the diagonal basis}.
\item{\it One cannot draw pictures of individual quantum processes}.
\item{\it One cannot duplicate an unknown quantum state}.
\end{enumerate}

This negative viewpoint on Quantum Physics, due to its contrast to classical physics, has only
recently been turned positive and QC is one of the best illustrations of this
{\it psychological revolution}. Actually, one could caricature
Quantum Information Processing
as the science of turning Quantum conundrums into potentially useful applications.

Let us illustrate this for QC.
One of the basic negative statement of Quantum Physics reads:
\beq
Every\ measurement\ perturbs\ the\ system
\label{MeasPert}
\eeq
(except if the quantum state is compatible with
the measurement). The positive side of this axiom can be seen when applied to a communication
between Alice and Bob (the conventional names of the sender and receiver, respectively),
provided the communication is quantum. The latter
means that the support of information are quantum systems, like, for example, individual photons.
Indeed, then axiom (\ref{MeasPert}) applies also to the
eavesdroppers, i.e. to a malicious Eve (the conventional name given to the adversary
in cryptology). Hence, Eve cannot get any information about the communication
without introducing perturbations which would reveal her presence.

To make this intuition more precise, imagine that Alice codes information in
individual photons which she sends to Bob. If Bob receives the photons unperturbed,
then, by the basic axiom (\ref{MeasPert}), the photons were not measured. No measurement implies
that Eve did not get any information about the photons (note that acquiring
information is synonymous to carrying out measurements). Consequently, after exchanging
the photons, Alice and Bob can check whether someone ``was listening'': they simply
compare a randomly chosen subset of their data using a public channel.
If Bob received the randomly chosen subset unperturbed then the logic goes as follows:\\
\beqa
No\ perturbation &\Rightarrow& No\ measurement \nonumber \\
&\Rightarrow& No\ eavesdropping
\eeqa
It is as simple as that! \\

Actually, there are two more points to add. First, in
order to ensure that axiom (\ref{MeasPert}) applies,
Alice encodes her information in non-orthogonal
states (we shall illustrate this in the sections \ref{BB84} and \ref{OtherProtocols}).
Second, as we have presented
it so far, Alice and Bob could discover any eavesdropper, but only after they exchanged
their message. It would of course be much better to ensure the privacy in advance, and not
afterwards! To achieve this, Alice and Bob complement
the above
simple idea with a second idea, again a very simple one, and one which is entirely
classical. Alice and Bob do not use the quantum channel to transmit information, but
only to transmit a random sequence of bits, i.e. a key. Now, if the key is unperturbed,
then Quantum Physics guarantees that no one got any information about this key by eavesdropping
(i.e. measuring) the quantum communication channel. In this case, Alice and Bob can safely
use this key to encode messages. If, on the contrary, the key turns out to be
perturbed, then Alice and Bob simply disregard it; since the key does not contain any information,
they did not lose any.

\begin{figure}
\infig{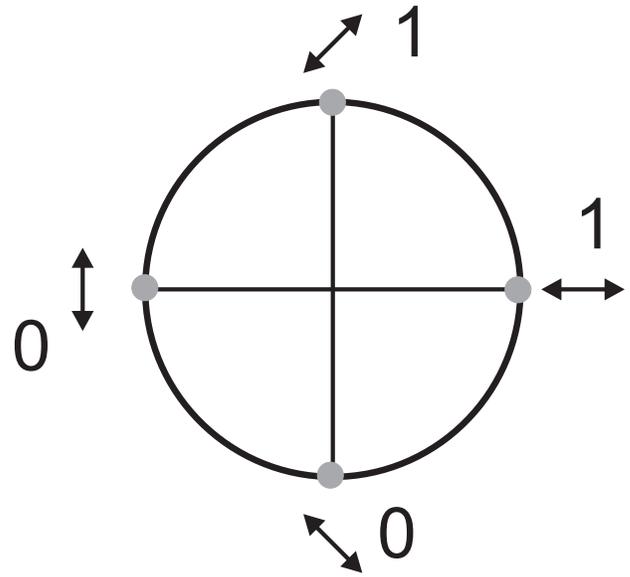}{0.95\columnwidth}
\caption{Implementation
of the BB84 protocol. The four states lie on the equator of the
Poincar\'{e} sphere.}
\label{fig2_1}
\end{figure}

Let us make this general idea somewhat more precise, anticipating
section \ref{BB84}. In practice, the individual quanta used by
Alice and Bob, often called qubits (for quantum bits), are encoded
in individual photons. For example, vertical and horizontal
polarization code for bit value zero and one, respectively. The
second basis, can then be the diagonal one ($\pm45^o$ linear
polarization), with $+45^o$ for bit 1 and $-45^o$ for bit 0,
respectively (see Fig. \ref{fig2_1}).  Alternatively, the circular
polarization basis could be used as second basis. For photons the
quantum communication channel can either be free space (see
section \ref{Satellite}) or optical fibers -- special fibers or
the ones used in standard telecommunication -- (section
\ref{QChannels}). The communication channel is thus not really
quantum. What is quantum are the information carriers.

But before continuing, we need to see how QC could fit in the existing
cryptosystems. For this purpose the next section briefly surveys some of the main
aspects of modern cryptology.

\subsection{Classical cryptography}\label{TodayCrypto}
Cryptography is the art of rendering a message unintelligible
to any unauthorized party. It is part of the broader field of cryptology,
which also includes crypto-analysis, the art of code breaking (for a historical
perspective, see Singh 1999). To achieve this
goal, an algorithm (also called a cryptosystem or cipher) is used to combine
a message with some additional information -- known as the ``key'' -- and
produce a cryptogram. This technique is known as ``encryption''. For a
cryptosystem to be secure, it should be impossible to unlock the cryptogram
without the key. In practice, this demand is often softened so that the
system is just extremely difficult to crack. The idea is that the message
should remain protected at least as long as the information it contains is valuable.
Although confidentiality is the traditional application of cryptography, it
is used nowadays to achieve broader objectives, such as authentication,
digital signatures and non-repudiation (Brassard 1988).

\subsubsection{Asymmetrical (public-key) cryptosystems}\label{PubKey}
Cryptosytems come in two main classes -- depending on whether Alice and Bob
use the same key. Asymmetrical systems involve the use of different keys for
encryption and decryption. They are commonly known as public-key
cryptosystems. Their principle was first proposed in 1976 by Whitfield
Diffie and Martin Hellman, who were then at Stanford University in the US.
The first actual implementation was then developed by Ronald Rivest, Adi
Shamir,and Leonard Adleman of the Massachusetts Institute of Technology in
1978\footnote{According to the British Government, public key cryptography was originally invented at the Government
Communications Headquarters in Cheltenham as early as in 1973. For an historical account,
see for example the book by Simon Singh (1999).}.
It is known as RSA and is still widely used. If Bob wants to be able
to receive messages encrypted with a public key cryptosystem, he must first
choose a ``private'' key, which he keeps secret. Then, he computes from this
private key a ``public'' key, which he discloses to any interested party.
Alice uses this public key to encrypt her message. She transmits
the encrypted message to Bob, who decrypts it with the private key.
Public-key cryptosystems are convenient and they have thus become very
popular over the last 20 years. The security of the internet, for example,
is partially based on such systems. They can be thought of as a mailbox,
where anybody can insert a letter. Only the legitimate owner can\ then
recover it, by opening it with his private key.

The security of public key cryptosystems is based on computational
complexity. The idea is to use mathematical objects called one-way functions.
By definition, it is
easy to compute the function $f(x)$ given the variable $x$, but difficult to
reverse the calculation and compute $x$ from $f(x)$. In the context of
computational complexity, the word ``difficult'' means that the time to do a
task grows exponentially with the number of bits in the input, while
``easy'' means that it grows polynomially.
Intuitively, it is easy to understand that it only takes a
few seconds to work out $67\times 71$, but it takes much longer to find the
prime factors of $4757$. However, factoring has a ``trapdoor'', which means
that it is easy to do the calculation in the difficult direction provided
that you have some additional information. For example, if you were told
that $67$ was one of the prime factors of $4757$, the calculation would be
relatively simple. The security of RSA is actually based on the
factorization of large integers.

In spite of its elegance suffers from a major flaw. Whether
factoring  is ``difficult'' or not could never be proven. This implies that the existence of a
fast algorithm for factorization cannot be ruled out. In addition, the discovery in 1994
by Peter Shor of a polynomial algorithm allowing fast factorization of integers with a
quantum computer puts additional doubts on the non-existence of a polynomial
algorithm for classical computers.

Similarly, all public-key cryptosystems rely on unproven
assumptions for their security, which could themselves be weakened or
suppressed by theoretical or practical advances.
So far, no one has proved the existence of any one-way
function with a trapdoor.
In other words, the existence of secure
asymmetric cryptosystems is not proven. This casts an
intolerable threat on these cryptosystems.

In a society where information
and secure communication is of utmost importance, as in ours, one cannot tolerate such a
threat. Think, for instance, that an overnight breakthrough in mathematics could
make electronic money instantaneously worthless. To limit such economical and social
risks, there is no possibility but to turn to symmetrical cryptosystems.
QC has a role to play in such alternative systems.

\subsubsection{Symmetrical (secret-key) cryptosystems}\label{SymCrypto}
Symmetrical ciphers require the use of a single key for both encryption and
decryption. These systems can be thought of as a safe, where the message is
locked by Alice with a key. Bob in turns uses a copy of this key to unlock
the safe. The ``one-time pad'', first proposed by Gilbert Vernam of AT\&T in
1926, belongs to this category. In this scheme, Alice encrypts her message,
a string of bits denoted by the binary number $m_{1}$, using a randomly
generated key $k$. She simply adds each bit of the message with the
corresponding bit of the key to obtain the scrambled text ($s=m_{1}\oplus k$%
, where $\oplus $ denotes the binary addition modulo 2 without carry). It is
then sent to Bob, who decrypts the message by subtracting the key ($%
s\ominus k=m_{1}\oplus k\ominus k=m_{1}$). Because the
bits of the scrambled text are as random as those of the key, they do not
contain any information. This cryptosystem is thus provably secure in the
sense of information theory (Shannon 1949). Actually, this is today the only
provably secure cryptosystem!

Although perfectly secure, the problem with this
system is that it is essential for Alice and Bob to possess a common secret
key, which must be at least as long as the message itself. They can only use
the key for a single encryption -- hence the name ``one-time pad''. If they
used the key more than once, Eve could record all of the scrambled messages
and start to build up a picture of the plain texts and thus also of the key.
(If Eve recorded two different
messages encrypted with the same key, she could add the scrambled text to
obtain the sum of the plain texts: $s_{1}\oplus s_{2}=m_{1}\oplus k\oplus
m_{2}\oplus k=m_{1}\oplus m_{2}\oplus k\oplus k=m_{1}\oplus m_{2}$, where
we used the fact that $\oplus $ is commutative.) Furthermore, the key has to
be transmitted by some trusted means, such as a courier, or through a
personal meeting between Alice and Bob. This procedure can be complex and
expensive, and may even amount to a loophole in the system.

Because of the problem of distributing long sequences of key bits,
the one-time pad is currently used only for the most critical applications.
The symmetrical cryptosystems in use for routine applications such as
e-commerce employ rather short keys.
In the case of the Data Encryption Standard (also known as DES, promoted by
the United States' National Institute of Standards and Technology), a 56 bits key is
combined with the plain text divided in blocks in a rather complicated way, involving
permutations and non-linear functions to produce the cipher text blocks (see Stallings 1999
for a didactic presentation).
Other cryptosystems (e.g. IDEA or AES) follow similar principles.
Like asymmetrical cryptosystems, they offer only computational security.
However for a given key length, symmetrical systems are more secure than their
asymmetrical counterparts.

In practical implementations, asymmetrical algorithms are not so much used for encryption, because of their slowness,
but to distribute session keys for symmetrical cryptosystems such as DES.
Because the security of those algorithms is not proven (see paragraph \ref{PubKey}),
the security of the whole implementation can be compromised.
If they were broken by mathematical advances, QC would constitute the only way to solve
the key distribution problem.

\subsubsection{The one-time-pad as ``classical teleportation''}\label{CTelep}
The one-time-pad has an interesting characteristic. Assume that Alice aims at
transferring to Bob a faithful copy of a classical system, without giving any information
to Eve about this system. For this purpose Alice
and Bob have only access to an insecure classical channel. This is possible provided
they share an arbitrary long secret key. Indeed, in principle Alice can measure the state
of her classical system with arbitrary high precision and then use the one-time-pad to
securely communicate this information to Bob who can then, in principle, reconstruct
(a copy of) the classical system. This somewhat artificial use of the one-time-pad has
an interesting quantum relative, (see section \ref{QTelep}).

\subsection{The example of the BB84 protocol}\label{BB84}
\subsubsection{Principle}
The first protocol for QC has been proposed in 1984 by Charles H. Bennett,
from IBM New-York, and Gilles Brassard, from the University of Montreal, hence the
name BB84 under which this protocol is recognized nowadays. They
published their work in a conference in India, totally unknown to
physicists. This underlines at once that QC needs the collaboration between different
communities, with different jargons and different habits and conventions\footnote{For instance,
it is amusing to note that physicists must publish in reputed journals
while conference proceedings
are of secondary importance. For computer science, on the contrary, the proceedings
of the best conferences are considered as the top, while journals are secondary!}.
The interdisciplinary character
of QC is the probable reason for its relatively slow start, but it certainly
contributes crucially to the vast and fast expansion over the recent years.

We shall explain the BB84 protocol using the language of spin
$\half$, but clearly any 2-level quantum system would do. The
protocol uses 4 quantum states that constitute 2 bases, think of
the states up $|\up\rangle$, down $|\down\rangle$, left
$|\lefta\rangle$ and right $|\righta\rangle$. The bases are
maximally conjugate in the sense that any pair of vectors, one
from each basis, has the same overlap, e.g.
$|\langle\up|\lefta\rangle|^2 =\half$. Conventionally, one
attributes the binary value 0 to states $|\up\rangle$ and
$|\righta\rangle$ and the value 1 to the other two states, and
calls the states qubits (for quantum bits). In the first step,
Alice sends individual spins to Bob in states chosen at random
among the 4 basic states (in Fig. \ref{fig2_1} the spin states
$|\up\>$,$|\down\>$, $|\righta\>$ and $|\lefta\>$ are identified
with the polarization states ``horizontal'', ``verical'',
``+45$^o$'' and ``-45$^o$'', respectively). How she ``chooses at
random'' is a delicate problem in practice (see section
\ref{QRNG}), but in principle she could use her free will. The
individual spins could be sent all at once, or one after the other
(much more practical); the only restriction being that Alice and
Bob can establish a one-to-one correspondence between the
transmitted and the received spins. Next, Bob measures the
incoming spins in one of the two bases, chosen at random (using a
random number generator independent from that of Alice). At this
point, whenever they used the same basis, they get perfectly
correlated results. However, whenever they used different basis,
they get uncorrelated results. Hence, on average, Bob obtains a
string of bits with 25\% errors, called the {\it raw key}. This
error rate is so large that standard error correction schemes
would fail. But in this protocol, as we shall see, Alice and Bob know which bits are
perfectly correlated (the ones for which Alice and Bob used the
same basis) and which ones are completely uncorrelated (all the
other ones). Hence, a straightforward error correction scheme is
possible: For each bit Bob announces publicly in which basis he
measured the corresponding qubit (but he does not tell the result
he obtained). Alice then only tells whether or not the state in
which she encoded that qubit is compatible with the basis
announced by Bob. If the state is compatible, they keep the bit,
if not they disregard it. In this way about 50\% of the bit string
is discarded. This shorter key obtained after bases reconciliation
is called the {\it sifted key}\footnote{This terminology has been
introduced by Ekert and Huttner in 1994.}. The fact that Alice and
Bob use a public channel at some stage of their protocol is very
common in crypto-protocols. This channel does not have to be
confidential, but has to be authentic. Hence, any adversary Eve
can listen to all the communication on the public channel, but
she can't modify it. In practice Alice and Bob may use the same
transmission channel to implement both the quantum and the classical
channels.

Note that neither Alice nor Bob can decide which key results from the protocol\footnote{Alice
and Bob can however determine the statistics of the key.}. Indeed, it is the conjunction of
both of their random choices which produces the key.

Let us now consider the security of the above ideal
protocol (ideal because so far we did not take into account unavoidable noise due
to technical imperfections). Assume that some adversary Eve intercepts a qubit propagating
from Alice to Bob. This is very easy, but if Bob does not receive an expected qubit, he will
simply inform Alice to disregard it. Hence, in this way Eve only lowers the bit rate
(possibly down to zero), but she does not gain any useful information. For real eavesdropping
Eve must send a qubit to Bob. Ideally she would like to send this qubit in its original
state, keeping a copy for herself.

\subsubsection{No cloning theorem}\label{NoQCM}
Following Wootters and Zurek (1982) it is easy
to prove that perfect copying is impossible in the quantum world
(see also Milonni and Hardies 1982, Dieks 1982,
and the anticipating intuition by Wigner in 1961). Let $\psi$ denote the
original state of the qubit, $|b\>$ the blank copy\footnote{$|b\>$ corresponds to the stock of white
paper in everyday's photocopy machine. We shall assume that exceptionally this stock is not empty,
a purely theoretical assumption, as is well known.}
and denote $|0\>\in \H_{QCM}$ the initial state of Eve's ``quantum copy
machine'', where the Hilbert space $\H_{QCM}$ of the quantum cloning machine is arbitrary.
The ideal machine would produce:
\beq
\psi\otimes|b\>\otimes|0\>\righta\psi\otimes\psi\otimes|f_\psi\>
\eeq
where $|f_\psi\>$ denotes the final state of Eve's machine which might depend on $\psi$.
Accordingly, using obvious notations,
\beqa
&&|\up,b,0\>\righta|\up,\up,f_\up\>  \\
and~~ &&|\down,b,0\>\righta|\down,\down,f_\down\>.
\eeqa
By linearity of quantum dynamics it follows that
\beqa
|\righta,b,0\>&=&\frac{1}{\sqrt{2}}(|\up\>+|\down\>)\otimes|b,0\>\\
&\righta&\frac{1}{\sqrt{2}}(|\up,\up,f_\up\>+|\down,\down,f_\down\>).
\eeqa
But the latter state differs from the ideal
copy $|\righta,\righta,f_\righta\>$, whatever the states $|f_\psi\>$ are.

Consequently,
Eve can't keep a perfect quantum copy, because perfect quantum copy machines can't exist.
The possibility to copy classical information is probably one of the most characteristic features
of information in the every day sense. The fact that quantum states, nowadays often called
quantum information, can't be copied is certainly one of the most specific attributes
which make this new kind of information so different, hence so attractive. Actually, this
``negative rule'' has clearly its positive
side, since it prevents Eve from perfect eavesdropping, and hence makes QC
potentially secure.

\subsubsection{Intercept-resend strategy}\label{IRstrat}
We have seen that the eavesdropper needs to send a qubit to Bob, while keeping a
necessarily imperfect copy for herself. How imperfect the copy has to be, according to
quantum theory, is a delicate problem that we shall address in chapter \ref{Eve}. Here,
let us develop a simple eavesdropping strategy, called intercept-resend.
This simple and even practical attack consists in Eve
measuring each qubit in one of the two basis, precisely as Bob does. Then,
she resends to Bob another qubit in the state corresponding to her measurement result.
In about half of the cases Eve will be lucky and choose the basis compatible with the
state prepared by Alice. In these cases she resends to Bob a qubit in the correct state and
Alice and Bob won't notice her intervention. However, in the other 50\% cases, Eve unluckily
uses the basis incompatible with the state prepared by Alice. This necessarily happens,
since Eve has no information on Alice's random generator (hence the importance that this
generator is truly random). In these cases the qubits sent out by Eve are in states
with overlap $\half$ with the correct states. Alice and Bob discover thus her intervention
in about half of these cases, since they get uncorrelated results.
Altogether, if Eve uses this intercept-resend strategy, she gets 50\% information, while
Alice and Bob have about 25\% of errors in their sifted key, i.e. after they eliminated
the cases in which they used incompatible states, there are still about 25\% errors.
They can thus easily detect the presence of Eve. If, however, Eve applies this strategy
to only a fraction of the communication, 10\% let's say, then the error rate will be
only $\approx$2.5\% while Eve's information would be $\approx$5\%. The next section explains
how Alice and Bob can counter such attacks.

\subsubsection{Error correction, privacy amplification and quantum secret growing}\label{ECPA}
At this point in the BB84 protocol, Alice and Bob share a so-called sifted key.
But this key contains errors. The errors are caused as well by technical imperfections, as
possibly by Eve's intervention.
Realistic error rates on the sifted key using today's technology are of a few percent. This
contrasts strongly with the $10^{-9}$ typical in optical communication. Of course, the
few percent errors will be corrected down to the standard $10^{-9}$ during the (classical)
error correction step of the protocol. In order to avoid confusion, especially among
the optical communication specialists, Beat Perny from Swisscom and Paul Townsend,
then with BT, proposed to name the error
rate on the sifted key QBER, for Quantum Bit Error Rate, to make it clearly distinct
from the BER used in standard communications.

Such a situation where the legitimate partners share
classical information, with high but not 100\% correlation and with possibly some
correlation to a third party is common to all quantum cryptosystems. Actually, it is also
a standard starting point for classical information based cryptosystems
where one assumes that somehow Alice,
Bob and Eve have random variables $\alpha$, $\beta$ and $\epsilon$, respectively, with joint probability
distribution $P(\alpha,\beta,\epsilon)$. Consequently, the last step in a QC protocol
uses classical algorithms, first to correct the errors, next to lower Eve's information
on the final key, a process called {\it privacy amplification}.

The first mention of privacy amplification appears in Bennett, Brassard and Robert
(1988). It was then extended in collaboration with C. Cr\'epeau and U. Maurer from the
University of Montreal and the ETH Z\"urich, respectively
(Bennett {\it et al.} 1995, see also Bennett {\it et al.} 1992a). Interestingly, this work
motivated by QC found applications in standard information-based
cryptography (Maurer 1993, Maurer and Wolf 1999).

Assume that such a joint probability distribution $P(\alpha,\beta,\epsilon)$
exists. Near the end of this section, we comment on this assumption.
Alice and Bob have access only to the marginal distribution  $P(\alpha,\beta)$. From
this and from the laws of quantum mechanics, they have to deduce constraints on the
complete scenario $P(\alpha,\beta,\epsilon)$,
in particular they have to bound Eve's information (see sections \ref{SymAttack}
and \ref{SecProofs}). Given
$P(\alpha,\beta,\epsilon)$, necessary and sufficient
conditions for a positive secret key rate between Alice and Bob,
$S(\alpha,\beta||\epsilon)$, are not yet known. However, a useful lower bound
is given by the difference
between Alice and Bob's mutual Shannon information $I(\alpha,\beta)$
and Eve's mutual information (Csisz\'{a}r and K\"orner 1978, and theorem 1
in section \ref{SecProofs}):
\beq
S(\alpha,\beta||\epsilon)\ge\max\{I(\alpha,\beta)-I(\alpha,\epsilon), I(\alpha,\beta)-I(\beta,\epsilon)\}
\label{Smin}
\eeq
Intuitively, this result states that secure key distillation (Bennett {\it et al.} 1992a) is possible
whenever Bob has more information than Eve.

The bound (\ref{Smin}) is tight if Alice and Bob are restricted to one-way communication,
but for two-way communication, secret key agreement might be possible even when (\ref{Smin})
is not satisfied (see next paragraph \ref{AdvDist}).

Without discussing any algorithm in  detail, let us give some intuition how Alice and Bob can
establish a secret key when condition (\ref{Smin}) is satisfied. First,
once the sifted key is obtained (i.e. after the bases have been announced),
Alice and Bob publicly compare a randomly chosen subset
of it. In this way they estimate the error rate (more generally, they estimate their marginal
probability distribution $P(\alpha,\beta)$). These publicly disclosed bits are then
discarded. Next, either condition (\ref{Smin}) is
not satisfied and they stop the protocol. Or condition (\ref{Smin}) is satisfied and they
use some standard error correction protocol to get a shorter key without errors.

With the simplest error correction protocol, Alice randomly chooses pairs of bits and announces their
XOR value (i.e. their sum modulo 2). Bob replies either ``accept'' if he has the same XOR value
for his corresponding bits, or ``reject'' if not. In the first case, Alice and Bob keep
the first bit of the pair and eliminate the second one, while in the second case they eliminate
both bits. In reality, more complex and efficient algorithms are used.

After error correction, Alice and Bob have identical copies of a key,
but Eve may still have some
information about it (compatible with condition (\ref{Smin})). Alice and Bob thus need to
lower Eve's information down to an arbitrarily low value using some privacy amplification protocols.
These classical protocols typically work as follows. Alice again randomly choses
pairs of bits and computes their XOR value. But, contrary to error correction she does not
announce this XOR value. She only announces which bits she chose (e.g. bit number 103 and 537).
Alice and Bob then replace the two bits by their XOR value. In this way they shorten their
key while keeping it error free, but if Eve has only partial
information on the two bits, her information on the XOR value is even lower. Consider for
example that Eve knows only the value of the first bit, and nothing about the second one. Then
she has no information at all on the XOR value. Also, if Eve knows the value of both bits with
60\% probability, then the probability that she guesses correctly the value of the XOR is
only of $0.6^2+0.4^2=52$\%. This process would have to be repeated several times;
more efficient algorithms use larger blocks (Brassard and Salvail 1993).

The error correction and privacy amplification algorithms sketched above are purely
classical algorithms. This illustrates that QC is a truly
interdisciplinary field.

Actually, the above presentation is incomplete. Indeed, in this presentation, we have
assumed that Eve has measured her probe before Alice and Bob run the error correction
and privacy amplification algorithms, hence that $P(\alpha,\beta,\epsilon)$ exists.
In practice this is a very reasonable assumption,
but, in principle, Eve could wait until the end of all the protocol, and then optimize her
measurements accordingly. Such ``delayed choice eavesdropping strategies\footnote{Note however that Eve
has to choose the interaction between her probe and the qubits before the public discussion
phase of the protocol.}'' are discussed in chapter \ref{Eve}.

It should now be clear that QC does not provide a complete
solution for all cryptographic purposes\footnote{For a while it was thought that {\it bit
commitment} (see, e.g., Brassard 1988), a powerful primitive in cryptology, could be realized using quantum principles.
However, Dominic Mayers (1996a and 1997) and Lo and Chau (1998) proved it to be impossible
(see also Brassard {\it et al.} 1998).}.
Actually, quite on the contrary, QC can only be
used as a complement to standard symmetrical cryptosystems. Accordingly, a more precise name for QC is {\it
Quantum Key Distribution}, since this is all QC does. Nevertheless, we prefer to
keep the well known terminology which gives its title to this review.

Finally, let us emphasize that every key distribution system must incorporate some authentification scheme:
the two parties must identify themselves. If not, Alice could actually be communicating
directly with Eve! A straightforward possibility is that Alice and Bob initially share
a short secret. Then QC provides them with a longer one and, for example, they each keep
a small portion for authentification at the next session (Bennett {\it et al.} 1992a).
From this perspective, QC is a {\it Quantum Secret Growing} protocol.

\subsubsection{Advantage distillation}\label{AdvDist}
QC has triggered and still triggers research in classical information
theory. The best known example is probably the development of privacy amplification algorithms
(Bennett {\it et al.} 1988 and 1995). This in turn triggered the development of new cryptosystems
based on weak but classical signals, emitted for instance by satellites (Maurer 1993)%
\footnote{Note that
here the confidentiality is not guaranteed by the laws of physics, but relies on the assumption that
Eve's technology is limited, e.g. her antenna is finite, her detectors have limited efficiencies.}.
These new developments required secret key agreement protocols that can be used even when the
condition (\ref{Smin}) doesn't apply. Such protocols, called {\it advantage distillation},
necessarily use two way communication and are much less efficient than privacy amplification. Usually,
they are not considered in the literature on QC. But, conceptually, they are remarkable from
at least two points of view. First it is somewhat surprising that secret key agreement is
possible even if Alice and Bob start with less mutual (Shannon) information than Eve. However,
they can take advantage of the authenticated public channel: Alice and Bob can decide which
series of realization to keep, whereas Eve can't influence this process\footnote{The idea is
that Alice picks out several instances where she got the same bit and communicates the
instances - but not the bit - to Bob. Bob replies yes only if it happens that for all these
instances he also has the same bit value. For large error rates this is unlikely, but when
it happens there is a large chance that both have the same bit. Eve can't influence the choice
of the instances. All she can do is
to use a majority vote for the cases accepted by Bob. The probability that Eve makes an error
can be much larger than the probability that Bob makes an error (i.e. that all his instances
are wrong), even if Eve's initial information is larger than Bob's.} (Maurer 1993,
Maurer and Wolf 1999).

Recently a second remarkable connection between quantum and classical secret key agreement
has been discovered (assuming they use the Ekert
protocol described in paragraph \ref{EkertProtocol}): If Eve follows the strategy which optimizes her Shannon information,
under the assumption that she attacks the qubit one at a time (the so-called individual
attacks, see section \ref{SymAttack}), then Alice and Bob can use advantage distillation
if and only if Alice and Bob's qubits are still entangled (they can thus use quantum privacy
amplification (Deutsch {\it et al.} 1996)) (Gisin and Wolf 1999). This connection
between the concept of {\it entanglement}, central to quantum information theory, and
the concept of {\it intrinsic classical information}, central to classical information
based cryptography (Maurer and Wolf 1999),
has been shown to be general (Gisin and Wolf 2000). The connection seems
even to extend to {\it bound entanglement} (Gisin {\it et al.} 2000).

\subsection{Other protocols}\label{OtherProtocols}
\subsubsection{2-state protocol}\label{2state}
In 1992 Charles H. Bennett noticed that actually 4 states is more than necessary for
QC: all what is really needed is 2 nonorthogonal states. Indeed the security
relies on the impossibility for any adversary to distinguish unambiguously and without
perturbation between the different states that Alice may send to Bob, hence 2 states are
necessary and if they are incompatible (i.e. not mutually orthogonal), then 2 states are
also sufficient. This is a conceptually important clarification. It also made several
of the first experimental demonstrations easier (this is further discussed in
section \ref{Besancon}). But in practice it is not a good solution.
Indeed, although 2 nonorthogonal states can't be distinguished unambiguously without
perturbation, one can unambiguously distinguish them at the cost of some losses
(Ivanovic 1987, Peres 1988).
This possibility has even been demonstrated in practice (Huttner {\it et al.} 1996,
Clarke {\it et al.} 2000). Hence, Alice and Bob would have to monitor the attenuation of the
quantum channel (and even this is not entirely safe if Eve could replace the channel
by a more transparent one, see section \ref{QND}).
The two-state protocol can also be implemented using an interference between a macroscopic bright
pulse and a dim pulse with less than one photon on average (Bennett, 1992).
The presence of the bright pulse makes
this protocol specially resistant to eavesdropping, even in settings with high attenuation.
Indeed Bob can monitor the bright pulses, to make sure that Eve does not remove any. In this case,
Eve cannot eliminate the dim pulse without revealing her presence, because the interference of
the bright pulse with vacuum would introduce errors.
A practical implementation of this protocol
is discussed in section \ref{Besancon}. Huttner {\it et al.} extended this reference beam monitoring to
the four-states protocol in 1995.

\subsubsection{6-state protocol}\label{6state}
\begin{figure}
\infig{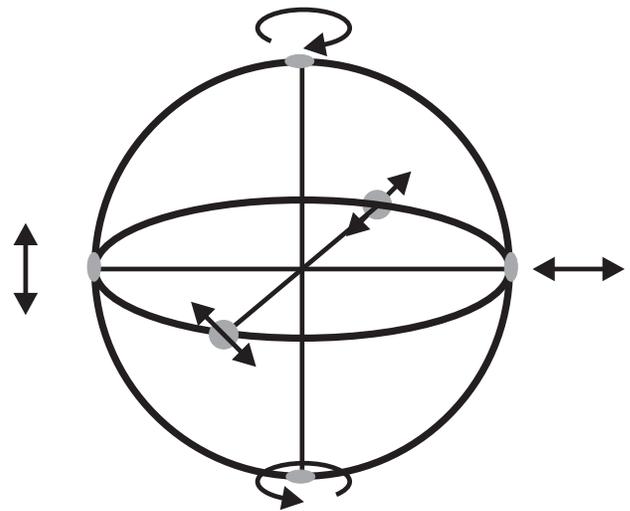}{0.95\columnwidth}
\caption{Poincar\'{e}
sphere with a representation of six states that can be used to
implement the generalization of the BB84 protocol.} \label{fig2_2}
\end{figure}

While two states are enough and four states are standard, a
6-state protocol respects much more the symmetry of the qubit
state space, see Fig. \ref{fig2_2} (Bruss 1998,
Bechmann-Pasquinucci and Gisin 1999). The 6 states constitute 3
bases, hence the probability that Alice and Bob chose the same
basis is only of $\frac{1}{3}$. But the symmetry of this protocol
greatly simplifies the security analysis and reduces Eve's optimal
information gain for a given error rate QBER.
If Eve measures every photon, the QBER is 33\%, compared to 25\% in the case of the BB84 protocol.

\subsubsection{EPR protocol}\label{EkertProtocol}
This variation of the BB84 protocol is of special conceptual, historical and practical
interest. The idea is due to Artur Ekert (1991) from Oxford University, who, while elaborating on
a suggestion of David Deutsch (1985), discovered QC independently of the BB84 paper.
Intellectually, it is very satisfactory to see this direct connection to the famous
EPR paradox (Einstein, Podolski and Rosen 1935): the initially philosophical debate turned to
theoretical physics with Bell's inequality (1964), then to experimental physics
(Freedmann and Clauser 1972, Fry and Thompson 1976, and Aspect, Dalibard and Roger 1982),
and is now -- thanks to Ekert's ingenious idea -- part of applied physics.

\begin{figure}
\infig{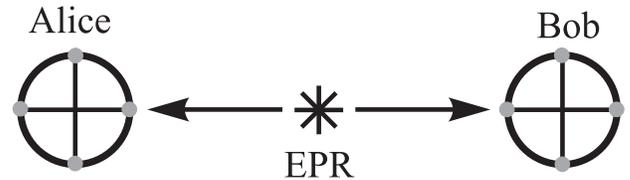}{0.95\columnwidth}
\caption{EPR protocol,
with the source and a Poincar\'{e} representation of the four
possible states measured independently by Alice and Bob.}
\label{fig2_3}
\end{figure}

The idea consists in replacing the quantum channel carrying qubits
from Alice to Bob by a channel carrying 2 qubits from a common
source, one qubit to Alice and one to Bob. A first possibility
would be that the source emits the two qubits always in the same
state chosen randomly among the 4 states of the BB84 protocol.
Alice and Bob would then both measure their qubit in one of the
two bases, again chosen independently and randomly. The source then announces the
bases and Alice and Bob keep the data only when they happen to have done their
measurements in the compatible basis. If the source
is reliable, this protocol is equivalent to the BB84 one: Every
thing is as if the qubit propagates backwards in time from Alice
to the source, and then forwards to Bob! But better than trusting
the source, which could be in Eve's hand, the Ekert protocol
assumes that the 2 qubits are emitted in a maximally entangled
state like: \beq
\phi^+=\frac{1}{\sqrt{2}}(|\up,\up\>+|\down,\down\>).
\label{entState} \eeq Then, when Alice and Bob happen to use the
same basis, both the x-basis or both the y-basis, i.e. in about half of the cases,
their results are
identical, providing them with a common key. Note the similarity
between the 1-qubit BB84 protocol illustrated in Fig. \ref{fig2_1}
and the 2-qubit Ekert protocol of Fig. \ref{fig2_3}.
The analogy can be even made stronger by noting that for all unitary evolutions
$U_1$ and $U_2$, the following equality hold:
\beq
U_1\otimes U_2\Phi^{(+)}=\opone\otimes U_2U_1^t\Phi^{(+)}
\eeq
where $U_1^t$ denotes the transpose.

\begin{figure}
\infig{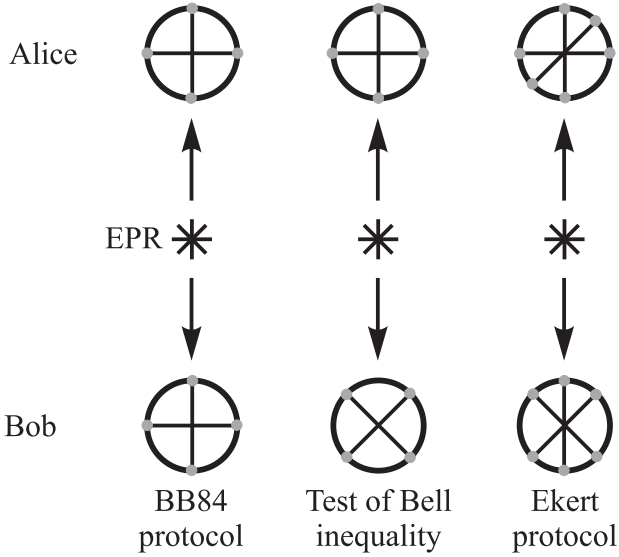}{0.95\columnwidth}
\caption{Illustration
of protocols exploiting EPR quantum systems. To implement the BB84
quantum cryptographic protocol, Alice and Bob use the same bases
to prepare and measure their particles. A representation of their
states on the Poincar\'{e} sphere is shown. A similar setup, but
with Bob's bases rotated by $45^{\circ}$, can be used to test the
violation of Bell inequality. Finally, in the Ekert protocol,
Alice and Bob may use the violation of Bell inequality to test for
eavesdropping.} \label{fig2_4}
\end{figure}

In his 1991 paper Artur Ekert suggested to base the security of
this 2-qubit protocol on Bell's inequality, an inequality which
demonstrates that some correlation predicted by quantum mechanics
can't be reproduced by any local theory (Bell 1964). For this,
Alice and Bob have a third choice of basis (see Fig. \ref{fig2_4}).
In this way the probability that they happen to choose the same
basis is reduced from $\half$ to $\frac{2}{9}$, but at the same
time as they establish a key they collect enough data to test Bell
inequality\footnote{A maximal violation of Bell inequality is necessary to rule out tampering by Eve. In this case, the QBER must necessarily be equal to zero.
With a non-maximal violation, as typically obtained in experimental systems, Alice and Bob can distil a secure key using error correction and privacy amplification.}.
They can thus check that the source really emits the
entangled state (\ref{entState}) and not merely product states.
The following year Bennett, Brassard and Mermin (1992b) criticized
Ekert's letter, arguing that the violation of Bell inequality is
not necessary for the security of QC and emphasizing the close
connection between the Ekert and the BB84 schemes. This criticism
might be missing an important point. Indeed, although the exact
relation between security and Bell inequality is not yet fully
known, there are clear results establishing fascinating
connections, (see section \ref{Bell}). In October 1992, an article
by Bennett, Brassard and Ekert demonstrated that the founding
fathers joined forces to develop the field in a pleasant
atmosphere (Bennett {\it et al.} 1992c)!

\subsubsection{Other variations}
There is a large collection of variations around the BB84 protocol. Let us mention a few,
chosen somewhat arbitrarily. First, one can assume that the two bases are not
chosen with equal probability (Ardehali {\it et al.} 1998). This has the nice consequence that the probability that
Alice and Bob choose the same basis is larger than $\half$, increasing thus the transmission
rate of the {\it sifted key}. However, this protocol makes Eve's job easier as she is more likely to guess
correctly the used basis. Consequently, it is not clear whether the final key rate, after error
correction and privacy amplification, is higher or not.

Another variation consists in using quantum systems of dimension larger than 2
(Bechmann-Pasquinucci and Tittel 2000, Bechmann-Pasquinucci and Peres 2000, Bourennane {\it et al.} 2001a). Again, the
practical value of this idea has not yet been fully determined.

A third variation worth mentioning is due to Goldenberg and Vaidman, from Tel-Aviv
University (1995). They suggested
to prepare the qubits in a superposition of two spatially separated states, then to send
one component of this superposition and to wait until Bob received it before sending the
second component. This doesn't sound of great practical value, but has the nice conceptual
feature that the minimal two states do not need to be mutually orthogonal.

\subsection{Quantum teleportation as ``Quantum one-time-pad''}\label{QTelep}
Since its discovery in 1993 by a surprisingly large group of physicists,
Quantum teleportation (Bennett {\it et al.} 1993) received
a lot of attention in the scientific community as well as in the general public. The dream
of beaming travellers through the Universe is exciting, but completely out of the realm
of any foreseeable technology. However, quantum teleportation can be seen as the fully
quantum version of the one-time-pad, see paragraph \ref{CTelep}, hence as the ultimate
form of QC. Similarly to ``classical
teleportation'', let's assume that Alice aims at transferring to Bob a faithful copy of
a quantum system. If Alice has full knowledge of the quantum state, the problem is not
really a quantum one (Alice information is classical). If, on the opposite, Alice does not
know the quantum state, she cannot send a copy, since quantum copying is impossible
according to quantum physics (see paragraph \ref{NoQCM}). Nor can she send classical instructions,
since this would allow the production of many copies. However, if Alice and Bob share
arbitrarily many entangled qubits, sometimes called a quantum key, and share a
classical communication channel then the quantum
teleportation protocol provides them with a mean to transfer the quantum state of the
system from Alice to Bob. In the course of running this protocol, Alice's quantum system is destroyed
without Alice learning anything about the quantum state, while Bob's qubit ends in a state
isomorphic to the state of the original system (but Bob doesn't learn anything about the
quantum state). If the initial quantum system is a
quantum message coded in the form of a sequence of qubits, then this quantum message
is faithfully and securely transferred to Bob, without any information leaking to the
outside world (i.e. to anyone not sharing the prior entanglement with Alice and Bob).
Finally, the quantum message could be formed of a 4 letter quantum alphabet constituted
by the 4 states of the BB84 protocol. With futuristic, but not impossible technology, Alice
and Bob could have their entangled qubits in appropriate wallets and could establish
a totally secure communication at any time, without even having to know where the partner
is located (provided they can communicate classically).

\subsection{Optical amplification, quantum nondemolition measurements
and optimal quantum cloning}\label{Ampli}
After almost every general talk on QC, two questions arise: what about
optical amplifiers? and what about quantum nondemolition measurements? In this section
we briefly address these questions.

Let us start with the second one, being the easiest. The terminology ``quantum
nondemolition measurement'' is simply a confusing one! There is nothing like a
quantum measurement that does not perturb (i.e. modify) the quantum state,
except if the state happens to be an eigenstate of the observable. Hence, if for some
reason one conjectures that a quantum system is in some state (or in a state among a set of mutually
orthogonal ones), this can be in principle tested
repeatedly (Braginsky and Khalili 1992). But if the state is only restricted to be in a finite set containing non-orthogonal states,
as in QC, then there is no way to
perform a measurement without ``demolishing'' (perturbing) the state.
Now, in QC the terminology ``nondemolition measurement'' is also used with a different meaning:
one measures the number of photons in a pulse without affecting the degree of freedom coding
the qubit (e.g. the polarization), (see section \ref{QND}), or one detects the presence of a
photon without destroying it (Nogues {\it et al.} 1999).
Such measurements are usually called ``ideal measurements'', or ``projective
measurements'', because they produce the least possible perturbation (Piron 1990) and because they
can be represented by projectors.
It is important to stress that these ``ideal measurements'' do not invalidate the security of QC.

Let us consider now optical amplifiers (a laser medium, but without
mirrors, so that amplification takes place in a single pass, see Desurvire 1994). They
are widely used in today's optical
communication networks. However, they are of no use for quantum communication. Indeed, as
seen in section \ref{BB84}, the copying of quantum information is impossible. Here we illustrate
this characteristic of quantum information with the example of
optical amplifiers: the necessary presence of spontaneous emission whenever there
is stimulated emission, prevents perfect copying. Let us clarify this important and often confusing point,
following the work of Simon  {\it et al.} (1999 and 2000;
see also Kempe {\it et al.} 2000, and De Martini {\it et al.} 2000).
Let the two basic qubit states $|0\rangle$ and $|1\rangle$ be physically implemented by two
optical modes: $|0\rangle\equiv|1,0\rangle$ and $|1\rangle\equiv|0,1\rangle$.
$|n,m\>_{ph}\otimes|k,l\>_a$ denotes thus the state of n photons in mode 1 and m in mode 2, and
$k,l=0~ (1)$ the ground (excited) state of 2-level atoms coupled to mode 1 and 2,
respectively. Hence spontaneous emission corresponds to
\beqa
\label{spontem}
|0,0\>_{ph}\otimes|1,0\>_a &\righta & |1,0\>_{ph}\otimes|0,0\>_a,\\
|0,0\>_{ph}\otimes|0,1\>_a &\righta & |0,1\>_{ph}\otimes|0,0\>_a
\eeqa
and stimulated emission to
\beqa
|1,0\>_{ph}\otimes|1,0\>_a &\righta & \sqrt{2}|2,0\>_{ph}\otimes|0,0\>_a,\\
|0,1\>_{ph}\otimes|0,1\>_a &\righta & \sqrt{2}|0,2\>_{ph}\otimes|0,0\>_a
\label{stimem}
\eeqa
where the $\sqrt{2}$ factor takes into account the ratio stimulated/spontaneous
emission.
Let the initial state of the atom be a mixture of the following two states
(each with equal weight 50\%):
\beq
|0,1\>_a \hspace{1 cm} |1,0\>_a
\eeq
By symmetry, it suffices to consider one possible initial
state of the qubit, e.g. 1 photon in the first mode $|1,0\>_{ph}$. The initial state
of the photon+atom system is thus a mixture:
\beq
|1,0\>_{ph}\otimes|1,0\>_a \hspace{0.5 cm} or \hspace{0.5 cm} |1,0\>_{ph}\otimes|0,1\>_a
\eeq
This corresponds to the first order term in an evolution with a Hamiltonian (in the
interaction picture):
$H=\chi(a_1^\dagger\sigma_1^-+a_1\sigma_1^\dagger + a_2^\dagger\sigma_2^-+a_2\sigma_2^\dagger)$.
After some time the 2-photon component of the evolved states reads:
\beq
\sqrt{2}|2,0\>_{ph}\otimes|0,0\>_a \hspace{0.5 cm} or \hspace{0.5 cm} |1,1\>_{ph}\otimes|0,0\>_a
\eeq
The correspondence with a pair of spin $\half$ goes as follows:
\beq
|2,0\rangle=|\up\up\rangle \hspace{1cm} |0,2\rangle=|\down\down\rangle
\eeq
\beq
|1,1\>_{ph}=\psi^{(+)}=\frac{1}{\sqrt{2}}\left(|\up\down\>+|\down\up\>\right)
\eeq
Tracing over the amplifier (i.e. the 2-level atom),
an (ideal) amplifier achieves the following transformation:
\beq
P_\up \righta 2P_{\up\up}+P_{\psi^{(+)}}
\eeq
where the $P$'s indicate projectors (i.e. pure state density matrices) and the lack of
normalization results from the first order expansion used in (\ref{spontem}) to (\ref{stimem}).
Accordingly, after normalization, each photon is in state :
\beq
Tr_{1-ph \; mode}\left(\frac{2P_{\up\up}+P_{\psi^{(+)}}}{3}\right)=\frac{2P_{\up}+\half\opone}{3}
\eeq
The corresponding fidelity is:
\beq
F=\frac{2+\half}{3}=\frac{5}{6}
\eeq
which is precisely the optimal fidelity compatible with quantum mechanics
(Bu\v{z}ek and Hillery 1996, Bruss et al 1998, Gisin and Massar 1997).
In other words, if we start with a single photon in an arbitrary state, and pass it through
an amplifier, then due to the effect of spontaneous emission the fidelity of the state exiting
the amplifier, in the cases where it consists of exactly two photons, with the initial state will be equal to at most 5/6.
Note that if it were possible to
make better copies, then, using EPR correlations between spatially separated systems,
signaling at arbitrarily fast speed would also be possible (Gisin 1998).

\newpage
\section{Technological challenges}\label{TechChallenges}
The very first demonstration of QC was a table top experiment
performed at the IBM laboratory in the early 1990's over a
distance of 30 cm (Bennett {\it et al.} 1992a), marking the start of
impressive experimental improvements during the last years.
The 30 cm distance is of little practical interest. Either the
distance should be even shorter, think of a credit card and the
ATM machine (Huttner {\it et al.} 1996b), but in this case all of
Alice's components should fit on the credit card. A nice idea, but
still impractical with present technology. Or the distance should be much longer, at least
in the km range. Most of the research so far uses optical fibers
to guide the photons from Alice to Bob and we shall mainly
concentrate here on such systems. There is, however, also some
very significant research on free space systems, (see section
\ref{Satellite}).

Once the medium is chosen,
there remain the questions of the source and detectors. Since they
have to be compatible, the crucial choice is the wavelength. There
are two main possibilities. Either one chooses a wavelength around
800 nm where efficient photon counters are commercially available,
or one chooses a wavelength compatible with today's
telecommunication optical fibers, i.e. near 1300 nm or 1550 nm.
The first choice requires free space transmission or the use of
special fibers, hence the installed telecommunication networks
can't be used. The second choice requires the improvement or
development of new detectors, not based on silicon semiconductors,
which are transparent above 1000 nm wavelength.

In case of transmission using optical fibers, it is still unclear
which of the two alternatives will turn out to be the best choice.
If QC finds niche markets, it is conceivable that special fibers
will be installed for that purpose. But it is equally conceivable
that new commercial detectors will soon make it much easier to
detect single photons at telecommunication wavelengths. Actually,
the latter possibility is very likely, as several research groups
and industries are already working on it. There is another good
reason to bet on this solution: the quality of telecommunication
fibers is much higher than that of any special fiber, in
particular the attenuation is much lower (this is why the
telecommunication industry chose these wavelengths): at 800 nm,
the attenuation is about 2 dB/km (i.e. half the photons are lost
after 1.5 km), while it is only of the order of 0.35 and 0.20
dB/km at 1300 nm and 1550 nm, respectively (50\% loss after about
9 and 15 km) \footnote{ The losses in dB ($l_{db}$) can be
calculated from the losses in percent ($l_{\%}$): $l_{dB}=-10
\log_{10}(1-\frac{l_\%}{100})$.}.

In case of free space transmission, the choice of wavelength is
straightforward since the region where good photon detectors exist
-- around 800 nm -- coincides with the one where absorption is
low. However, free space transmission is restricted to line-of
sight links and is very weather dependent.

In the next sections we successively consider the questions ``how
to produce single photons?'' (section \ref{PhotonSources}), ``how to
transmit them?'' (section \ref{QChannels}), ``how to detect single
photons?'' (section \ref{PhotonCounting}), and finally ``how to
exploit the intrinsic randomness of quantum processes to build
random generators?'' (section \ref{QRNG}).

\subsection{Photon sources}\label{PhotonSources}
Optical quantum cryptography is based on the use of single photon
Fock states. Unfortunately, these states are difficult to realize
experimentally. Nowadays, practical implementations rely on faint
laser pulses or entangled photon pairs, where both the photon as
well as the photon-pair number distribution obeys Poisson
statistics. Hence, both possibilities suffer from a small
probability of generating more than one photon or photon pair at
the same time. For large losses in the quantum channel even small
fractions of these multi-photons can have important consequences
on the security of the key (see section \ref{QND}), leading to
interest in ``photon guns'', see paragraph \ref{PhotonGun}). In
this section we briefly comment on sources based on faint pulses
as well as on entangled photon-pairs, and we compare their
advantages and drawbacks.

\subsubsection{Faint laser pulses}\label{WeakPulse}
There is a very simple solution to approximate single photon Fock
states: coherent states with an ultra-low mean photon number $\mu
$. They can easily be realized using only standard semiconductor
lasers and calibrated attenuators. The probability to find $n$
photons in such a coherent state follows the Poisson statistics:
\begin{equation}
P(n,\mu )=\frac{\mu ^{n}}{n!}e^{-^{\mu }}.
\end{equation}
Accordingly, the probability that a non-empty weak coherent pulse
contains more than 1 photon,
\beqa
P(n>1|n>0,\mu)&=&\frac{1-P(0,\mu)-P(1,\mu)}{1-P(0,\mu)} \nonumber \\
&=&\frac{1-e^{-\mu }(1+\mu )}{1-e^{-\mu }}%
\cong \frac{\mu }{2}
\eeqa
can be made arbitrarily small. Weak pulses are thus extremely
practical and have indeed been used in the vast majority of
experiments. However, they have one major drawback. When $\mu $ is
small, most pulses are empty: $P(n=0)\approx 1-\mu $. In
principle, the resulting decrease in bit rate could be compensated
for thanks to the achievable GHz modulation rates of
telecommunication lasers. But in practice the problem comes from
the detectors' dark counts (i.e. a click without a photon
arriving). Indeed, the detectors must be active for all pulses,
including the empty ones. Hence the total dark counts increase
with the laser's modulation rate and the ratio of the detected
photons over the dark counts (i.e. the signal to noise ratio)
decreases with $\mu $ (see section
\ref{SQBER}). The problem is especially severe for longer
wavelengths where photon detectors based on Indium Gallium
Arsenide semiconductors (InGaAs) are needed (see section
\ref{PhotonCounting}) since the noise of these detectors explodes
if they are opened too frequently (in practice with a rate larger
than a few MHz). This prevents the use of really low photon
numbers, smaller than approximately 1\%. Most experiments to date
relied on $\mu = 0.1$, meaning that 5\% of the nonempty pulses
contain more than one photon. However, it is important to stress that, as pointed out by
L\"utkenhaus (2000), there is an optimal $\mu $ depending on the
transmission losses
\footnote{Contrary to a frequent misconception, there is nothing special about a $\mu$ value of 0.1, eventhough
it has been selected by most experimentalists. The optimal value -- i.e. the value that yields the highest key exchange rate after distillation -- depends on the optical losses in the channel and on assumptions
about Eve's technology (see \ref{QND} and \ref{beamsplitter}).}.
After key distillation, the security is just as good with
faint laser pulses as with Fock states. The price to pay for using such states lies in a reduction of the bit rate.

\subsubsection{Photon pairs generated by parametric downconversion}\label{PDC}
Another way to create pseudo single-photon states is the
generation of photon pairs and the use of one photon as a trigger
for the other one (Hong and Mandel 1986). In contrast to the
sources discussed before, the second detector must be activated
only whenever the first one detected a photon, hence when $\mu$ =
1, and not whenever a pump pulse has been emitted, therefore
circumventing the problem of empty pulses.

The photon pairs are generated by spontaneous parametric down
conversion in a $\chi ^{(2)}$ non-linear crystal\footnote{ For a
review see Rarity and Tapster 1988, and for latest developments
Tittel {\it et al.} 1999, Kwiat {\it et al.} 1999, Jennewein {\it et al.} 2000b,
Tanzilli {\it et al.} 2001.}. In this process, the inverse of the
well-known frequency doubling, one photon spontaneously splits
into two daughter photons -- traditionally called signal and idler
photon -- conserving total energy and momentum. In this context,
momentum conservation is called phase matching, and can be
achieved despite chromatic dispersion by exploiting the
birefringence of the nonlinear crystal. The phase matching allows
to choose the wavelength, and determines the bandwidth of the
downconverted photons. The latter is in general rather large and
varies from a few nanometers up to some tens of nanometers. For
the non degenerate case one typically gets 5-10 nm, whereas in the
degenerate case (central frequency of both photons equal) the bandwidth can be as large as 70 nm.

This photon pair creation process is very inefficient, typically
it needs some 10$^{10}$ pump photons to create one pair in a given
mode\footnote{Recently we achieved a conversion rate of 10$^{-6}$
using an optical waveguide in a periodically poled LiNbO$_{3}$
crystal (Tanzilli {\it et al.} 2001).}. The number of photon pairs per
mode is thermally distributed within the coherence time of the
photons, and follows a poissonian distribution for larger time
windows (Walls and Milburn 1995). With a pump power of 1 mW, about
10$^{6}$ pairs per second can be collected in single mode fibers.
Accordingly, in a time window of roughly 1ns the conditional
probability to find a second pair having detected one is
$10^{6}\cdot 10^{-9}\approx0.1\%$. In case of continuous pumping,
this time window is given by the detector resolution. Tolerating,
e.g. 1\% of these multi-pair events, one can generate 10$^{7}$
pairs per second, using a realistic 10 mW pump. Detecting for
example 10 \% of the trigger photons, the second detector has to
be activated 10$^{6}$ times per second. In comparison, the example
of 1\% of multi-photon events corresponds in the case of faint
laser pulses to a mean photon number of $\mu = 0.02$. In order to
get the same number 10$^{6}$ of non-empty pulses per second, a
pulse rate of 50 MHz is needed. For a given photon statistics,
photon pairs allow thus to work with lower pulse rates (e.g. 50
times lower) and hence reduced detector-induced errors. However,
due to limited coupling efficiency into optical fibers, the
probability to find the sister photon after detection of the
trigger photon in the respective fiber is in practice lower than
1. This means that the effective photon number is not one, but
rather $\mu \approx 2/3$ (Ribordy {\it et al.} 2001), still well above
$\mu = 0.02$.

Photon pairs generated by parametric down conversion offer a
further major advantage if they are not merely used as pseudo
single-photon source, but if their entanglement is exploited.
Entanglement leads to quantum correlations which can be used for
key generation, (see paragraph \ref{EkertProtocol} and chapter
\ref{QCPhotonPair}). In this case, if two photon pairs are emitted
within the same time window but their measurement basis is choosen
independently, they produce completely uncorrelated results.
Hence, depending on the realization, the problem of multiple
photon can be avoided, see section \ref{Passivebasis}.

\begin{figure}
\infig{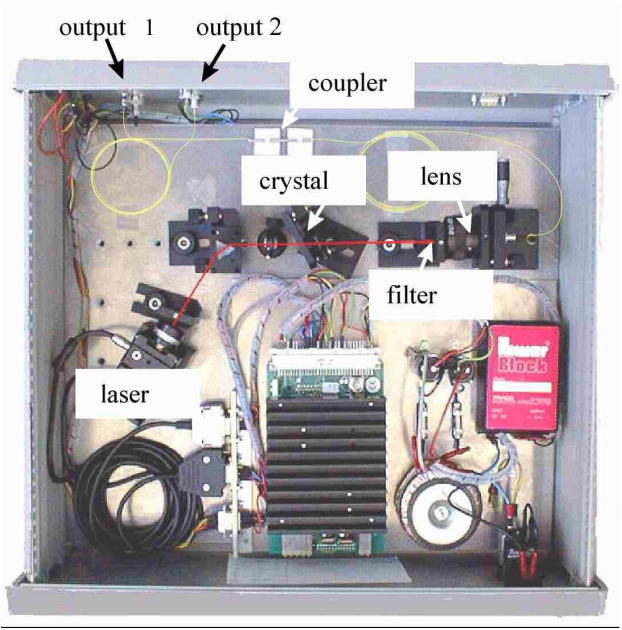}{0.95\columnwidth} \caption{Photo of our
entangled photon-pair source as used in the first long-distance
test of Bell inequalities (Tittel {\it et al.} 1998). Note that the
whole source fits in a box of only $40 \times 45 \times 15 cm^{3}$
size, and that neither special power supply nor water cooling is
necessary.} \label{fig3_1}
\end{figure}

Figure \ref{fig3_1} shows one of our sources creating entangled
photon pairs at 1310 nm wavelength as used in tests of Bell
inequalities over 10 kilometers (Tittel {\it et al.} 1998). Although not
as simple as faint laser sources, diode pumped photon pair sources
emitting in the near infrared can be made compact, robust and
rather handy.

\subsubsection{Photon guns}\label{PhotonGun}
The ideal single photon source is a device that when one pulls the
trigger, and only then, emits one and only one photon. Hence the
name photon gun. Although photon anti-bunching has been
demonstrated already years ago (Kimble {\it et al.} 1977), a practical
and handy device is still awaited. At present, there are
essentially three different experimental approaches that come more
or less close to this ideal.

A first idea is to work with a single two-level quantum system
that can obviously not emit two photons at a time. The
manipulation of single trapped atoms or ions requires a much too
involved technical effort. Single organics dye molecules in
solvents (S.C. Kitson {\it et al.} 1998) or solids (Brunel {\it et al.} 1999,
Fleury {\it et al.} 2000) are easier to handle but only offer limited
stability at room temperature. Promising candidates, however, are
nitrogen-vacancy centers in diamond, a substitutional nitrogen
atom with a vacancy trapped at an adjacent lattice position
(Kurtsiefer {\it et al.} 2000, Brouri {\it et al.} 2000). It is possible to
excite individual nitrogen atoms with a 532 nm laser beam, which
will subsequently emit a fluorescence photon around 700 nm (12ns
decay time). The fluorescence exhibits strong photon anti-bunching
and the samples are stable at room temperature. However, the big
remaining experimental challenge is to increase the collection
efficiency (currently about 0.1\%) in order to obtain mean photon numbers close to 1. To
obtain this, an optical cavity or a photonic bandgap structure
must suppress the emission in all spatial modes but one.
In addition, the spectral bandwith of this type of source is broad (of the order of 100 nm), enhancing the effect of
pertubations in a quantum channel.

A second approach is to generate photons by single electrons in a mesoscopic
p-n junction. The idea is to take profit of the fact that thermal electrons
show anti-bunching (Pauli exclusion principle) in contrast to photons
(Imamoglu and Yamamoto, 1994). First experimental results have been
presented (Kim {\it et al.} 1999), however with extremely low
efficiencies, and only at a temperature of 50mK!

Finally, another approach is to use the photon emission of
electron-hole pairs in a semiconductor quantum dot. The frequency
of the emitted photon depends on the number of electron-hole pairs
present in the dot. After one creates several such pairs by
optical pumping, they will sequentially recombine and hence emit
photons at different frequencies. Therefore, by spectral filtering
a single-photon pulse can be obtained (G\'{e}rard {\it et al.} 1999,
Santori {\it et al.} 2000, and  Michler {\it et al.} 2000). These dots can be
integrated in solid-states microcavities with strong enhancements
of the spontaneous emission (G\'{e}rard {\it et al.} 1998).

In summary, today's photon guns are still too complicated to be used in a
QC-prototype. Moreover, due to their low quantum efficiencies they do not offer
an advantage with respect to faint laser pulses with extremely low mean
photon numbers $\mu$.

\subsection{Quantum channels}\label{QChannels}
The single photon source and the detectors must be connected by a
``quantum channel''. Such a channel is actually nothing specially
quantum, except that it is intended to carry information encoded
in individual quantum systems. Here ``individual'' doesn't mean
``non-decomposible'', it is meant in opposition to ``ensemble''. The
idea is that the information is coded in a physical system only
once, contrary to classical communication where many photons carry
the same information. Note that the present day limit for
fiber-based classical optical communication is already down to a
few tens of photons, although in practice one usually uses many
more. With the increasing bit rate and the limited mean power --
imposed to avoid nonlinear effects in silica fibers -- these
figures are likely to get closer and closer to the quantum domain.

The individual quantum systems are usually 2-level systems, called
qubits. During their propagation they must be protected from
environmental noise. Here ``environment'' refers to everything
outside the degree of freedom used for the encoding, which is not
necessarily outside the physical system. If, for example, the
information is encoded in the polarization state, then the optical
frequencies of the photon is part of the environment. Hence,
coupling between the polarization and the optical frequency has to
be mastered\footnote{Note that, as we will see in chapter \ref{QCPhotonPair},
using entangled photons prevents such information leakage.}
(e.g. avoid wavelength sensitive polarizers and
birefringence). Moreover, the sender of the qubits
should avoid any correlation between the polarization and the
spectrum of the photons.

Another difficulty is that the bases used by Alice to code the
qubits and the bases used by Bob for his measurements must be
related by a known and stable unitary transformation. Once this
unitary transformation is known, Alice and Bob can compensate for
it and get the expected correlation between their preparations and
measurements. If it changes with time, they need an active
feedback to track it, and if the changes are too fast the
communication must be interrupted.

\subsubsection{Singlemode fibers}\label{smFibers}
\begin{figure}
\infig{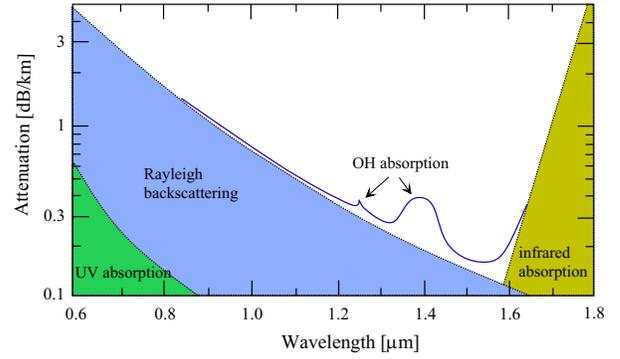}{0.95\columnwidth} \caption{Transmission
losses versus wavelength in optical fibers. Electronic transitions
in $SiO_2$ lead to absorption at lower wavelengths, excitation of
vibrational modes to losses at higher wavelength. Superposed is
the absorption due to Rayleigh backscattering and to transitions
in OH groups. Modern telecommunication is based on wavelength
around 1.3 $\mu$m (second telecommunication window) and around 1.5
$\mu$m (third telecommunication window).} \label{fig3_2}
\end{figure}

Light is guided in optical fibers thanks to the refractive index
profile $n(x,y)$ across the section of the fibers (traditionally,
the z-axis is along the propagation direction). Over the last 25
years, a lot of effort has been made to reduce transmission losses
-- initially several dB per km --, and nowadays, the attenuation
is as low as 2dB/km at 800nm wavelength, 0.35 dB/km at 1310 nm,
and 0.2 dB/km at 1550 nm (see Fig. \ref{fig3_2}). It is amusing to
note that the dynamical equation describing optical pulse
propagation (in the usual slowly varying envelope aproximation) is
identical to the Schr\"odinger equation, with $V(x,y)=-n(x,y)$
(Snyder 1983). Hence a positive bump in the refractive index
corresponds to a potential well. The region of the well is called
the fiber core. If the core is large, many bound modes exist,
corresponding to many guided modes in the fiber. Such fibers are
called multimode fibers, their core being usually 50 micrometer in
diameter. The modes couple easily, acting on the qubit like a
non-isolated environment. Hence multimode fibers are not
appropriate as quantum channels (see however Townsend 1998a and
1998b). If, however, the core is small enough (diameter of the order of a few wavelengths)
then a single spatial mode is guided. Such fibers are
called singlemode fibers. For telecommunications wavelength (i.e.
1.3 and 1.5 $\mu$m), their core is typically 8 $\mu$m in diameter.
Singlemode fibers are very well suited to carry single quanta. For
example, the optical phase at the output of a fiber is in a stable
relation with the phase at the input, provided the fiber doesn't
get elongated. Hence, fiber interferometers are very stable, a
fact exploited in many instruments and sensors (see, e.g.,
Cancellieri 1993).

Accordingly, a singlemode fiber with perfect cylindric symmetry
would provide an ideal quantum channel. But all real fibers have
some asymmetries and then the two polarization modes are no longer
degenerate but each has its own propagation constant. A similar
effect is caused by chromatic dispersion, where the group delay
depends on the wavelength. Both dispersion effects are the subject
of the next paragraphs.

\subsubsection{Polarization effects in singlemode fibers}\label{PolEffects}
Polarization effects in singlemode fibers are a common source of
problems in all optical communication schemes, as well classical
as quantum ones. In recent years this has been a major topic for
R\&D in classical optical communication (Gisin {\it et al.} 1995). As a
result, today's fibers are much better than the fibers a decade
ago. Nowadays, the remaining birefringence is small enough for the
telecom industry, but for quantum communication, any
birefringence, even extremely small, will always remain a concern.
All fiber based implementations of QC have to face this problem.
This is clearly true for polarization based systems; but it is
equally a concern for phase based systems, since the interference
visibility depends on the polarization states. Hence, although
polarization effects are not the only source of difficulties, we
shall describe them in some detail, distinguishing between 4
effects: the geometrical one, birefringence, polarization mode
dispersion and polarization dependent losses.

The {\bf Geometric phase} as encountered when guiding light in an
optical fiber is a special case of the Berry
phase\footnote{Introduced by Michael Berry in 1984, then observed
in optical fiber by Tomita and Chiao (1986), and on the single
photon level by Hariharan {\it et al.} (1993), studied in connection to
photon pairs by Brendel {\it et al.} (1995).} which results when any
parameter describing a property of the system under concern, here
the $k$-vector characterizing the propagation of the light field,
undergoes an adiabatic change. Think first of a linear
polarization state, let's say vertical at the input. Will it still
be vertical at the output? Vertical with respect to what?
Certainly not the gravitational field! One can follow that linear
polarization by hand along the fiber and see how it may change
even along a closed loop. If the loop stays in a plane, the state
after a loop coincides with the input state. But if the loop
explores the 3 dimensions of our space, then the final state will
differ from the initial one by an angle. Similar reasoning holds
for the axes of elliptical polarization states. The two circular
polarization states are the eigenstates: during parallel transport
they acquire opposite phases, called the Berry phase. The presence
of a geometrical phase is not fatal for quantum communication, it
simply means that initially Alice and Bob have to align their
systems by defining for instance the vertical and diagonal
directions (i.e. performing the unitary transformation mentioned
before). If these vary slowly, they can be tracked, though this
requires an active feedback. However, if the variations are too
fast, the communication might be interrupted. Hence, aerial cables
that swing in the wind are not appropriate (except with
selfcompensating configurations, see paragraph \ref{PandP1}).

{\bf Birefringence} is the presence of two different phase
velocities for two orthogonal polarization states. It is caused by
asymmetries in the fiber geometry and in the residual stress
distribution inside and around the core. Some fibers are made
birefringent on purpose. Such fibers are called polarization
maintaining (PM) fibers because the birefringence is large enough
to effectively uncouple the two polarization eigenmodes. But note
that only these two orthogonal polarization modes are maintained;
all the other modes, on the contrary, evolve very quickly, making
this kind of fiber completely unsuitable for polarization-based QC
systems\footnote{PM fibers might be of use for phase based QC
systems. However, this requires the whole setup -- transmission
lines as well as interferometers at Alice's and Bob's -- to be
made of PM fibers. While this is principally possible, the need of
installing a completely new fiber network makes this solution not
very practical.}. The global effect of the birefringence is
equivalent to an arbitrary combination of two waveplates, that is,
it corresponds to a unitary transformation. If this transformation
is stable, Alice and Bob can compensate for it. The effect of
birefringence is thus similar to the geometrical effect, though,
in addition to a rotation, it may also affect the ellipticity.
Stability of birefringence requires slow thermal and mechanical
variations.

{\bf Polarization Mode Dispersion (PMD)} is the presence of two
different group velocities for two orthogonal polarization modes.
It is due to a delicate combination of two causes. First,
birefringence produces locally two group velocities. For optical
fibers, this local modal dispersion is in good approximation equal
to the phase dispersion, of the order of a few ps/km. Hence,
locally an optical pulse tends to split into a fast mode and a
slow mode. But because the birefringence is small, the two modes
couple easily. Hence any small imperfection along the fiber
produces polarization mode coupling: some energy of the fast mode
couples into the slow mode and vice-versa. PMD is thus similar to
a random walk\footnote{In contrast to Brownian motion describing
particles diffusion in space as time passes, here photons diffuse
in time as they propagate along the fiber.} and grows only with
the square root of the fiber length. It is expressed in
$\frac{ps}{\sqrt{km}}$, with values as low as
0.1$\frac{ps}{\sqrt{km}}$ for modern fibers and possibly as high
as 0.5 or even 1$\frac{ps}{\sqrt{km}}$ for older ones.

Typical lengths for the polarization mode coupling vary from a few
meters up to hundreds of meters. The stronger the coupling, the
weaker the PMD (the two modes do not have time to move away between
the couplings). In modern fibers, the couplings are even
artificially increased during the drawing process of the fibers
(Hart {\it et al.} 1994, Li and Nolan 1998). Since the couplings are
exceedingly sensitive, the only reasonable description is a
statistical one, hence PMD is described as a statistical
distribution of delays $\delta\tau$. For long enough fibers, the
statistics is Maxwellian and PMD is related to the fiber length
$\ell$, the mean coupling length $h$, the mean modal birefringence
$B$ and to the RMS delay as follows (Gisin {\it et al.} 1995):
PMD$\equiv\sqrt{\ml\delta\tau^2\mr}=Bh\sqrt{\ell/h}$. PMD could
cause depolarization which would be devastating for quantum
communication, similar to any decoherence in quantum information
processing. But fortunately, for quantum communication the remedy
is easy, it suffices to use a source with a coherence time larger
than the largest delay $\delta\tau$. Hence, when laser pulses are
used (with typical spectral widths $\Delta\lambda\le 1$ nm, corresponding to a coherence time $\ge 3$ ps, see
paragraph \ref{WeakPulse}), PMD is no real problem. For photons
created by parametric down conversion, however, PMD can impose
severe limitations since $\Delta\lambda\ge 10$ nm (coherence time $\le 300$ fs) is not unusual.

{\bf Polarization Dependent Losses (PDL)} is a differential
attenuation between two orthogonal polarization modes. This effect
is negligible in fibers, but can be significant in components like
phase modulators. In particular, some integrated optics waveguides
actually guide only one mode and thus behave almost like
polarizers (e.g. proton exchange waveguides in LiNbO$_3$). PDL is
usually stable, but if connected to a fiber with some
birefringence, the relation between the polarization state and the
PDL may fluctuate, producing random outcomes (Elamari et al.
1998). PDL cannot be described by a unitary operator acting in the
polarization state space (but it is of course unitary in a larger
space (Huttner {\it et al.} 1996a). It does thus not preserve the scalar
product. In particular, it can turn non-orthogonal states into
orthogonal ones which can then be distinguished unambiguously (at
the cost of some loss) (Huttner {\it et al.} 1996a, Clarke {\it et al.} 2000).
Note that this could be used by Eve, specially to eavesdrop on the
2-state protocol (paragraph \ref{2state}).

Let us conclude this paragraph on polarization effects in fibers
by mentioning that they can be passively compensated, provided one
uses a go-\&-return configuration, using Faraday mirrors, as
described in section \ref{PandP1}.

\subsubsection{Chromatic dispersion effects in singlemode fibers}\label{CD}
In addition to polarization effects, chromatic dispersion (CD) can
cause problems for quantum cryptography as well. For instance, as
explained in sections \ref{PhaseExp} and \ref{ETent}, schemes implementing phase- or
phase-and-time-coding rely on photons arriving at well defined
times, that is on photons well localized in space. However, in
dispersive media like optical fibers, different group velocities
act as a noisy environment on the localization of the photon as
well as on the phase acquired in an interferometer. Hence, the
broadening of photons featuring non-zero bandwidth, or, in other
words, the coupling between frequency and position must be
circumvented or controlled. This implies working with photons of
small bandwidth, or, as long as the bandwidth is not too large,
operating close to the wavelength $\lambda_0$ where chromatic
dispersion is zero, i.e. for standard fibers around 1310 nm.
Fortunately, fiber losses are relatively small at this wavelength
and amount to $\approx$0.35 dB/km. This region is called the
second telecommunication window\footnote{The first one, around 800
nm, is almost no longer used. It was motivated by the early
existence of sources and detectors at this wavelength. The third
window is around 1550 nm where the attenuation reaches an absolute
minimum (Thomas {\it et al.} 2000) and where erbium doped fibers provide
convenient amplifiers (Desurvire 1994).}. There are also special
fibers, called dispersion-shifted, with a refractive index profile
such that the chromatic dispersion goes to zero around 1550 nm,
where the attenuation is minimal (Neumann 1988)\footnote{Chromatic
dispersion in fibers is mainly due to the material, essentially
silicon, but also to the refractive index profile. Indeed, longer
wavelengths feel regions further away from the core where the
refractive index is lower. Dispersion-shifted fibers have,
however, been abandoned by today's industry, because it turned out
to be simpler to compensate for the global chromatic dispersion by
adding an extra fiber with high negative dispersion. The
additional loss is then compensated by an erbium doped fiber
amplifier.}.

\begin{figure}
\infig{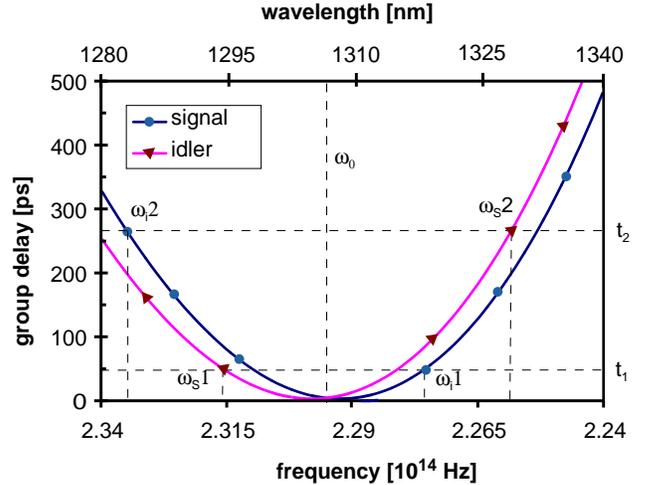} {0.95\columnwidth}\caption{Illustration
of cancellation of chromatic dispersion effects in the fibers
connecting an entangled-particle source and two detectors. The
figure shows differential group delay (DGD) curves for two
slightly different, approximately 10 km long fibers. Using
frequency correlated photons with central frequency $\omega_0$ --
determined by the properties of the fibers --, the difference of
the propagation times $t_{2}-t_{1}$ between signal (at $\omega_s1,
\omega_s2$) and idler photon (at $\omega_i1, \omega_i2$) is the
same for all $\omega_s,\omega_i$. Note that this cancellation
scheme is not restricted to signal and idler photons at nearly
equal wavelengths. It applies also to asymmetrical setups where
the signal photon (generating the trigger to indicate the presence
of the idler photon) is at a short wavelength of around 800 nm and
travels only a short distance. Using a fiber with appropriate zero
dispersion wavelength $\lambda_0$, it is still possible to achieve
equal DGD with respect to the energy-correlated idler photon at
telecommunication wavelength, sent through a long fiber.}
\label{fig3_3}\end{figure}

CD does not constitute a problem in case of faint laser pulses
where the bandwidth is small. However, it becomes a serious issue
when utilizing photon pairs created by parametric downconversion.
For instance, sending photons of 70 nm bandwidth (as used in our
long-distance Bell inequality tests, Tittel {\it et al.} 1998) down 10 km of
optical fibers leads to a temporal spread of around 500 ps
(assuming photons centered at $\lambda_0$ and a typical dispersion slope of
0.086 $\frac{ps}{nm^2 km}$). However, this can be compensated for
when using energy-time entangled photons (Franson 1992, Steinberg
{\it et al.} 1992a and 1992b, Larchuk {\it et al.} 1995). In contrast to
polarization coding where frequency and the physical property used
to implement the qubit are not conjugate variables, frequency and
time (thus position) constitute a Fourier pair. The strict energy
anti-correlation of signal and idler photon enables one to achieve
a dispersion for one photon which is equal in magnitude but
opposite in sign to that of the sister photon, corresponding thus
to the same delay\footnote{Assuming a predominantly linear dependence of CD in
function of the optical frequency, a realistic assumption.} (see
Fig. \ref{fig3_3}). The effect of broadening of the two wave
packets then cancels out and two simultaneously emitted photons
stay coincident. However, note that the arrival time of the pair
varies with respect to its emission time. The frequency
anticorrelation provides also the basis for avoiding decrease of
visibility due to different wavepacket broadening in the two arms
of an interferometer. And since the CD properties of optical
fibers do not change with time -- in contrast to birefringence --
no on-line tracking and compensation is required. It thus turns
out that phase and phase-time coding is particularly suited to
transmission over long distances in optical fibers: nonlinear
effects decohering the qubit ``energy'' are completely negligible,
and CD effects acting on the localization can be avoided or
compensated for in many cases.

\subsubsection{Free-space links}\label{freespacetransmission}

Although telecommunication based on optical fibers is very
advanced nowadays, such channels may not always be available.
Hence, there is also some effort in developing free space
line-of-sight communication systems - not
only for classical data transmission but for quantum cryptography
as well (see Hughes {\it et al.} 2000a and Gorman {\it et al.} 2000).

\begin{figure}
\infig{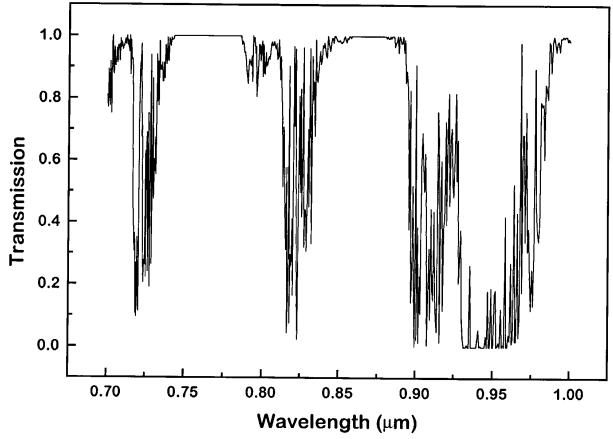}{0.95\columnwidth} \caption{Transmission
losses in free space as calculated using the LOWTRAN code for
earth to space transmission at the elevation and location of Los
Alamos, USA. Note that there is a low loss window at around 770 nm -- a
wavelength where high efficiency Silicon APD's can be used for
single photon detection (see also Fig. \ref{fig3_5} and compare to
Fig. \ref{fig3_2}).} \label{fig3_4}
\end{figure}

Transmission over free space features some advantages compared to
the use of optical fibers. The atmosphere has a high transmission
window at a wavelength of around 770 nm (see Fig. \ref{fig3_4})
where photons can easily be detected using commercial, high
efficiency photon counting modules (see chapter
\ref{PhotonCountSi}). Furthermore, the atmosphere is only weakly
dispersive and essentially non-birefringent\footnote{In contrast
to an optical fiber, air is not subject to stress, hence
isotropic.} at these wavelengths. It will thus not alter the
polarization state of a photon.

However, there are some drawbacks concerning free-space links as
well. In contrast to transmitting a signal in a guiding medium
where the energy is ``protected'' and remains localized in a small
region in space, the energy transmitted via a free-space link
spreads out, leading to higher and varying transmission losses. In
addition to loss of energy, {\bf ambient daylight}, or even light
from the moon at night, might couple into the receiver, leading to
a higher error rate. However, the latter errors can be maintained
at a reasonable level by using a combination of spectral filtering
($\le 1$ nm interference filters), spatial filtering at the receiver and
timing discrimination using a coincidence window of typically a
few ns. Finally, it is clear that the performance of free-space
systems depends dramatically on {\bf atmospheric conditions} and
is possible only with clear weather.

Finally, let us briefly comment on the different sources leading
to {\bf coupling losses}. A first concern is the transmission of
the signals through a turbulent medium, leading to
{\it{arrival-time jitter}} and {\it{beam wander}} (hence problems
with {\it{beam pointing}}). However, as the time-scales for {\bf
atmospheric turbulences} involved are rather small -- around 0.1
to 0.01 s --, the time jitter due to a variation of the effective
refractive index can be compensated for by sending a reference
pulse at a different wavelength at short time (around 100 ns)
before each signal pulse. Since this reference pulse experiences
the same atmospheric conditions as the subsequent one, the signal
will arrive essentially without jitter in the time-window defined
by the arrival of the reference pulse. In addition, the reference
pulse can be reflected back to the transmitter and used to correct
the direction of the laser beam by means of adaptive optics, hence
to compensate for {\it{beam wander}} and to ensure good {\it{beam
pointing}}

Another issue is the beam divergence, hence increase of spot size
at the receiver end caused by {\bf diffraction} at the transmitter
aperture. Using for example 20 cm diameter optics, the diffraction
limited spot size after 300 km is of $\approx$ 1 m. This effect
can in principle be kept small taking advantage of larger optics.
However, it can also be of advantage to have a spot size large
compared to the receiver's aperture in order to ensure constant
coupling in case of remaining beam wander.
In their 2000 paper, Gilbert and Hamrick provide a comprehensive discussion
of free-space channels in the context of QC.

\subsection{Single-photon detection}\label{PhotonCounting}
With the availability of pseudo single-photon and photon-pair
sources, the success of quantum cryptography is essentially
dependent on the possibility to detect single photons. In
principle, this can be achieved using a variety of techniques, for
instance photo-multipliers, avalanche-photodiodes, multichannel
plates, superconducting Josephson junctions. The ideal detector
should fulfill the following requirements:

\vspace{0.5cm}
\begin{itemize}
\item{
it should feature a high quantum detection efficiency over a large
spectral range,}

\item{the probability of generating noise, that is a signal without a photon
arriving, should be small,}

\item{to ensure a good timing resolution, the time between detection of a
photon and generation of an electrical signal
should be as constant as possible, i.e. the time jitter should be small,}

\item{the recovery time (i.e. the deadtime) should be small to allow high data rates.}

\end{itemize}

In addition, it is important to keep the detectors handy. For
instance, a detector which needs liquid helium or even nitrogen
cooling would certainly render a commercial development difficult.

Unfortunately, it turns out that it is impossible to meet all
mentioned points at the same time. Today, the best choice is
avalanche photodiodes (APD). Three different semiconductor
materials are used: either Silicon, Germanium or Indium Gallium
Arsenide, depending on the wavelengths.

APDs are usually operated in so-called {\it{Geiger mode}}. In this
mode, the applied voltage exceeds the breakdown voltage, leading
an absorbed photon to trigger an electron avalanche consisting of
thousands of carriers. To reset the diode, this
macroscopic current must be quenched -- the emission of charges
stopped and the diode recharged (Cova {\it et al.} 1996). Three main
possibilities exist:

\begin{itemize}
\item{
In {\it{passive-quenching}} circuits, a large (50-500 k$\Omega$)
resistor is connected in series with the APD (see e.g. Brown {\it et
al.} 1986). This causes a decrease of the voltage across the APD as
soon as an avalanche starts. When it drops below breakdown
voltage, the avalanche stops and the diode recharges. The recovery
time of the diode is given by its capacitance and by the value of
the quench resistor. The maximum count rate varies from some
hundred kHz to a few MHz.}
\item{
In {\it{active quenching}} circuits, the bias voltage is actively
lowered below the breakdown voltage as soon as the leading edge of
the avalanche current is detected (see e.g. Brown {\it et al.} 1987). This mode
enables higher count rates compared to passive quenching (up to
tens of MHz), since the deadtime can be as short as some tens of
ns. However, the fast electronic feedback system renders active
quenching circuits much more complicated than passive ones.}
\item{
Finally, in {\it{gated mode}} operation, the bias voltage is kept
below the breakdown voltage and is raised above only for a short
time when a photon is expected to arrive, typically a few ns.
Maximum count-rates similar to active quenching circuits can be
obtained using less complicated electronics. Gated mode operation
is commonly used in quantum cryptography based on faint laser
pulses where the arrival-times of the photons are well known.
However, it only applies if prior timing information is available.
For 2-photon schemes, it is most often combined with one passive
quenched detector, generating the trigger signal for the gated
detector.}
\end{itemize}

Apart from Geiger mode, Brown {\it et al.} also investigated the
performance of Silicon APDs operated in {\it{sub-Geiger}} mode
(Brown {\it et al.} 1989). In this mode, the bias voltage is kept
slightly smaller than the breakdown voltage such that the
multiplication factor -- around 100 -- already enables to detect
an avalanche, however, is still small enough to prevent real
breakdowns. Unfortunately, the single-photon counting performance
in this mode is rather bad and initial efforts have not been
continued, the major problem being the need for extremely
low-noise amplifiers.

\vspace{0.5cm} An avalanche engendered by carriers created in the
conduction band of the diode can not only be caused by an
impinging photon, but also by unwanted causes. These might be
thermal or band-to-band tunneling processes, or emissions from
trapping levels populated while a current transits through the
diode. The first two causes produce avalanches not due to photons
and are referred to as darkcounts. The third process depends on
previous avalanches and its effect is called afterpulses. Since
the number of trapped charges decreases exponentially with time,
these afterpulses can be limited by applying large deadtimes.
Thus, there is a trade-off between high count rates and low
afterpulses. The time-constant of the exponential decrease of
afterpulses shortens for higher temperatures of the diode.
Unfortunately, operating APDs at higher temperature leads to a
higher fraction of thermal noise, that is higher dark counts.
There is thus again a tradeoff to be optimized. Finally,
increasing the bias voltage leads to a larger quantum efficiency
and a smaller time jitter, at the cost of an increase in the
noise.

\begin{figure}
\infig{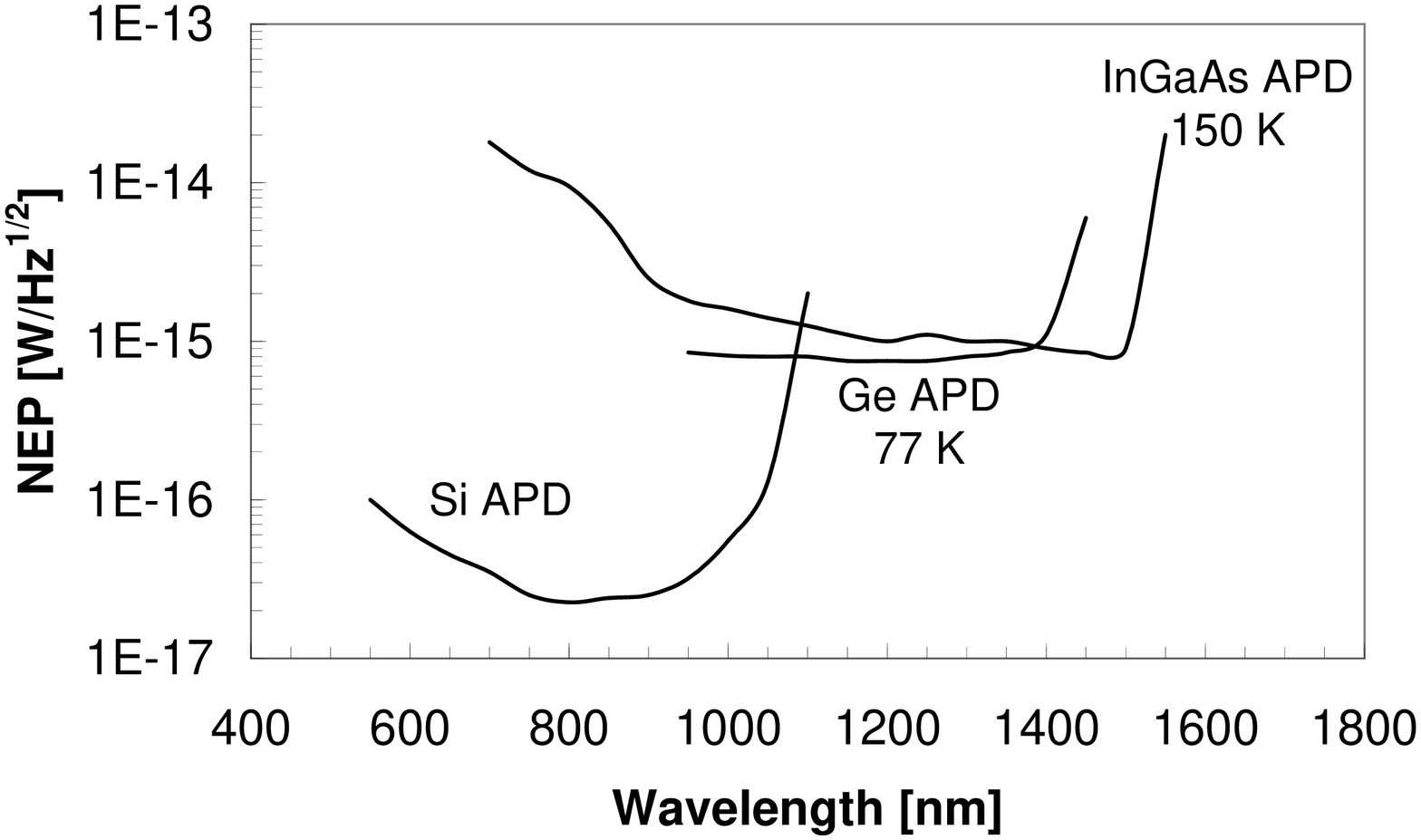}{0.95\columnwidth} \caption{Noise
equivalent power as a function of wavelength for Silicon,
Germanium, and InGaAs/InP APD's.} \label{fig3_5}
\end{figure}

We thus see that the optimal operating parameters, voltage,
temperature and dead time (i.e. maximum count rate) depend on the
very application. Besides, since the relative magnitude of
efficiency, thermal noise and after pulses varies with the type of
semiconductor material used, no general solution exists. In the
two next paragraphs we briefly present the different types of
APDs. The first paragraph focuses on Silicon APDs which enable the
detection of photons at wavelengths below 1$\mu$m, the second one
comments on Germanium and on Indium Gallium Arsenide APDs for
photon counting at telecommunication wavelength. The different
behaviour of the three types is shown in Fig. \ref{fig3_5}.
Although the best figure of merit for quantum cryptography is the
ratio of dark count rate $R$ per time unit to detection efficiency
$\eta$, we depict here the better-known noise equivalent power NEP
which shows similar behaviour. The NEP is defined as the optical
power required to measure a unity signal-to-noise ratio, and is
given by
\begin{equation}
NEP=\frac{h\nu}{\eta}\sqrt{2R}
\label{NEP}.
\end{equation}
\noindent Here, $h$ is  Planck's constant and $\nu$ is the
frequency of the impinging photons.

\subsubsection{Photon counting at wavelengths below 1.1 $\mu$m}\label{PhotonCountSi}

Since the beginning of the 80's, a lot of work has been done to
characterize Silicon APDs for single photon counting (Ingerson
1983, Brown 1986, Brown 1987, Brown 1989, Spinelli 1996), and the
performance of Si-APDs has continuously been improved. Since the
first test of Bell inequality using Si-APDs by Shih and Alley in
1988, they have completely replaced the photo-multipliers used
until then in the domain of fundamental quantum optics, known now
as quantum communication. Today, quantum efficiencies of up to
76\% (Kwiat {\it et al.} 1993) and time jitter down to 28 ps (Cova {\it et
al.} 1989) have been reported. Commercial single photon counting
modules are available (EG$\&$G SPCM-AQ-151), featuring quantum
efficiencies of 70 \% at a wavelength of 700 nm, a time jitter of
around 300 psec and maximum count rates larger than 5 MHz.
Temperatures of -20$^o$C -- sufficient to keep thermally generated
dark counts as low as 50 Hz -- can easily be achieved using
Peltier cooling. Single photon counters based on Silicon APDs thus
offer an almost perfect solution for all applications where
photons of a wavelength below 1 $\mu$m can be used. Apart from
fundamental quantum optics, this includes quantum cryptography in
free space and in optical fibers, however, due to high losses, the
latter one only over short distances.

\subsubsection{Photon counting at telecommunication wavelengths}\label{PhotonCountTelecom}

When working in the second telecommunication window (1.3$\mu$m),
one has to take advantage of APDs made from Germanium or
InGaAs/InP semiconductor materials. In the third window (1.55
$\mu$m), the only option is InGaAs/InP APDs.

Photon counting with Germanium APDs, although known for 30 years
(Haecker, Groezinger and Pilkuhn 1971), started to be used in the
domain of quantum communication with the need of transmitting
single photons over long distances using optical fibers, hence
with the necessity to work at telecommunications wavelength. In
1993, Townsend, Rarity and Tapster (Townsend {\it et al.} 1993a)
implemented a single photon interference scheme for quantum
cryptography over a distance of 10 km, and in 1994, Tapster,
Rarity and Owens (1994) demonstrated a violation of Bell
inequalities over 4 km. These experiments where the first ones to
take advantage of Ge APDs operated in passively quenched Geiger
mode. At a temperature of 77K which can be achieved  using either
liquid nitrogen or Stirling engine cooling, typical quantum
efficiencies of about 15 \% at dark count rates of 25 kHz can be
found (Owens {\it et al.} 1994), and time jitter down to 100 ps have
been observed (Lacaita {\it et al.} 1994) -- a normal value being
200-300 ps.

Traditionally, Germanium APDs have been implemented in the
domain of long-distance quantum communication. However, this type
of diode is currently getting replaced by InGaAs APDs and it is
more and more difficult to find Germanium APDs on the market.
Motivated by pioneering research reported already in 1985 (Levine,
Bethea and Campbell 1985), latest research focusses on InGaAs
APDs, allowing single photon detection in both telecommunication
windows. Starting with work by Zappa {\it et al.} (1994), InGaAs APDs as
single photon counters have meanwhile been characterized
thoroughly (Lacaita {\it et al.} 1996, Ribordy {\it et al.} 1998, Hiskett {\it et
al.} 2000, Karlsson {\it et al.} 1999, and Rarity {\it et al.} 2000, Stucki {\it et al.} 2001), and first
implementations for quantum cryptography have been reported
(Ribordy 1998, Bourennane {\it et al.} 1999, Bethune and Risk 2000, Hughes {\it et al.} 2000b, Ribordy {\it et al.} 2000).
However, if operating Ge APDs is already inconvenient compared to
Silicon APDs, the handiness of InGaAs APDs is even worse, the
problem being a extremely high afterpulse fraction. Therefore,
operation in passive quenching mode is impossible for applications
where noise is crucial. In gated mode, InGaAs APDs feature a
better performance for single photon counting at 1.3 $\mu m$
compared to Ge APDs. For instance, at a temperature of 77 K and a
dark count probability of $10^{-5}$ per 2.6 ns gate, quantum
efficiencies of around 30\% and of 17\% have been reported for
InGaAs and Ge APDs, respectively (Ribordy {\it et al.} 1998), while the
time jitter of both devices is comparable. If working at a
wavelength of 1.55 $\mu m$, the temperature has to be increased
for single photon detection. At 173 K and a dark count rate of now
$10^{-4}$, a quantum efficiency of 6\%  can still be observed
using InGaAs/InP devices while the same figure for Germanium APDs
is close to zero.

To date, no industrial effort has been done to optimize APDs
operating at telecommunication wavelength for photon counting, and
their performance is still far behind the one of Silicon
APDs\footnote{The first commercial photon counter at
telecommunication wavelengths came out only this year (Hamamatsu
photomultiplier R5509-72). However, the efficiency does not yet
allow an implementation for quantum cryptography.}. However, there
is no fundamental reasons why photon counting at wavelengths above
1 $\mu m$ should be more delicate than below, except that the
photons are less energetic. The real reasons for the lack of
commercial products are, first, that Silicon, the most common
semiconductor, is not sensitive (the band gap is too large), and
secondly that the market for photon counting is not yet mature.
But, without great risk, one can forecast that good commercial
photon counters will become available in the near future, and that
this will have a major impact on quantum cryptography.

\subsection{Quantum random number generators}\label{QRNG}
The key used in the one-time-pad must be secret and used only
once. Consequently, it must be as long as the message and must be
perfectly random. The later point proves to be a delicate and
interesting one. Computers are deterministic systems  that cannot
create truly random numbers. But all secure cryptosystems, both
classical and quantum ones, require truly random
numbers\footnote{The pin number that the bank attributes to your
credit card must be random. If not, someone knows it!}! Hence, the
random numbers must be created by a random physical process.
Moreover, to make sure that the random process is not merely
looking random with some hidden deterministic pattern, it is
necessary that it is completely understood. It is thus of interest
to implement a simple process in order to gain confidence in its
proper operation.

A natural solution is to rely on the random choice of a single
photon at a beamsplitter\footnote{Strictly speaking, the choice is
made only once the photons are detected at one of the outports.}
(Rarity {\it et al.} 1994). In this case the randomness is in principle
guaranteed by the laws of quantum mechanics, though, one still has
to be very careful not to introduce any experimental artefact that
could correlate adjacent bits. Different experimental realizations
have been demonstrated (Hildebrand 2001, Stefanov {\it et al.} 2000,
Jennewein {\it et al.} 2000a) and prototypes are commercially available
(www.gap-optique.unige.ch). One particular problem is the deadtime
of the detectors, that may introduce a strong anticorrelation
between neighboring bits. Similarly, afterpulses may provoke a
correlation.
These detector-related effects increase with higher pulse rates,
limiting the bit rate of quantum number generator to some MHz.

In the BB84 protocol Alice has to choose randomly between four
different states and Bob between two bases. The limited random
number generation rate may force Alice to produce her numbers in
advance and store them, opening a security weakness. On Bob's side
the random bit creation rate can be lower since, in principle, the
basis must be changed only after a photon has been detected, which
normally happens at rates below 1 MHz. However, one has to make
sure that this doesn't give the spy an opportunity for a Trojan
horse attack (see section \ref{Trojan})!

An elegant configuration integrating the random number generator
into the QC system consists in using a passive choice of bases, as
discussed in chapter \ref{QCPhotonPair} (Muller {\it et al.} 1993).
However, the problem of detector induced correlation remains.

\subsection{Quantum repeaters}\label{Qrepeaters}
\begin{figure}
\infig{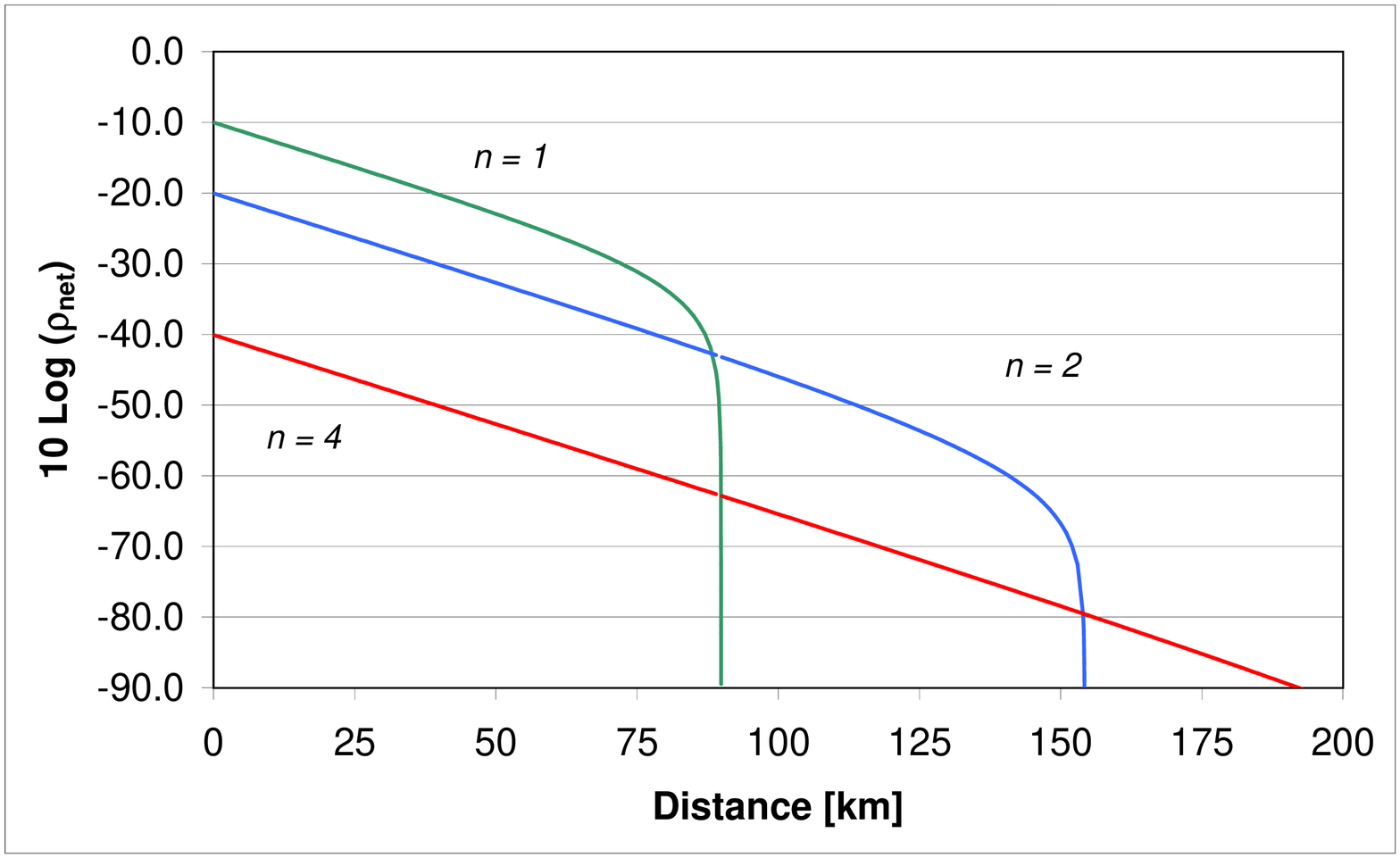}{0.95\columnwidth} \caption{Normalized net key creation
rate $\rho_{net}$ as a function of the distance in optical fibers. For $n=1$, Alice
uses a perfect single photon source. For $n>1$, the link is
divided into $n$ equal length sections and $n/2$ 2-photon sources
are distributed between Alice and Bob. Parameters: detection
efficiency $\eta=10\%$, dark count probability $p_{dark}=10^{-4}$,
fiber attenuation $\alpha=0.25$ dB/km.} \label{fig3_6}
\end{figure}

Todays fiber based QC systems are limited to tens of kilometers.
This is due to the combination of fiber losses and detectors'
noise. The losses by themselves do only reduce the bit rate
(exponentially with the distance), but with perfect detectors the
distance would not be limited. However, because of the dark
counts, each time a photon is lost there is a chance that a dark
count produces an error. Hence, when the probability of a dark
count becomes comparable to the probability that a photon is
correctly detected, the signal to noise ratio tends to 0 (more
precisely the mutual information $I(\alpha,\beta)$ tends to a
lower bound\footnote{The absolute lower bound is 0, but
dependening on the assumed eavesdropping strategy, Eve could take
advantage of the losses. In the latter case, the lower bound is
given by her mutual information $I(\alpha,\epsilon)$.}). In this
section we briefly explain how the use of entangled photons and of
entanglement swapping (\.Zukowski {\it et al.} 1993) could open
ways to extend the achievable distances in a foreseeable future
(some prior knowledge of entanglement swapping is assumed). Let us
denote $t_{link}$ the transmission coefficient (i.e.
$t_{link}$=probability that a photon sent by Alice gets to one of
Bob's detectors), $\eta$ the detectors' efficiency and $p_{dark}$
the dark count probability per time bin. With a perfect single
photon source, the probability $P_{raw}$ of a correct qubit
detection reads: $P_{raw}=t_{link}\eta$, while the probability
$P_{det}$ of an error is: $P_{det}=(1-t_{link}\eta)p_{dark}$.
Accordingly, the QBER=$\frac{P_{det}}{P_{raw}+P_{det}}$ and the
normalized net rate reads: $\rho_{net}=(P_{raw}+P_{det})\cdot
fct(QBER)$ where the function $fct$ denotes the fraction of bits
remaining after error correction and privacy amplification. For
the sake of illustration we simply assume a linear dependence
dropping to zero for QBER$\ge 15\%$ (This simplification does not
affect the qualitative results of this section. For a more precise calculation, see L\"utkenhaus 2000.):
$fct(QBER)=1-\frac{QBER}{15\%}$. The corresponding net rate
$\rho_{net}$ is displayed on Fig. \ref{fig3_6}. Note that it drops
to zero near 90 km.

Let us now assume that instead of a perfect single-photon source,
Alice and Bob use a (perfect) 2-photon source set in the middle of
their quantum channel. Each photon has then a probability
$\sqrt{t_{link}}$ to get to a detector. The probability of a
correct joined detection is thus $P_{raw}=t_{link}\eta^2$, while
an error occurs with probability
$P{det}=(1-\sqrt{t_{link}}\eta)^2p_{dark}^2+2\sqrt{t_{link}}\eta(1-\sqrt{t_{link}}\eta)p_{dark}$
(both photon lost and 2 dark counts, or one photon lost and one
dark count). This can be conveniently rewritten as:
$P_{raw}=t_{link}\eta^n$ and
$P_{det}=(t_{link}^{1/n}\eta+(1-t_{link}^{1/n}\eta)p_{dark})^n-t_{link}\eta^n$
valid for any division of the link into $n$ equal-length sections
and $n$ detectors. Note that the measurements performed at the
nodes between Alice and Bob do transmit (swap) the entanglement to
the twin photons, without revealing any information about the
qubit (these measurements are called Bell-measurements and are the
core of entanglement swapping and of quantum teleportation). The
corresponding net rates are displayed in Fig. \ref{fig3_6}.
Clearly, the rates for short distances are smaller when several
detectors are used, because of their limited efficiencies (here we
assume $\eta=10\%$). But the distance before the net rate drops to
zero is extended to longer distances! Intuitively, this can be
understood as follows. Let's consider that a logical qubit
propagates from Alice to Bob (although some photons propagate in
the opposite direction). Then, each 2-photon source and each
Bell-measurement acts on this logical qubit as a kind of QND
measurement: they test whether the logical qubit is still there!
In this way, Bob activates his detectors only when there is a
large chance $ t_{link}^{1/n}$ that the photon gets to his
detectors.

Note that if in addition to the detectors' noise there is noise
due to decoherence, then the above idea can be extended, using
entanglement purification. This is essentially the idea of quantum
repeaters (Briegel {\it et al.} 1998, Dur {\it et al.} 1999).

\newpage
\section{Experimental quantum cryptography with Faint laser pulses}\label{QCFaintPulses}
Experimental quantum key distribution was demonstrated for the first time in
1989 (it was published only in 1992 by Bennett {\it et al.} 1992a). Since then,
tremendous progress has been made. Today, several groups have shown that
quantum key distribution is possible, even outside the laboratory. In
principle, any two-level quantum system could be used to implement QC.
In
practice, all implementations have relied on photons. The reason is that their
interaction with the environment, also called decoherence, can be controlled
and moderated. In addition, researchers can benefit from all the tools
developed in the past two decades for optical telecommunications. It is
unlikely that other carriers will be employed in the foreseeable future.

Comparing different QC-setups is a difficult task, since several criteria must
be taken into account. What matters in the end is of course the rate of
corrected secret bits (distilled bit rate, R$_{dist}$) that can be transmitted
and the transmission distance. One can already note that with present and near
future technology, it will probably not be possible to achieve rates of the order of
gigahertz, nowadays common with conventional optical communication systems
(in their comprehensive paper published in 2000, Gilbert and Hamrick discuss
practical methods to achieve high bit rate QC).
This implies that encryption with a key exchanged through QC is to be limited
to highly confidential information. While the determination of the
transmission distance and rate of detection (the raw bit rate, R$_{raw}$) is
straightforward, estimating the net rate is rather difficult. Although in
principle errors in the bit sequence follow only from tampering by a
malevolent eavesdropper, the situation is rather different in reality.
Discrepancies in the keys of Alice and Bob also always happen because of
experimental imperfections. The error rate (here called quantum bit error
rate, or QBER) can be easily
determined. Similarly, the error correction procedure is rather simple. Error
correction leads to a first reduction of the key rate that depends strongly on
the QBER. The real problem consist in estimating the information obtained by
Eve, a quantity necessary for privacy amplification. It does not only depend
on the QBER, but also on other factors, like the photon number
statistics of the source, or the way the choice of the measurement basis is
made. Moreover in a pragmatic approach, one might also accept restrictions on
Eve's technology, limiting her strategies and therefore also the information
she can obtain per error she introduces. Since the efficiency of privacy
amplification rapidly decreases when the QBER increases, the distilled bit rate depends dramatically
on Eve's information \ and hence on the assumptions made. One can define as
the maximum transmission distance, the distance where the distilled rate
reaches zero. This can give an idea of the difficulty to evaluate a QC system
from a physical point of view.

Technological aspects must also be taken into account. In this article we do not focus
on all the published performances (in particular not on the
key rates), which strongly depend on present technology and the financial
possibilities of the research teams having carried out the experiments. On the
contrary, we try to weight the intrinsic technological difficulties associated
with each set-up and to anticipate certain technological advances. And last
but not least the cost of the realization of a prototype should also be
considered.

In this chapter, we first deduce a general formula for the QBER and consider
its impact on the distilled rate. We then review faint pulses implementations.
We class them according to the property used to encode the qubits value and
follow a rough chronological order. Finally, we assess the possibility to
adopt the various set-ups for the realization of an industrial prototype.
Systems based on entangled photon pairs are presented in the next chapter.

\subsection{Quantum Bit Error Rate}\label{SQBER}
The QBER is defined as the number of wrong bits to the total
number of received bits\footnote{In the followin we are considering systems
implementing the BB84 protocol. For other protocols some of the formulas have
to be slightly adapted.} and is normally in the order of a few percent. In the
following we will use it expressed as a function of rates:
\begin{equation}
QBER=\frac{N_{wrong}}{N_{right}+N_{wrong}}=\frac{R_{error}}{R_{sift}%
+R_{error}}\approx\frac{R_{error}}{R_{sift}}%
\end{equation}
where the sifted key corresponds to the cases in which Alice and Bob made compatible
choices of bases, hence its rate is half that of the raw key.

The raw rate is essentially the product of the pulse rate $f_{rep}$, the mean
number of photon per pulse $\mu$, the probability $t_{link}$ of a photon to
arrive at the analyzer and the probability $\eta$ of the photon being
detected:
\begin{equation}
R_{sift}=\frac{1}{2}R_{raw}=\frac{1}{2}q~f_{rep}~\mu~t_{link}~\eta
\label{R-sift}%
\end{equation}
The factor q (q$\leq$1, typically 1 or
${\frac12}$%
) must be introduced for some phase-coding setups in order to correct for
noninterfering path combinations (see, e.g., sections \ref{PhaseExp} and \ref{ETent}).

One can distinguish three different contributions to $R_{error}$. The first one arises
because of photons ending up in the wrong detector, due to unperfect
interference or polarization contrast. The rate R$_{opt}$ is given by the
product of the sifted key rate and the probability p$_{opt}$ of a photon going in the
wrong detector:
\begin{equation}
R_{opt}=R_{sift}~p_{opt}=\frac{1}{2}q~f_{rep}~\mu~t_{link}~p_{opt}~\eta
\end{equation}
This contribution can be considered, for a given set-up, as an intrinsic error
rate indicating the suitability to use it for QC. We will discuss it below in the case
of each particular system.

The second contribution, $R_{det}$, arises from the detector dark counts (or from
remaining environmental stray light in free space setups). This rate is
independent of the bit rate\footnote{This is true provided that afterpulses
(see section \ref{PhotonCounting}) do not contribute to the dark counts.}. Of
course, only dark counts falling in a short time window when a photon is
expected give rise to errors.%
\begin{equation}
R_{det}=\frac{1}{2}\frac{1}{2}f_{rep}p_{dark}n
\label{Rdet}
\end{equation}
where p$_{dark}$ is the probability of registering a dark count per
time-window and per detector, and n is the number of detectors. The two
$\frac{1}{2}$-factors are related to the fact that a dark count has a 50\%
chance to happen with Alice and Bob having chosen incompatible bases (thus
eliminated during sifting) and a 50\% chance to arise in the correct detector.

Finally error counts can arise from uncorrelated photons, because of imperfect
photon sources:
\begin{equation}
R_{acc}=\frac{1}{2}\frac{1}{2}p_{acc}f_{rep}t_{link}n\eta
\end{equation}
This factor appears only in systems based on entangled photons, where the
photons belonging to different pairs but arriving in the same time window are
not necessarily in the same state. The quantity $p_{acc}$ is the
probability to find a second pair within the time window, knowing that a first
one was created\footnote{Note that a passive choice of
measurement basis implies that four detectors (or two detectors during
two time windows) are activated for every pulse, leading thus to a doubling of
R$_{det}$ and R$_{acc}$.}.

The QBER can now be expressed as follows:%

\beqa
QBER  &=&\frac{R_{opt}+R_{det}+R_{acc}}{R_{sift}}\\
&=&p_{opt}+\frac{p_{dark}\cdot n}{t_{link}\cdot\eta\cdot2\cdot q\cdot\mu
}+\frac{p_{acc}}{2\cdot q\cdot\mu} \label{QBERdetail}\\
&=&QBER_{opt}+QBER_{det}+QBER_{acc}%
\label{QBERss}
\eeqa

We analyze now these three contributions. The first one, $QBER_{opt}$, is
independent on the transmission distance (it is independent of $t_{link}$). It
can be considered as a measure of the optical quality of the setup, depending
only on the polarisation or interference fringe contrast. The technical effort
needed to obtain, and more important, to maintain a given $QBER_{opt}$ is an
important criterion for evaluating different QC-setups. In polarization based
systems, it's rather simple to achieve a polarisation contrast of 100:1,
corresponding to a $QBER_{opt}$ of 1\%. In fiber based QC, the problem is to maintain this value
in spite of polarisation fluctuations and depolarisation in the fiber link.
For phase coding setups, QBER$_{opt}$ and the interference visibility are
related by%
\begin{equation}
QBER_{opt}=\frac{1-V}{2}%
\end{equation}
A visibility of 98\% translates thus into an optical error rate of 1\%. Such a
value implies the use of well aligned and stable interferometers. In bulk
optics perfect mode overlap is difficult to achieve, but the polarization is
stable. In single-mode fiber interferometers, on the contrary, perfect mode
overlap is automatically achieved, but the polarisation must be controlled and
chromatic dispersion can constitute a problem.

The second contribution, $QBER_{det}$, increases with distance, since the
darkcount rate remains constant while the bit rate goes down like $t_{link}$.
It depends entirely on the ratio of the dark count rate to the quantum
efficiency.
At present, good single-photon detectors are not commercially available for
telecommunication wavelengths. The span of QC is not limited by decoherence.
As $QBER_{opt}$ is essentially independent of the fiber length, it is the detector
noise that limits the transmission distance.

Finally, the $QBER_{acc}$ contribution is present only in some 2-photon
schemes in which multi-photon pulses are processed in such a way that they do
not necessarily encode the same bit value (see e.g. paragraphs \ref{PhCoding} and
\ref{PhaseTimeCoding}).
Indeed, although in all systems there is a probability for multi-photon
pulses, in most these contribute only to the information available to Eve (see
section \ref{QND}) and not to the QBER. But for implementations featuring
passive choice by each photon, the multi-photon pulses do not contribute to
Eve's information but to the error rate (see section \ref{Passivebasis}).

\begin{figure}
\infig{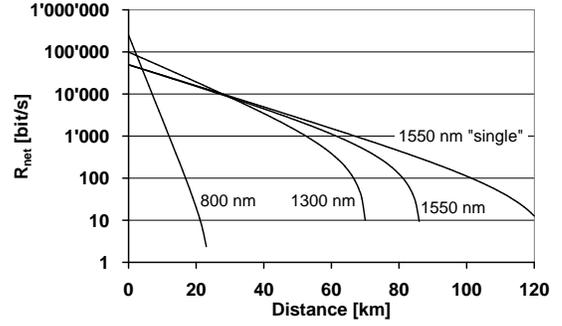}{0.95\columnwidth} \caption{Bit rate after
error correction and privacy amplification vs. fiber length. The
chosen parameters are: pulse rates 10 Mhz for faint laser pulses
($\mu = 0.1$) and 1 MHz for the case of ideal single photons (1550
nm ``single''); losses 2, 0.35 and 0.25 dB/km, detector efficiencies
50\%, 20\% and 10\%, and dark count probabilities $10^{-7}$,
$10^{-5}$, $10^{-5}$ for 800nm, 1300nm and 1550 nm respectively.
Losses at Bob and QBER$_{opt}$ are neglected.} \label{fig4_1}
\end{figure}

Now, let us calculate the useful bit rate as a function of the
distance. R$_{sift}$ and QBER are given as a function of
$t_{link}$ in eq. (\ref{R-sift}) and (\ref{QBERdetail})
respectively. The fiber link transmission decreases exponentially
with the length. The fraction of bits lost due to error correction
and privacy amplification is a function of QBER and depends on
Eve's strategy. The number of remaining bits $R_{net}$ is given by
the sifted key rate multiplied by the difference of the Alice-Bob
mutual Shannon information $I(\alpha,\beta)$ and Eve's maximal
Shannon information $I^{max}(\alpha,\epsilon)$:
\begin{equation}
R_{net}=R_{sift}\bigg(I(\alpha,\beta)-I^{max}(\alpha,\epsilon)\bigg)
\end{equation}
\noindent The latter are calculated here according to eq.
(\ref{Ginfo}) and (\ref{Ginfo2}) (section \ref{SymAttack}),
considering only individual attacks and no multiphoton pulses. We
obtain R$_{net}$ (useful bit rate after error correction and
privacy amplification) for different wavelengths as shown in Fig.
\ref{fig4_1}. There is first an exponential decrease, then, due to
error correction and privacy amplification, the bit rates fall
rapidly down to zero. This is most evident comparing the curves
1550 nm and 1550 nm ``single'' since the latter features 10 times
less QBER. One can see that the maximum range is about 100 km.  In
practice it is closer to 50 km, due to non-ideal error correction
and privacy amplification, multiphoton pulses and other optical
losses not considered here.
Finally, let us mention that typical key creation rates of the order of
a thousand bits per second over distances of a few tens of kilometers
have been demonstrated experimentally (see, for example, Ribordy {\it et al.} 2000 or  Townsend 1998b).

\subsection{Polarization coding}\label{1hvPolExp}

Encoding the qubits in the polarization of photons is a natural solution. The
first demonstration of QC by Charles Bennett and his coworkers (Bennett
{\it et al.} 1992a) made use of this choice.
They realized a system where
Alice and Bob exchanged faint light pulses produced by a LED and containing less than one photon on
average over a distance of 30 cm in air.
In spite of the small scale of this
experiment, it had an important impact on the community in the sense that it
showed that it was not unreasonable to use single photons instead of classical
pulses for encoding bits.

\begin{figure}
\infig{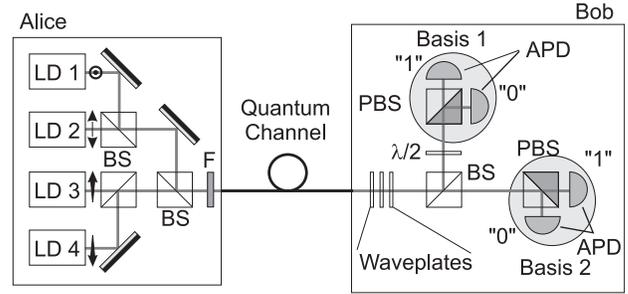}{0.95\columnwidth} \caption{Typical system
for quantum cryptography using polarization coding (LD: laser
diode, BS:\ beamsplitter, F: neutral density filter, PBS:
polarizing beam splitter, $\lambda/2$: half waveplate, APD:
avalanche photodiode).} \label{fig4_2}
\end{figure}

A typical system for QC with the BB84 four states protocol using
the polarization of photons is shown in Fig. \ref{fig4_2}. Alice's
system consists of four laser diodes. They emit short classical
photon pulses ($\approx1ns$)
polarized at $-45%
{{}^\circ}%
$, $0%
{{}^\circ}%
$, $+45%
{{}^\circ}%
$, and $90%
{{}^\circ}%
$. For a given qubit, a single diode is triggered. The pulses are then
attenuated by a set of filters to reduce the average number of photons well
below 1, and sent along the quantum channel to Alice.

It is essential that the pulses remain polarized for Bob to be able to extract
the information encoded by Alice. As discussed in paragraph \ref{PolEffects},
polarization mode dispersion may depolarize the photons, provided the delay it
introduces between both polarization modes is larger than the coherence time.
This sets a constraint on the type of lasers used by Alice.

When reaching Bob, the pulses are extracted from the fiber. They travel
through a set of waveplates used to recover the initial polarization states by
compensating the transformation induced by the optical fiber (paragraph
\ref{PolEffects}). The pulses reach then a symmetric beamsplitter,
implementing the basis choice. Transmitted photons are analyzed in the
vertical-horizontal basis with a polarizing beamsplitter and two photon
counting detectors. The polarization state of the reflected photons is first
rotated with a waveplate by $45%
{{}^\circ}%
$ ($-45%
{{}^\circ}%
$ to $0%
{{}^\circ}%
$). The photons are then analyzed with a second set of polarizing beamsplitter
and photon counting detectors. This implements the diagonal basis. For illustration, let us follow a
photon polarized at $+45%
{{}^\circ}%
$, we see that its state of polarization is arbitrarily transformed in the
optical fiber. At Bob's end, the polarization controller must be set to bring
it back to $+45%
{{}^\circ}%
$. If it chooses the output of the beamsplitter corresponding to the
vertical-horizontal basis, it will experience equal reflection and
transmission probability at the polarizing beamsplittter, yielding a random
outcome. On the other hand, if it chooses the diagonal basis, its state will
be rotated to $90%
{{}^\circ}%
$. The polarizing beamsplitter will then reflect it with unit probability,
yielding a deterministic outcome.

Instead of Alice using four lasers and Bob two polarizing beamsplitters, it is
also possible to implement this system with active polarization modulators
such as Pockels cells. For emission, the modulator is randomly activated for
each pulse to rotate the state of polarization to one of the four states,
while, at the receiver, it randomly rotates half of the incoming pulses by $45%
{{}^\circ}%
$. It is also possible to realize the whole system with fiber optics components.

\begin{figure}
\infig{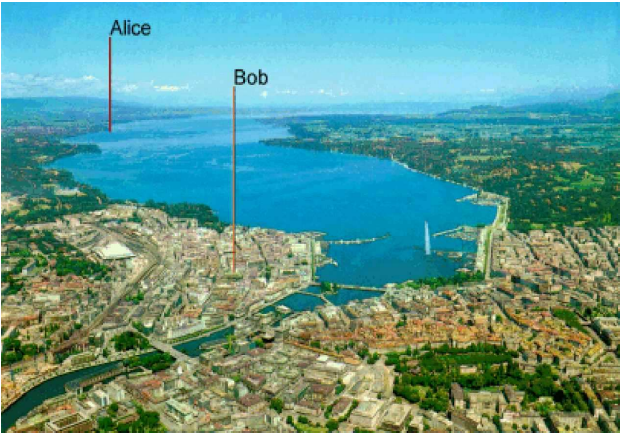}{0.95\columnwidth} \caption{Geneva and
Lake Geneva. The Swisscom optical fiber cable used for quantum
cryptography experiments runs under the lake between the town of
Nyon, about 23 km north of Geneva, and the centre of the city.}
\label{fig4_3}
\end{figure}

Antoine Muller and his coworkers at the University of Geneva used
such a system to perform QC experiments over optical fibers (1993,
see also Br\'{e}guet {\it et al.} 1994). They created a key over a
distance of 1100 meters with photons at 800 nm. In order to
increase the transmission distance, they repeated the experiment
with photons at 1300nm (Muller et al.1995 and 1996) and created a
key over a distance of 23 kilometers. An interesting feature of
this experiment is that the quantum channel connecting Alice and
Bob consisted in an optical fiber part of an installed cable, used
by the telecommunication company Swisscom for carrying phone
conversations. It runs between the Swiss cities of Geneva and
Nyon, under Lake Geneva (Fig. \ref{fig4_3}). This was the first
time QC was performed outside of a physics laboratory. It had a
strong impact on the interest of the wider public for the new
field of quantum communication.

These two experiments highlighted the fact that the polarization
transformation induced by a long optical fiber was unstable over time. Indeed,
when monitoring the QBER of their system, Muller noticed that, although it
remained stable and low for some time (of the order of several minutes), it
would suddenly increase after a while, indicating a modification of the
polarization transformation in the fiber. This implies that a real fiber based QC system
requires active alignment to compensate for this evolution. Although not
impossible, such a procedure is certainly difficult. James Franson did indeed
implement an active feedback aligment system ( 1995), but did not pursue along
this direction. It is interesting to note that replacing
standard fibers with polarization maintaining fibers does not solve the
problem. The reason is that, in spite of their name, these fibers do not
maintain polarization, as explained in paragraph \ref{PolEffects}.

Recently, Paul Townsend of BT Laboratories also investigated such polarization
encoding systems for QC on short-span links up to 10 kilometers (1998a and
1998b) with photons at 800nm. It is interesting to note that, although he used
standard telecommunications fibers which can support more than one spatial
mode at this wavelength, he was able to ensure single-mode propagation by
carefully controlling the launching conditions. Because of the problem
discussed above, polarization coding does not seem to be the best choice for
QC in optical fibers. Nevertheless, this
problem is drastically improved when considering free space key exchange,
as the air has essentially no birefringence at all (see section
\ref{Satellite}).

\subsection{Phase coding}\label{PhaseExp}
The idea of encoding the value of qubits in the phase of photons was first
mentioned by Bennett in the paper where he introduced the two-states protocol
(1992). It is indeed a very natural choice for optics specialists. State
preparation and analysis are then performed with interferometers, that can be
realized with single-mode optical fibers components.

\begin{figure}
\infig{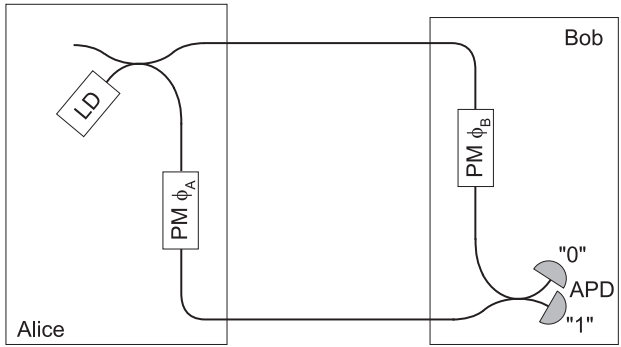}{0.95\columnwidth}
\caption{Conceptual interferometric set-up for quantum cryptography using an
optical fiber Mach-Zehnder interferometer (LD: laser diode, PM:
phase modulator, APD: avalanche photodiode).}
\label{fig4_4}
\end{figure}

Fig. \ref{fig4_4} presents an optical fiber version of a
Mach-Zehnder interferometer. It is made out of two symmetric
couplers -- the equivalent of beamsplitters -- connected to each
other, with one phase modulator in each arm. One can inject light
in the set-up using a continuous and classical source, and monitor
the intensity at the output ports. Provided that the coherence
length of the light used is larger than the path mismatch in the
interferometers, interference fringes can be recorded. Taking into
account the $\pi/2$-phase shift experienced upon reflection at a
beamsplitter, the effect of the phase modulators ($\phi_{A}$ and
$\phi_{B}$) and the path length difference ($\Delta
L$), the intensity in the output port labeled ``0'' is given by:%

\begin{equation}
I_{0}=\overline{I}\cdot\cos^{2}\left(  \frac{\phi_{A}-\phi_{B}+k\Delta L}%
{2}\right)  \label{PhaseCodingInterf}%
\end{equation}
where $k$\ is the wave number and $\overline{I}$\ the intensity of the source.
If the phase term is equal to $\pi/2+n\pi$\ where $n$\ is an integer,
destructive interference is obtained. Therefore the intensity registered in
port ``0'' reaches a minimum and all the light exits in port ``1''. When the
phase term is equal to $n\pi$, the situation is reversed: constructive
interference is obtained in port ``0'', while the intensity in port ``1'' goes
to a minimum. With intermediate phase settings, light can be recorded in both
ports. This device acts like an optical switch. It is essential to keep the
path difference stable in order to record stationary interferences.

Although we discussed the behavior of this interferometer for classical light,
it works exactly the same when a single photon is injected. The probability to
detect the photon in one output port can be varied by changing the phase. It
is the fiber optic version of Young's slits experiment, where the arms of the
interferometer replace the apertures.

This interferometer combined with a single photon source and photon counting
detectors can be used for QC. Alice's set-up consists of the source, the first
coupler and the first phase modulator, while Bob takes the second modulator
and coupler, as well as the detectors. Let us consider the implementation of
the four-states BB84 protocol. On the one hand, Alice can apply one of four
phase shifts ($0,\pi/2,\pi,3\pi/2$) to encode a bit value. She associates
$0$\ and $\pi/2$\ to bit $0$, and $\pi$\ and $3\pi/2$\ to bit $1$. On the
other hand, Bob performs a basis choice by applying randomly a phase shift of
either $0$\ or $\pi/2$, and he associates the detector connected to the output
port ``0'' to a bit value of 0, and the detector connected to the port ``1''
to 1. When the difference of their phase is equal to $0$\ or $\pi$, Alice and
Bob are using compatible bases and they obtain deterministic results. In such
cases, Alice can infer from the phase shift she applied, the output port
chosen by the photon at Bob's end and hence the bit value he registered. Bob,
on his side, deduces from the output port chosen by the photon, the phase that
Alice selected. When the phase difference equals $\pi/2$ or $3\pi/2$, the bases
are incompatible and the photon chooses randomly which port it takes at
Bob's coupler. This is summarized in Table 1. We must stress that it is
essential with this scheme to keep the path difference stable during a key
exchange session. It should not change by more than a fraction of a wavelength
of the photons. A drift of the length of one arm would indeed change the phase
relation between Alice and Bob, and induce errors in their bit sequence.\newline %

\begin{tabular}
[c]{|c|c|c|c|c|}\hline\hline
\multicolumn{2}{|c|}{Alice} & \multicolumn{3}{|c|}{Bob}\\\hline
Bit value & $\phi_{A}$ & $\phi_{B}$ & $\phi_{A}-\phi_{B}$ & Bit
value\\\hline\hline
0 & 0 & \multicolumn{1}{||c|}{0} & 0 & 0\\\cline{3-5}%
0 & 0 & \multicolumn{1}{||c|}{$\pi/2$} & $3\pi/2$ & ?\\\cline{3-5}%
1 & $\pi$ & \multicolumn{1}{||c|}{0} & $\pi$ & 1\\\cline{3-5}%
1 & $\pi$ & \multicolumn{1}{||c|}{$\pi/2$} & $\pi/2$ & ?\\\cline{3-5}%
0 & $\pi/2$ & \multicolumn{1}{||c|}{0} & $\pi/2$ & ?\\\cline{3-5}%
0 & $\pi/2$ & \multicolumn{1}{||c|}{$\pi/2$} & 0 & 0\\\cline{3-5}%
1 & $3\pi/2$ & \multicolumn{1}{||c|}{0} & $3\pi/2$ & ?\\\cline{3-5}%
1 & $3\pi/2$ & \multicolumn{1}{||c|}{$\pi/2$} & $\pi$ & 1\\\hline
\end{tabular}

\bigskip Table 1: Implementation of the BB84 four-states protocol with
phase encoding. \bigskip

\begin{figure}
\infig{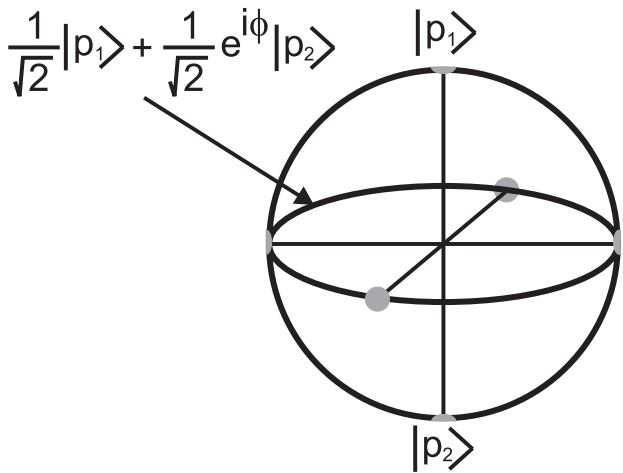}{0.95\columnwidth}
\caption{Poincar\'{e}
sphere representation of two-levels quantum states generated by
two-paths interferometers. The states generated by an
interferometer where the first coupler is replaced by a switch
correspond to the poles. Those generated with a symetrical
beamsplitter are on the equator. The azimuth indicates the phase
between the two paths.}
\label{fig4_5}
\end{figure}

It is interesting to note that encoding qubits with 2-paths
interferometers is formally isomorphic to polarization encoding.
The two arms correspond to a natural basis, and the weights
$c_{j}$ of each qubit state $\psi=\left(
c_{1}e^{-i\phi/2},c_{2}e^{i\phi/2}\right)  $ are determined by the
coupling ratio of the first beam splitter while the relative phase
$\phi$ is introduced in the interferometer. The Poincar\'{e}
sphere representation, which applies to all two-levels quantum
systems, can also be used to represent phase-coding states. In
this case, the azimuth angle represents the relative phase between the
light having propagated along the two arms. The elevation
corresponds to the coupling ratio of the first beamsplitter.
States produced by a switch are on the poles, while those
resulting from the use of a 50/50 beamsplitter lie on the equator.
Figure \ref{fig4_5} illustrates this analogy. Consequently, all
polarization schemes can also be implemented using phase coding.
Similarly, every coding using 2-path interferometers can be
realized using polarization. However, in practice one choice is
often more convenient than the other, depending on circumstances
like the nature of the quantum channel\footnote{Note, in addition,
that using many-path interferometers opens up the possibility to
code quantum systems of dimensions larger than 2, like qutrits,
ququarts, etc. (Bechmann-Pasquinucci and Tittel 2000,
Bechmann-Pasquinucci and Peres 2000, Bourennane {\it et al.} 2001a).}.

\subsubsection{The double Mach-Zehnder implementation}\label{DoubleMZ}
\begin{figure}
\infig{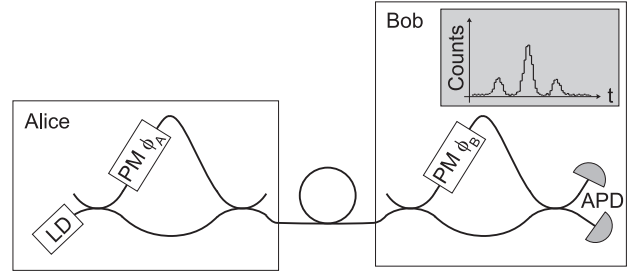}{0.95\columnwidth}
\caption{Double
Mach-Zehnder implementation of an interferometric system for
quantum cryptography (LD: laser diode, PM: phase modulator, APD:
avalanche photodiode). The inset represents the temporal count
distribution recorded as a function of the time passed since the
emission of the pulse by Alice.
Interference is observed in the central peak.}
\label{fig4_6}
\end{figure}

Although the scheme presented in the previous paragraph works
perfectly well on an optical table, it is impossible to keep the
path difference stable when Alice and Bob are separated by more
than a few meters. As mentioned above, the relative length of the
arms should not change by more than a fraction of a wavelength.
Considering a separation between Alice and Bob of 1 kilometer for
example, it is clear that it is not possible to prevent path
difference changes smaller than $1\mu m$ caused by environmental
variations. In his 1992 letter, Bennett also showed how to get
round this problem. He suggested to use two unbalanced
Mach-Zehnder interferometers connected in series by a single
optical fiber (see Fig. \ref{fig4_6}), both Alice and Bob being
equipped with one. When monitoring counts as a function of the
time since the emission of the photons, Bob obtains three peaks
(see the inset in Fig. \ref{fig4_6}). The first one corresponds to
the cases where the photons chose the short path both in Alice's
and in Bob's interferometers, while the last one corresponds to
photons taking twice the long paths. Finally, the central peak
corresponds to photons choosing the short path in Alice's
interferometer and the long one in Bob's, and to the opposite. If
these two processes are indistinguishable, they produce
interference. A timing window can be used to discriminate between
interfering and non-interfering events. Disregarding the latter,
it is then possible for Alice and Bob to exchange a key.

The advantage of this set-up is that both ``halves'' of the photon travel in
the same optical fiber. They experience thus the same optical length in the
environmentally sensitive part of the system, provided that the variations in
the fiber are slower than their temporal separations, determined by the
interferometer's imbalance ($\approx5ns$). This condition is much less
difficult to fulfill. In order to obtain a good interference visibility, and
hence a low error rate, the imbalancements of the interferometers must be
equal within a fraction of the coherence time of the photons. This implies
that the path differences must be matched within a few millimeters, which does
not constitute a problem. Besides, the imbalancement must be chosen so that it
is possible to clearly distinguish the three temporal peaks and thus
discriminate interfering from non-interfering events. It must then typically
be larger than the pulse length and than the timing jitter of the photon
counting detectors. In practice, the second condition is the most stringent
one. Assuming a time jitter of the order of $500ps$, an
imbalancement of at least $1.5ns$ keeps the overlap between the peaks low.

The main difficulty associated with this QC scheme is that the imbalancements
of Alice's and Bob's interferometers must be kept stable within a fraction of
the wavelength of the photons during a key exchange to maintain correct phase
relations. This implies that the interferometers must lie in containers whose
temperature is stabilized. In addition, for long key exchanges an active
system is necessary to compensate the drifts\footnote{Polarization coding
requires the optimization of three parameters (three parameters are necessary
for unitary polarization control). In comparison, phase coding requires
optimization of only one parameter. This is possible because the coupling
ratios of the beamsplitters are fixed. Both solutions would be equivalent if
one could limit the polarization evolution to rotations of the elliptic
states, without changes in the ellipticity.}. Finally, in order to ensure the
indistinguishability of both interfering processes, one must make sure that in
each interferometer the polarization transformation induced by the short path
is the same as the one induced by the long one. Alice as much as Bob must then
use a polarization controller to fulfill this condition. However, the
polarization transformation in short optical fibers whose temperature is kept
stable, and which do not experience strains, is rather stable. This adjustment
does thus not need to be repeated frequently.

Paul Tapster and John Rarity from DERA working with Paul Townsend were the
first ones to test this system over a fiber optic spool of 10 kilometers
(1993a and 1993b). Townsend later improved the interferometer by replacing
Bob's input coupler by a polarization splitter to suppress the lateral
non-interfering peaks (1994). In this case, it is unfortunately again
necessary to align the polarization state of the photons at Bob's, in addition
to the stabilization of the interferometers imbalancement.
He later thoroughly
investigated key exchange with phase coding and improved the transmission
distance (Marand and Townsend 1995, Townsend 1998b).
He also tested the possibility to multiplex at two different wavelengths a
quantum channel with conventional data transmission over a single optical
fiber (Townsend 1997a).
Richard Hughes and his
co-workers from Los Alamos National Laboratory also extensively tested such an
interferometer (1996 and 2000b), up to distances of 48 km of installed
optical fiber
\footnote{Note that in this experiment Hughes and his coworkers used an unusually
high mean number of photons per pulse (They used a mean photon number of approximately 0.6 in the central interference peak, corresponding to a $\mu \approx 1.2$ in the pulses leaving Alice.
The latter value is the relevant one for an eavesdropping analysis, since Eve could use an interferometer -- conceivable with present technology --
where the first coupler is replaced by an optical switch and which allows her to exploit all the photons sent by Alice.).
In the light of this high $\mu$ and of the optical losses (22.8 dB), one may argue that this implementation was
not secure, even when taking into account only so-called realistic eavesdropping strategies (see \ref{beamsplitter}).
Finally, it is possible to estimate the results that other groups would have obtained if they had
used a similar value of $\mu$. One then finds that key distribution distances of the same order could have been achieved.
This illustrates that the distance is a somewhat arbitrary figure of merit for a QC system.}.

\subsubsection{The ``Plug-\&-Play'' systems}\label{PandP1}
As discussed in the two previous sections, both polarization and
phase coding require active compensation of optical path fluctuations. A
simple approach would be to alternate between adjustment periods, where
pulses containing large numbers of photons are exchanged between Alice and Bob
to adjust the compensating system correcting for slow drifts in phase or
polarization, and qubits transmission periods, where the number of photons is
reduced to a quantum level.

\begin{figure}
\infig{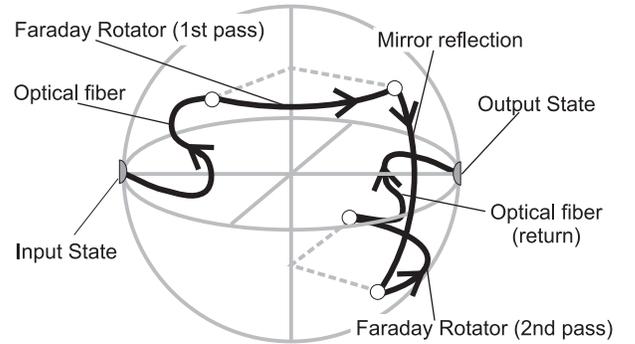}{0.95\columnwidth}
\caption{Evolution of
the polarization state of a light pulse represented on the
Poincar\'{e} sphere over a round trip propagation along an optical
fiber terminated by a Faraday mirror.}
\label{fig4_7}
\end{figure}

An approach invented in 1989 by Martinelli, then at CISE Tecnologie Innovative in
Milano, allows to automatically and passively compensate all polarization fluctuations
in an optical fiber (see also Martinelli, 1992). Let us consider first what
happens to the state of polarization of a pulse of light travelling through an
optical fiber, before being reflected by a Faraday mirror -- a mirror with a
$\frac{\lambda}{4}$ Faraday rotator\footnote{These components, commercially
available, are extremely compact and convenient when using telecommunications
wavelengths, which is not true for other wavelengths.} -- in front, and coming
back. We must first define a convenient description of the change in
polarization of light reflected by a mirror under perpendicular incidence. Let
the mirror be in the x-y plane and z be the optical axis. Clearly, all linear
polarization states are unchanged by a reflection. But right-handed circular
polarization is changed into left-handed and vice-versa. Actually, after a
reflection the rotation continues in the same sense, but since the propagation
direction is reversed, right-handed and left-handed are swapped. The same
holds for elliptic polarization states: the axes of the ellipse are unchanged,
but right and left are exchanged. Accordingly, on the Poincar\'{e} sphere the
polarization transformation upon reflection is described by a symmetry through
the equatorial plane: the north and south hemispheres are exchanged: $\vec
{m}\rightarrow(m_{1},m_{2},-m_{3})$. Or in terms of the qubit state vector:
\begin{equation}
T:\pmatrix{\psi_1\cr\psi_2}\rightarrow\pmatrix{\psi_2^*\cr\psi_1^*}%
\end{equation}
This is a simple representation, but some attention has to be
paid. Indeed this transformation is not a unitary one! Actually,
the above description switches from a right-handed reference frame
$XYZ$ to a left handed one $XY\tilde{Z}$, where $\tilde{Z}=-Z$.
There is nothing wrong in doing so and this explains the
non-unitary polarization transformation\footnote{Note that this
transformation is positive, but not completely positive. It is
thus closely connected to the partial transposition map (Peres
1996). If several photons are entangled, then it is crucial to
describe all of them in frames with the same chirality. Actually
that this is necessary is the content of the Peres-Horodecki
entanglement witness (Horodecki {\it et al.} 1996).}. Note that other
descriptions are possible, but they require to artificially break
the $XY$ symmetry. The main reason for choosing this particular
transformation is that the description of the polarization
evolution in the optical fiber before and after the reflection is
then straightforward. Indeed, let $U=e^{-i\omega
\vec{B}\vec{\sigma}\ell/2}$ describe this evolution under the
effect of some modal birefringence $\vec{B}$ in a fiber section of
length $\ell$ ($\vec{\sigma}$ is the vector whose components
are the Pauli matrices). Then, the evolution after reflection is simply
described by the inverse operator
$U^{-1}=e^{i\omega\vec{B}\vec{\sigma}\ell/2}$. Now that we have a
description for the mirror, let us add the Faraday rotator. It
produces a $\frac{\pi}{2}$ rotation of the Poincar\'{e} sphere
around the north-south axis: $F=e^{-i\pi\sigma_{z}/4}$ (see Fig.
\ref{fig4_7}). Because the Faraday effect is non-reciprocal
(remember that it is due to a magnetic field which can be thought
of as produced by a spiraling electric current), the direction of
rotation around the north-south axis is independent of the light
propagation direction. Accordingly, after reflection on the
mirror, the second passage through the Faraday rotator rotates the
polarization in the same direction (see again Fig. \ref{fig4_7})
and is described by the same operator $F$. Consequently, the total
effect of a Faraday mirror is to change any incoming polarization
state into its orthogonal state $\vec{m}\rightarrow-\vec{m}$. This
is best seen on Fig. \ref{fig4_7}, but can also be expressed
mathematically:
\begin{equation}
FTF:\pmatrix{\psi_1\cr\psi_2}\rightarrow\pmatrix{\psi_2^*\cr-\psi_1^*}%
\end{equation}
Finally, the whole optical fiber can be modelled as consisting of a discrete
number of birefringent elements. If there are N such elements in front of the
Faraday mirror, the change in polarization during a round trip can be
expressed as (recall that the operator FTF only changes the sign of the
corresponding Bloch vector $\vec{m}=\langle\psi|\vec{\sigma}|\psi\rangle$):%
\begin{equation}
U_{1}^{-1}...U_{N}^{-1}FTFU_{N}...U_{1}=FTF
\end{equation}
The output polarization state is thus orthogonal to the input one, regardless
of any birefringence in the fibers. This approach can thus correct for time
varying birefringence changes, provided that they are slow compared to the
time required for the light to make a round trip (a few hundreds of microseconds).

By combining this approach with time-multiplexing in a long path
interferometer, it is possible to implement a quantum cryptography
system based on phase coding where all optical and mechanical
fluctuations are automatically and passively compensated (Muller
{\it et al.} 1997). We performed a first experiment in early 1997
(Zbinden {\it et al.}, 1997), and a key was exchanged over an
installed optical fiber cable of 23 km (the same one as in the
case of polarization coding mentioned before). This setup features
a high interference contrast (fringe visibility of 99.8\%) and an excellent long term stability
and clearly established the value of the approach for QC. The fact
that no optical adjustments are necessary earned it the nickname
of ``plug \& play'' set-up.
It is interesting to note that the idea of combining time-multiplexing with
Faraday mirrors was first used to implement an ``optical microphone''
(Br\'{e}guet and Gisin, 1995)\footnote{Note that since then, we have used this
interferometer for various other applications: non-linear index of refraction
measurement \ in fibers (Vinegoni {\it et al.}, 2000a), optical switch
(Vinegoni {\it et al.}, 2000b).}.

\begin{figure}
\infig{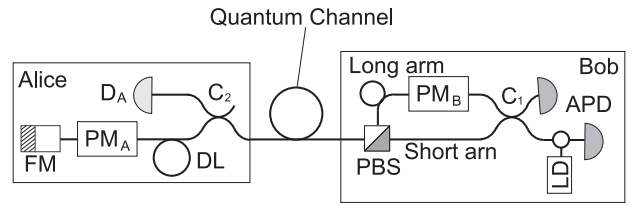}{0.95\columnwidth}
\caption{Self-aligned
``Plug \& Play'' system (LD: laser diode, APD: avalanche
photodiode, C$_{i}$: fiber coupler, PM$_{j}$: phase modulator,
PBS: polarizing beamsplitter, DL: optical delay line, FM: Faraday
mirror, D$_{A}$: classical detector).}
\label{fig4_8}
\end{figure}

However, our first realization still suffered from certain optical
inefficiencies, and has been improved since then. Similar to the
setup tested in 1997, the new system is based on time multiplexing
as well, where the interfering pulses travel along the same
optical path, however, in different time ordering. A schematic is
shown in Fig. \ref{fig4_8}. Briefly, to understand the general
idea, pulses emitted at Bobs can travel either via the short arm
at Bob's, be reflected at the Faraday mirror FM at Alice's and
finally, back at Bobs, travel via the long arm. Or, they travel
first via the long arm at Bob's, get reflected at Alice's, travel
via the short arm at Bob's and then superpose with the first
mentioned possibility on beamsplitter $C_1$. We now explain the
realization of this scheme more in detail: A short and bright
laser pulse is injected in the system through a circulator. It
splits at a coupler. One of the half pulses, labeled P$_{1}$,
propagates through the short arm of Bob's set-up directly to a
polarizing beamsplitter. The polarization transformation in this
arm is set so that it is fully transmitted. P$_{1}$ is then sent
onto the fiber optic link. The second half pulse, labeled P$_{2}$,
takes the long arm to the polarizing beamsplitter. The
polarization evolution is such that it is reflected. A phase
modulator present in this long arm is left inactive so that it
imparts no phase shift to the outgoing pulse. P$_{2}$ is also sent
onto the link, with a delay of the order of 200 ns. Both half
pulses travel to Alice. P$_{1}$ goes through a coupler. The
diverted light is detected with a classical detector to provide a
timing signal. This detector is also important in preventing so
called Trojan Horse attacks discussed in section \ref{Trojan}. The
non-diverted light propagates then through an attenuator and a
optical delay line -- consisting simply of an optical fiber spool
-- whose role will be explained later. Finally it passes a phase
modulator, before being reflected by Faraday mirror. P$_{2}$
follows the same path. Alice activates briefly her modulator to
apply a phase shift on P$_{1}$ only, in order to encode a bit
value exactly like in the traditional phase coding scheme. The
attenuator is set so that when the pulses leave Alice, they do not
contain more than a fraction of a photon. When they reach the PBS
after their return trip through the link, the polarization state
of the pulses is exactly orthogonal to what it was when they left,
thanks to the effect of the Faraday mirror. P1 is then reflected
instead of being transmitted. It takes the long arm to the
coupler. When it passes, Bob activates his modulator to apply a
phase shift used to implement his basis choice. Similarly, P2 is
transmitted and takes the short arm. Both pulses reach the coupler
at the same time and they interfere. Single-photon detectors are
then use to record the output port chosen by the photon.

We implemented with this set-up the full four states BB84
protocol. The system was tested once again on the same installed
optical fiber cable linking Geneva and Nyon (23 km, see Fig.
\ref{fig4_3}) at 1300 nm and observed a very low
QBER$_{opt}\approx1.4\%$ (Ribordy {\it et al.} 1998 and 2000).
Proprietary electronics and software were developed to allow fully
automated and user-friendly operation of the system. Because of
the intrinsically bi-directional nature of this system, great
attention must be paid to Rayleigh backscattering. The light
traveling in an optical fiber undergoes scattering by
inhomogeneities. A small fraction ($\approx$1\%) of this light is
recaptured by the fiber in the backward direction. When the
repetition rate is high enough, pulses traveling to Alice and back
from her must intersect at some point along the line. Their
intensity is however strongly different. The pulses are more than
a thousand times brighter before than after reflection from Alice.
Backscattered photons can accompany a quantum pulse propagating
back to Bob and induce false counts. We avoided this problem by
making sure that pulses traveling from and to Bob are not present
in the line simultaneously. They are emitted in the form of trains
by Bob. Alice stores these trains in her optical delay line, which
consists of an optical fiber spool. Bob waits until all the pulses
of a train have reached him, before sending the next one. Although
it completely solves the problem of Rayleigh backscattering
induced errors, this configuration has the disadvantage of
reducing the effective repetition frequency. A storage line half
long as the transmission line amounts to a reduction of the bit
rate by a factor of approximately three.

Researchers at IBM developed a similar system simultaneously and
independently (Bethune and Risk, 2000), also working at 1300 nm.
However, they avoided the problems associated with Rayleigh
backscattering, by reducing the intensity of the pulses emitted by
Bob. As these cannot be used for synchronization purposes any
longer, they added a classical channel wavelength multiplexed
(1550 nm) in the line, to allow Bob and Alice to synchronize their
systems. They tested their set-up on a 10 km long optical fiber
spool. Both of these systems are equivalent and exhibit similar
performances. In addition, the group of Anders Karlsson at the
Royal Institute of Technology in Stockholm verified in 1999 that
this technique also works at a wavelength of 1550 nm (Bourennane
{\it et al.}, 1999 and Bourennane
{\it et al.}, 2000). These experiments demonstrate the potential
of ``plug \& play''-like systems for real world quantum key
distribution. They certainly constitute a good candidate for
the realization of prototypes.

Their main disadvantage with respect to the other systems
discussed in this section is that they are more sensitive to
Trojan horse strategies (see section \ref{Trojan}). Indeed, Eve
could send a probe beam and recover it through the strong
reflection by the mirror at the end of Alice's system. To prevent
such an attack, Alice adds an attenuator to reduce the amount of
light propagating through her system. In addition, she must
monitor the incoming intensity using a classical linear detector.
Besides, systems based on this approach cannot be operated with a true
single-photon source, and will thus not benefit from the progress
in this field \footnote{The fact that the pulses travel along a
round trip implies that losses are doubled, yielding a reduced
counting rate.}.

\subsection{Frequency coding}\label{Besancon}
Phase based systems for QC require phase synchronization and stabilization.
Because of the high frequency of optical waves (approximately 200 THz at 1550
nm), this condition is difficult to fulfill. One solution is to use
self-aligned systems like the ``plug\&play'' set-ups discussed in the previous
section. Prof. Goedgebuer and his team from the University of Besan\c{c}on, in
France, introduced an alternative solution (Sun {\it et al.} 1995, Mazurenko {\it et al.} 1997, M\'{e}rolla
et al. 1999; see also Molotkov 1998). Note that the title of this section is
not completely correct in the sense that the value of the qubits is not coded
in the frequency of the light, but in the relative phase between sidebands of a central optical frequency.

\begin{figure}
\infig{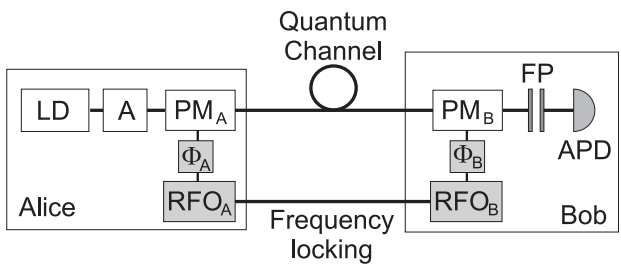}{0.95\columnwidth}
\caption{Implementation
of sideband modulation (LD: laser diode, A: attenuator, PM$_{i}$:
optical phase modulator, $\Phi_{j}$: electronic phase controller,
RFO$_{k}$: radio frequency oscillator, FP:\ Fabry-Perot filter,
APD: avalanche photodiode).}
\label{fig4_9}
\end{figure}

Their system is depicted in Fig. \ref{fig4_9}. A source emits
short pulses of classical monochromatic light with angular
frequency $\omega_{S}$. A first phase modulator $PM_{A}$ modulates
the phase of this beam with a frequency $\Omega\ll\omega_{S}$ and
a small modulation depth. Two sidebands are thus generated at
frequencies $\omega_{S}\pm\Omega$. The phase modulator is driven
by a radio-frequency oscillator $RFO_{A}$ whose phase $\Phi_{A}$
can be varied. Finally, the beam is attenuated so that the
sidebands contain much less than one photon per pulse, while the
central peak remains classical. After the transmission link, the
beam experiences a second phase modulation applied by $PM_{B}$.
This phase modulator is driven by a second radio-frequency
oscillator $RFO_{B}$ with the same frequency $\Omega$\ and a phase
$\Phi_{B}$. These oscillators must be synchronized. After passing
through this device, the beam contains the original central
frequency $\omega_{S}$, the sidebands created by Alice, and the
sidebands created by Bob. The sidebands at frequencies
$\omega_{S}\pm\Omega$ are mutually coherent and thus yield
interference. Bob can then record the interference pattern in
these sidebands, after removal of the central frequency and the
higher order sidebands with a spectral filter.

To implement the B92 protocol (see paragraph \ref{2state}), Alice randomly chooses the value of the phase
$\Phi_{A}$, for each pulse. She associates a bit value of ``0'' to the phase
$0$ and the bit ``1'' to phase $\pi$. Bob also chooses randomly whether to
apply a phase $\Phi_{B}$ of $0$ or $\pi$. One can see that if $\left|
\Phi_{A}-\Phi_{B}\right|  =0$, the interference is constructive and Bob's
single-photon detector has a non-zero probability of recording a
count. This probability depends on the number of photons present initially in the
sideband, as well as the losses induced by the channel. On the other hand, if
$\left|  \Phi_{A}-\Phi_{B}\right|  =\pi$, interference is destructive and no
count will ever be recorded. Consequently, Bob can infer, everytime he records
a count, that he applied the same phase as Alice. When a given pulse does not
yield a detection, the reason can be that the phases applied were different
and destructive interference took place. It can also mean that the phases were
actually equal, but the pulse was empty or the photon got lost. Bob cannot
decide between these two possibilities. From a conceptual point of view, Alice
sends one of two non-orthogonal states. There is then no way for Bob to
distinguish between them deterministically. However he can perform a
generalized measurement, also known as a {\it positive operator value
measurement}, which will sometimes fail to give an answer, and at all other
times gives the correct one.

Eve could perform the same measurement as Bob. When she obtains an
inconclusive result, she could just block both the sideband and the central
frequency so that she does not have to guess a value and does not risk
introducing an error. To prevent her from doing that, Bob verifies the presence
of this central frequency.
Now if Eve tries to conceal her presence by blocking only the sideband,
the reference central frequency will still have a certain probability of
introducing an error.
It is thus possible to catch Eve in both cases. The
monitoring of the reference beam is essential in all two-states protocol to
reveal eavesdropping. In addition, it was shown that this reference beam
monitoring can be extended to the four-states protocol (Huttner {\it et
al.}, 1995).

The advantage of this set-up is that the interference is controlled by the
phase of the radio-frequency oscillators. Their frequency is 6 orders of
magnitude smaller than the optical frequency, and thus considerably easier to
stabilize and synchronize. It is indeed a relatively simple task that can be achieved by
electronic means. The Besan\c{c}on group performed key distribution with such
a system. The source they used was a DBR laser diode at a wavelength of 1540 nm and
a bandwidth of 1 MHz. It was externally modulated to obtain 50 ns pulses, thus increasing the
bandwidth to about 20 MHz. They
used two identical LiNbO$_{3}$ phase modulators operating at a frequency
$\Omega/2\pi=300MHz$. Their spectral filter was a Fabry-Perot cavity with a
finesse of 55. Its resolution was 36 MHz. They performed key distribution over
a 20 km long single-mode optical fiber spool, recording a $QBER_{opt}$
contribution of approximately 4\%. They estimated that 2\% can be attributed
to the transmission of the central frequency by the Fabry-Perot cavity. Note also that the
detector noise is relatively large due to the large pulse durations. Both these errors
could be lowered by increasing the separation between the central peak and the
sidebands, allowing reduced pulse widths, hence shorter detection
times and lower darkcounts.
Nevertheless, a compromise must be found since, in addition to technical drawbacks of
high speed modulation, the polarization
transformation in an optical fiber depends on the wavelength. The remaining
2\% of the $QBER_{opt}$ is due to polarization effects in the set-up.

This system is another possible candidate. It's main
advantage is the fact that it could be used with a true single-photon source,
if it existed. On the other hand, the contribution of imperfect interference
visibility to the error rate is significantly higher than that measured with
``plug\&play'' systems. In addition, if this system is to be truly independent
of polarization, it is essential to ensure that the phase modulators have very
low polarization dependency. In addition, the stability of the frequency
filter may constitute a practical difficulty.

\subsection{Free space line-of-sight applications}\label{Satellite}

Since optical fiber channels
may not always be available, several groups are trying to develop
free space line-of-sight QC systems, capable for
example to distribute a key between buildings rooftops in an urban setting.

It may of course sound difficult to detect single photons amidst background light,
but the first experiments demonstrated the possibility of free space QC.
Besides, sending photons through the atmosphere also has advantages, since this medium is
essentially not birefringent (see paragraph \ref{freespacetransmission}). It is then possible to use
plain polarization coding. In addition, one can ensure a very high channel transmission
over large distances by choosing carefully the wavelength of the photons (see again paragraph \ref{freespacetransmission}).
The atmosphere has for example a high transmission ``window'' in the vicinity of 770 nm
(transmission as high as 80\% between a ground station and a satellite), which
happens to be compatible with commercial silicon APD photon counting modules
(detection efficiency as high as 65\% and low noise).

The systems developed for free space applications are actually very similar to
the one shown in Fig. \ref{fig4_2}. The main difference is that the emitter and
receiver are connected to telescopes pointing at each other, instead of an optical fiber.
The contribution of background light to errors can be maintained at a reasonable level by
using a combination of timing discrimination (coincidence windows of typically a few ns),
spectral filtering ($\le 1$ nm interference filters) and spatial filtering (coupling into an optical fiber).
This can be illustrated
with the following simple calculation. Let us suppose that the isotropic spectral background
radiance is $10^{-2}$ W/m$^{2}$ nm sr at 800 nm. This corresponds to the spectral radiance
of a clear zenith sky with a sun elevation of 77$^{\circ}$ (Zissis and Larocca, 1978). The divergence $\theta$
of a Gaussian beam with radius $w_{0}$ is given by $\theta = \lambda/w_{0}\pi$. The product of
beam (telescope) cross-section and solid angle, which is a constant,
is therefore $\pi w_{0}^{2}\pi \theta^{2}=\lambda^{2}$.
By multiplying the radiance by $\lambda^{2}$, one obtains the spectral power density.
With an interference filter of 1 nm width, the power on the detector is $6 \cdot 10^{-15}$ W,
corresponding to $2 \cdot 10^{4}$ photons per second or $2 \cdot 10^{-5}$ photons per ns time window.
This quantity is approximately two orders of magnitude larger than the dark count probability
of Si APD's, but still compatible with the requirements of QC.
Besides the performance of free space QC systems depends dramatically on
atmospheric conditions and air quality.
This is problematic for urban applications where pollution
and aerosols degrade the transparency of air.

The first free space QC experiment over a distance of more than a few centimeters
\footnote{Remember that Bennett and his coworkers performed the first demonstration
of QC over 30 cm in air (Bennett {\it et al.} 1992a).} was performed by Jacobs and
Franson in 1996. They exchanged a key over a distance of 150 m in a hallway illuminated
with standard fluorescent lighting and 75 m outdoor in bright daylight without excessive QBER.
Hughes and his team were the first to exchange a key over more than one kilometer
under outdoor nighttime conditions (Buttler {\it et al.} 1998, and Hughes {\it et al.} 2000a). More recently, they
even improved their system to reach a distance of 1.6 km under daylight conditions
(Buttler {\it et al.} 2000).
Finally Rarity and his coworkers performed a similar experiment where they exchanged
a key over a distance of 1.9 km under nighttime conditions (Gorman {\it et al.} 2000).

Before quantum repeaters become available and allow to overcome
the distance limitation of fiber based QC, free space systems seem
to offer the only possibility for QC over distances of more than a
few dozens kilometers. A QC link could be established between
ground based stations and a low orbit (300 to 1200 km) satellite.
The idea is first to exchange a key $k_{A}$ between Alice and a
satellite, using QC, next to establish another key $k_{B}$ between
Bob and the same satellite. Then the satellite publicly announces
the value $K=k_{A} \oplus k_{B}$ obtained after an XOR of the two
keys ($\oplus$ represents here the XOR operator or equivalently
the binary addition modulo 2 without carry). Bob subtracts then
his key from this value to recover Alice's key ($k_{A}=K \ominus
k_{B}$) \footnote{This scheme could also be used with optical
fiber implementation provided that secure nodes exist. In the case
of a satellite, one tacitly assumes that it constitutes such a
secure node.}. The fact that the key is known to the satellite
operator may be at first sight seen as a disadvantage. But this
point might on the contrary be a very positive one for the
development of QC, since governments always like to keep control
of communications! Although this has not yet been demonstrated,
Hughes as well as Rarity have estimated - in view of their free
space experiments - that the difficulty can be mastered. The main
difficulty would come from beam pointing - don't forget that the
satellites will move with respect to the ground - and wandering
induced by turbulences. In order to reduce this latter problem the
photons would in practice probably be sent down from the
satellite. Atmospheric turbulences are indeed almost entirely
concentrated on the first kilometer above the earth surface.
Another possibility to compensate for beam wander is to use
adaptative optics. Free space QC experiments over distances of the
order of 2 km constitute major steps towards key exchange with a
satellite. According to Buttler {\it et al.} (2000), the optical
depth is indeed similar to the effective atmospheric thickness
that would be encountered in a surface-to-satellite application.

\subsection{Multi-users implementations}
\begin{figure}
\infig{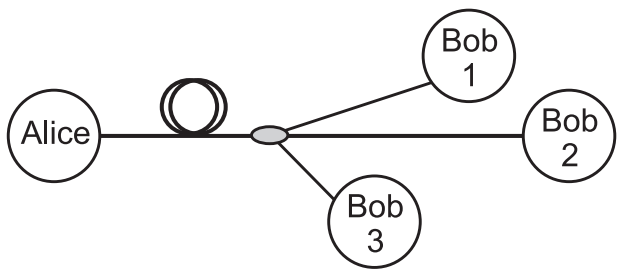}{0.95\columnwidth}
\caption{Multi-users
implementation of quantum cryptography with one Alice connected to
three Bobs by optical fibers. The photons sent by Alice randomly
choose to go to one or the other Bob at a coupler.}
\label{fig4_10}
\end{figure}

Paul Townsend and colleagues investigated the application of QC
over multi-user optical fiber networks (Phoenix et al 1995,
Townsend {\it et al.} 1994, Townsend 1997b). They used a passive optical
fiber network architecture where one Alice -- the network manager
-- is connected to multiple network users (i.e. many Bobs, see
Fig. \ref{fig4_10}). The goal is for Alice to establish a
verifiably secure and unique key with each Bob. In the classical
limit, the information transmitted by Alice is gathered by all
Bobs. However, because of their quantum behavior, the photons are
effectively routed at the beamsplitter to one, and only one, of
the users. Using the double Mach-Zehnder configuration discussed
above, they tested such an arrangement with three Bobs.
Nevertheless, because of the fact that QC requires a direct
and low attenuation optical channel between Alice and Bob, the possibility to implement it over large and
complex networks appears limited.

\newpage
\section{Experimental quantum cryptography with photon pairs}\label{QCPhotonPair}
The possibility to use entangled photon pairs for quantum cryptography was
first proposed by Ekert in 1991. In a subsequent paper, he
investigated, with other researchers, the feasibility of a practical system
(Ekert {\it et al.}, 1992). Although all tests of Bell inequalities (for a
review, see for example, Zeilinger 1999) can be seen as experiments of quantum
cryptography, systems specifically designed to meet the special requirements
of QC, like quick change of bases, were first implemented only recently
\footnote{This definition of quantum cryptography applies to the famous
experiment by Aspect and his co-workers testing Bell inequalities with time
varying analyzers (Aspect {\it et al.}, 1982). QC had however not yet been
invented. It also applies to the more recent experiments closing the
{\it locality loopholes}, like the one performed in Innsbruck using fast
polarization modulators (Weihs {\it et al.} 1998) or the one performed in Geneva
using two analyzers on each side (Tittel {\it et al.} 1999; Gisin and Zbinden
1999).}. In 1999, three groups demonstrated quantum cryptography based on the
properties of entangled photons. They were reported in the same issue of Phys.
Rev. Lett. (Jennewein {\it et al.} 2000b, Naik {\it et al.} 2000, Tittel {\it et al.} 2000),
illustrating the fast progress in the still new field of quantum communication.

When using photon pairs for QC, one advantage lies in the fact that one
can remove empty pulses, since the detection of one photon of a pair reveals
the presence of a companion. In principle, it is thus possible to have a
probability of emitting a non-empty pulse equal to one\footnote{Photon pair
sources are often, though not always, pumped continuously. In these cases, the
time window determined by a trigger detector and electronics defines an
effective pulse.}. It is beneficial only because presently available
single-photon detector feature high dark count probability. The difficulty to
always collect both photons of a pair somewhat reduces this advantage.
One frequently hears that photon-pairs have also the advantage of
avoiding multi-photon pulses, but this is not correct. For a given mean photon
number, the probability that a non-empty pulse contains more than one photon
is essentially the same for weak pulses and for photon pairs (see paragraph \ref{PDC}).
Second, using entangled photons pairs prevents unintended information leakage in
unused degrees of freedom (Mayers and Yao 1998). Observing a QBER smaller than
approximately 15\%, or equivalently
that Bell's inequality is violated, indeed guarantees that the photons are entangled and so that
the different states are not fully distinguishable through other degrees of freedom.
A third
advantage was indicated recently by new
and elaborate eavesdropping analyses. The fact that passive state preparation
can be implemented prevents multiphoton splitting attacks (see section \ref{Passivebasis}).

The coupling between the optical frequency and the property used to encode the
qubit, i.e. decoherence, is rather easy to master when using faint laser
pulses. However, this issue is more serious when using photon pairs, because
of the larger spectral width. For example, for a spectral width of 5 nm FWHM
-- a typical value, equivalent to a coherence time of 1 ps -- and a fiber with
a typical PMD of 0.2 $ps/\sqrt{km}$, transmission over a few kilometers
induces significant depolarization, as discussed in paragraph \ref{PolEffects}.
In case of polarization-entangled photons, this gradually destroys their
correlation. Although it is in principle possible to compensate this effect,
the statistical nature of the PMD makes this impractical\footnote{In the case
of weak pulses we saw that a full round trip together with the use of Faraday
mirrors circumvents the problem (see paragraph \ref{PandP1}). However, since the
channel loss on the way from the source to the Faraday mirror inevitably
increases the empty pulses fraction, the main advantage of photon pairs
vanishes in such a configuration.}. Although perfectly fine for free-space QC
(see section \ref{Satellite}), polarization entanglement is thus not adequate
for QC over long optical fibers. A similar effect arises when dealing with
energy-time entangled photons. Here, the chromatic dispersion destroys the
strong time-correlations between the photons forming a pair. However, as
discussed in paragraph \ref{CD}, it is possible to passively compensate for this
effect using either additional fibers with opposite dispersion, or exploiting
the inherent energy correlation of photon pairs.

Generally speaking, entanglement based systems are far more complex than faint
laser pulses set-ups. They will most certainly not be used in the short term
for the realization of industrial prototypes.
In addition the current experimental key creation rates obtained with these systems are at least two orders of magnitude smaller
than those obtained with faint laser pulses set-ups (net rate in the order of a few tens of bits per second rather than a few thousands bits per second for a 10 km distance).
Nevertheless, they offer
interesting possibilities in the context of cryptographic optical networks The
photon pairs source can indeed be operated by a key provider and situated
somewhere in between potential QC customers. In this case, the operator of the
source has no way to get any information about the key obtained by Alice and Bob.

It is interesting to emphasize the close analogy between 1 and 2-photon
schemes, which was first noted by Bennett, Brassard and Mermin (1992).
Indeed, in a 2-photon scheme, one can always consider that when Alice
detects her photon, she effectively prepares Bob's photon in a given state. In
the 1-photon analog, Alice's detectors are replaced by sources, while the
photon pair source between Alice and Bob is bypassed. The difference between
these schemes lies only in practical issues, like the spectral widths of the
light. Alternatively, one can look at this analogy from a different point of
view: in 2-photon schemes, everything is as if Alice's photon propagated
backwards in time from Alice to the source and then forwards from the source
to Bob.

\subsection{Polarization entanglement}
\begin{figure}
\infig{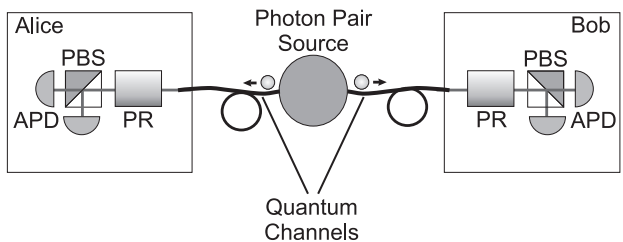}{0.95\columnwidth} \caption{Typical system
for quantum cryptography exploiting photon pairs entangled in
polarization (PR: active polarization rotator, PBS: polarizing
beamsplitter, APD: avalanche photodiode).} \label{fig5_1}
\end{figure}

A first class of experiments takes advantage of
polarization-entangled photon pairs. The setup, depicted in Fig.
\ref{fig5_1}, is similar to the scheme used for polarization
coding based on faint pulses. A two-photon source emits pairs of
entangled photons flying back to back towards Alice and Bob. Each
photon is analyzed with a polarizing beamsplitter whose
orientation with respect to a common reference system can be
changed rapidly. Two experiments, have been reported in the spring
of 2000 (Jennewein {\it et al.} 2000b, Naik {\it et al.} 2000).
Both used photon pairs at a wavelength of 700 nm, which were
detected with commercial single photon detectors based on Silicon
APD's. To create the photon pairs, both groups took advantage of
parametric downconversion in one or two BBO crystals pumped by an
argon-ion laser. The analyzers consisted of fast modulators, used
to rotate the polarization state of the photons, in front of
polarizing beamsplitters.

The group of Anton Zeilinger, then at the University of Innsbruck,
demonstrated such a crypto-system, including error correction, over a distance
of 360 meters (Jennewein {\it et al.} 2000b). Inspired by a test of Bell
inequalities performed with the same set-up a year earlier (Weihs {\it et
al.}, 1998), the two-photon source was located near the center between the two
analyzers. Special optical fibers, designed for guiding only a single mode at
700 nm, were used to transmit the photons to the two analyzers. The results of
the remote measurements were recorded locally and the processes of key sifting
and of error correction implemented at a later stage, long after the
distribution of the qubits. Two different protocols were implemented: one
based on Wigner's inequality (a special form of Bell inequalities), and the
other one following BB84.

The group of Paul Kwiat then at Los Alamos National Laboratory, demonstrated
the Ekert protocol (Naik {\it et al.} 2000). This
experiment was a table-top realization with the source and the analyzers only
separated by a few meters. The quantum channel consisted of a short free space
distance. In addition to performing QC, the researchers simulated different
eavesdropping strategies as well. As predicted by the theory, they observed a
rise of the QBER with an increase of the information obtained by the eavesdropper.
Moreover, they also recently implemented the six-state protocol
described in paragraph \ref{6state}, and observed the predicted QBER increase to 33\%
(Enzer {\it et al.} 2001).

The main advantage of polarization entanglement is the fact that analyzers are
simple and efficient. It is therefore relatively easy to obtain high
contrast. Naik and co-workers, for example, measured a polarization
extinction of 97\%, mainly limited by electronic imperfections of the fast
modulators. This amounts to a $QBER_{opt}$ contribution of only 1.5\%. In
addition, the constraint on the coherence length of the pump laser is not very
stringent (note that if it is shorter than the length of the crystal some
difficulties can appear, but we will not mention them here).

In spite of their qualities, it would be difficult to reproduce these experiments
on distances of more than a few kilometers of optical fiber. As mentioned in
the introduction to this chapter, polarization is indeed not robust enough to
decoherence in optical fibers. In addition, the polarization state transformation induced by an
installed fiber frequently fluctuates, making an active alignment system
absolutely necessary. Nevertheless, these experiments are very interesting in
the context of free space QC.

\subsection{Energy-time entanglement}\label{ETent}

\subsubsection{Phase-coding}\label{PhCoding}
\begin{figure}
\infig{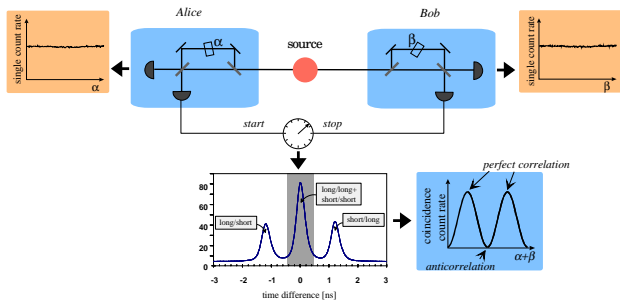}{0.95\columnwidth} \caption{Principle of
phase coding quantum cryptography using energy-time entangled
photons pairs.} \label{fig5_2}
\end{figure}

\begin{figure}
\infig{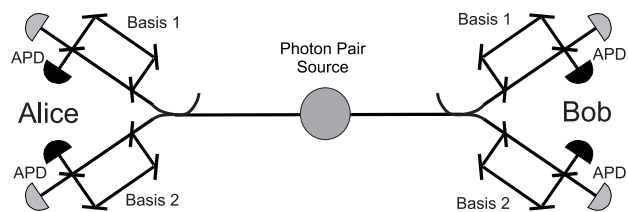}{0.95\columnwidth} \caption{System for
phase-coding entanglement based quantum cryptography (APD:
avalanche photodiode). The photons choose their bases randomly at Alice and
Bob's couplers.} \label{fig5_3}
\end{figure}

The other class of experiments takes advantage of energy-time
entangled photon pairs. The idea originates from an arrangement
proposed by Franson in 1989 to test Bell inequalities. As we will
see below, it is comparable to the double Mach-Zehnder
configuration discussed in section \ref{DoubleMZ}. A source emits
pairs of energy-correlated photons with both particles created at
exactly the same, however uncertain time (see Fig. \ref{fig5_2}).
This can be achieved by pumping a non-linear crystal with a pump
of large coherence time. The pairs of down-converted photons are
then split, and one photon is sent to each party down quantum
channels. Both Alice and Bob possess a widely, but identically
unbalanced Mach-Zehnder interferometer, with photon counting
detectors connected to the outputs. Locally, if Alice or Bob
change the phase of their interferometer, no effect on the count
rates is observed, since the imbalancement prevents any
single-photon interference. Looking at the detection-time at Bob's
with respect to the arrival time at Alice's, three different
values are possible for each combination of detectors. The
different possibilities in a time spectrum are shown in Fig.
\ref{fig5_2}. First, both photons can propagate through the short
arms of the interferometers. Next, one can take the long arm at
Alice's, while the other one takes the short one at Bob's. The
opposite is also possible. Finally, both photons can propagate
through the long arms. When the path differences of the
interferometers are matched within a fraction of the coherence
length of the down-converted photons, the short-short and the
long-long processes are indistinguishable, provided that the
coherence length of the pump photon is larger than the path-length
difference. Conditioning detection only on the central time peak,
one observes two-photon interferences which depends on the sum of
the relative phases in Alice's and Bob's interferometer --
non-local quantum correlation (Franson 1989)\footnote{The
imbalancement of the interferometers must be large enough so that
the middle peak can easily be distinguished from the satellite
ones. This minimal imbalancement is determined by the convolution
of the detector's jitter (tens of ps), the electronic jitter (from
tens to hundreds of ps) and
the single-photon coherence time (%
$<$%
1ps).} -- see Fig. \ref{fig5_2}. The phase in the interferometers
at Alice's and Bob's can, for example, be adjusted so that both
photons always emerge from the same output port. It is then
possible to exchange bits by associating values to the two ports.
This is, however, not sufficient. A second measurement basis must
be implemented, to ensure security against eavesdropping attempts.
This can be done for example by adding a second interferometer
to the systems (see Fig. \ref{fig5_3}). In the latter case, when
reaching an analyzer, a photon chooses randomly to go to one or
the other interferometer. The second set of interferometers can be
adjusted to also yield perfect correlations between output ports.
The relative phase between their arms should however be chosen so
that when the photons go to interferometers not associated, the
outcomes are completely uncorrelated.

\begin{figure}
\infig{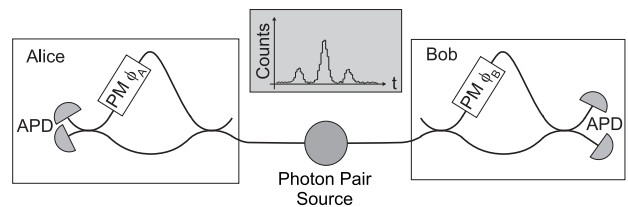}{0.95\columnwidth} \caption{Quantum
cryptography system exploiting photons entangled in energy-time
and active basis choice. Note the similarity with the faint laser
double Mach-Zehnder implementation depicted in Fig. \ref{fig4_6}.}
\label{fig5_4}
\end{figure}

Such a system features a passive state preparation by Alice,
yielding security against multiphoton splitting attacks (see
section \ref{Passivebasis}). In addition, it also features a
passive basis choice by Bob, which constitutes an elegant
solution: neither a random number generator, nor an active
modulator are necessary. It is nevertheless clear that
QBER$_{det}$ and QBER$_{acc}$ (defined in eq. (\ref{QBERss})) are
doubled since the number of activated detectors is twice as high.
This disadvantage is however not as important as it first appears
since the alternative, a fast modulator, introduces losses close
to 3dB, also resulting in an increase of these error
contributions.
The striking similarity between this scheme and the double
Mach-Zehnder arrangement discussed in the context of faint laser
pulses in section \ref{DoubleMZ} is obvious when comparing Fig.
\ref{fig5_4} and Fig. \ref{fig4_6}!

\begin{figure}
\infig{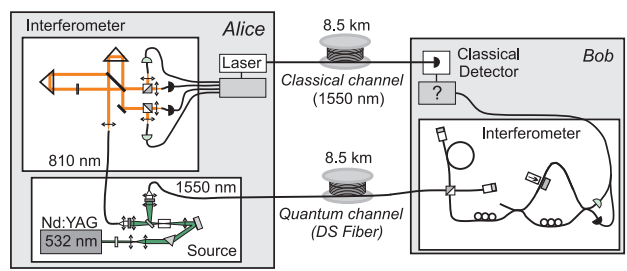}{0.95\columnwidth} \caption{Schematic
diagram of the first system designed and optimized for
long distance quantum cryptography and exploiting phase
coding of entangled photons.} \label{fig5_5}
\end{figure}

This scheme has been realized in the first half of 2000 by our
group at Geneva University (Ribordy {\it et al.}, 2001).
It constitutes the first experiment in which an asymmetric setup, optimized for QC
was used instead of a system designed for
tests of Bell inequality and having a source located in the center between
Alice and Bob (see Fig. \ref{fig5_5}).
The two-photon source (a
KNbO$_3$ crystal pumped by a doubled Nd-YAG laser) provides
energy-time entangled photons at non-degenerate wavelengths -- one
around 810 nm, the other one centered at 1550 nm. This choice
allows to use high efficiency silicon based single photon counters
featuring low noise to detect the photons of the lower wavelength.
To avoid the high transmission losses at this wavelength in
optical fibers, the distance between the source and the
corresponding analyzer is very short, of the order of a few
meters. The other photon, at the wavelength where fiber losses are
minimal, is sent via an optical fiber to Bob's interferometer and
is then detected by InGaAs APD's. The decoherence induced by
chromatic dispersion is limited by the use of dispersion-shifted
optical fiber (see section \ref{CD}).

Implementing the BB84 protocols in the way discussed above, with a
total of four interferometers, is difficult. They must indeed be
aligned and their relative phase kept accurately stable during the
whole key distribution session. To simplify this problem, we
devised birefringent interferometers with polarization
multiplexing of the two bases. Consequently, the constraint on the
stability of the interferometers is equivalent to that encountered
in the faint pulses double Mach-Zehnder system. We obtained
interference visibilities of typically 92\%, yielding in turn a
$QBER_{opt}$ contribution of about 4\%.
We demonstrated QC over a transmission distance of 8.5 km in a
laboratory setting using a fiber on a spool and generated several
Mbits of key in hour long sessions. This is the largest span
realized to date for QC with photon pairs.

As already mentioned, it is essential for this scheme to have a
pump laser whose coherence length is larger than the path
imbalancement of the interferometers. In addition, its wavelength
must remain stable during a key exchange session. These
requirements imply that the pump laser must be somewhat more
elaborate than in the case of polarization entanglement.

\subsubsection{Phase-time coding}\label{PhaseTimeCoding}
We have mentioned in section \ref{PhaseExp} that states generated by two-paths
interferometers are two-levels quantum systems. They can also be represented
on a Poincar\'{e} sphere. The four-states used for phase coding in the
previous section would lie on the equator of the sphere, equally distributed.
The coupling ratio of the beamsplitter is indeed 50\%, and they differ only by
a phase difference introduced between the components propagating through
either arm. In principle, the four-state protocol can be equally well
implemented with only two states on the equator and the two other ones on the
poles. In this section, we present a system exploiting such a set of states.
Proposed by our group in 1999 (Brendel {\it et al.}, 1999), the scheme
follows in principle the Franson configuration described in the context of
phase coding. However, it is based on a pulsed source emitting entangled
photons in so-called energy-time Bell states (Tittel {\it et al.} 2000).
The emission time of the photon
pair is therefore given by a superposition of only two discrete terms, instead
of a wide and continuous range bounded only by the large coherence length of
the pump laser (see paragraph \ref{PhCoding}).

\begin{figure}
\infig{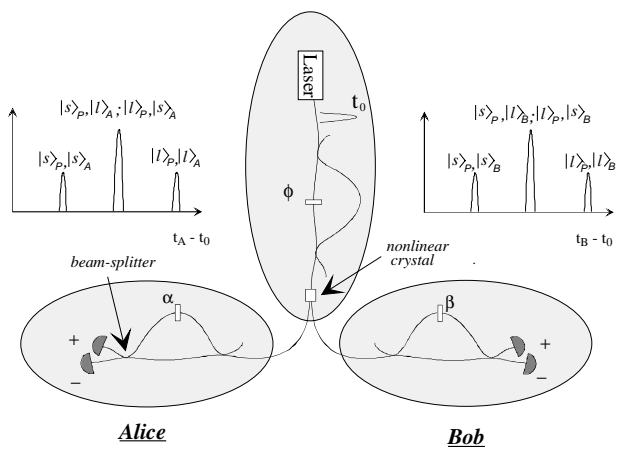}{0.95\columnwidth} \caption{Schematics of
quantum cryptography using entangled photons phase-time coding.}
\label{fig5_6}
\end{figure}

Consider Fig. \ref{fig5_6}. If Alice registers the arrival times
of the photons with respect to the emission time of the pump pulse
$t_{0}$, she finds the photons in one of three time slots (note
that she has two detectors to take into account). For instance,
detection of a photon in the first slot corresponds to ``pump
photon having traveled via the short arm and downconverted photon
via the short arm''. To keep it short, we refer to this process as
$|\,s\,\rangle _{P},|\,s\,\rangle_{A}$, where $P$ stands for the
pump- and $A$ for Alice's photon\footnote{Note that it does not
constitute a product state.}. However, the characterization of the
complete photon pair is still ambiguous, since, at this point, the
path of the photon having traveled to Bob (short or long in his
interferometer) is unknown to Alice. Figure \ref{fig5_6} illustrates
all processes leading to a detection in the different time slots
both at Alice's and at Bob's detector. Obviously, this reasoning
holds for any combination of two detectors. In order to build up
the secret key, Alice and Bob now publicly agree about the events
where both detected a photon in one of the satellite peaks --
without revealing in which one -- or both in the central peak --
without revealing the detector. This procedure corresponds to
key-sifting. For instance, in the example discussed above, if Bob
tells Alice that he also detected his photon in a satellite peak,
she knows that it must have been the left peak as well. This is
due to the fact that the pump photon has traveled via the short
arm -- hence Bob can detect his photon either in the left
satellite or in the central peak. The same holds for Bob who now
knows that Alice's photon traveled via the short arm in her
interferometer. Therefore, in case of joint detection in a
satellite peak, Alice and Bob must have correlated detection
times. Assigning a bit value to each side peak, Alice and Bob can
exchange a sequence of correlated bits.

The cases where both find the photon in the central time slot are used to
implement the second basis. They correspond to the $|\,s\,\rangle
_{P},|\,l\,\rangle_{A}|\,l\,\rangle_{B}$ and $|\,l\,\rangle_{P},|\,s\,\rangle
_{A}|\,s\,\rangle_{B}$ possibilities. If these are indistinguishable, one
obtains two-photon interferences, exactly as in the case discussed in the
previous paragraph on phase coding. Adjusting the phases, and maintaining them
stable, perfect correlations between output ports chosen by the photons at
Alice's and Bob's interferometers are used to establish the key bits in this
second basis.

Phase-time coding has recently been implemented in a laboratory experiment by
our group (Tittel et al., 2000) and was reported at the same time as the two
polarization entanglement-based schemes mentioned above. A contrast of
approximately 93\% was obtained, yielding a $QBER_{opt}$ contribution of
3.5\%, similar to that obtained with the phase coding scheme. This experiment
will be repeated over long distances, since losses in optical fibers are low
at the downconverted photons' wavelength (1300 nm).

An advantage of this set-up is that coding in the time basis is particularly
stable. In addition, the coherence length of the pump laser is not critical
anymore. It is however necessary to use relatively short pulses ($\approx$ 500
ps) powerful enough to induce a significant downconversion probability.

Phase-time coding, as discussed in this section, can also be realized with
faint laser pulses (Bechmann-Pasquinucci and Tittel, 2000). The 1-photon
configuration has though never been realized. It would be similar to the
double Mach-Zehnder discussed in paragraph \ref{DoubleMZ}, but with the first
coupler replaced by an active switch. For the time-basis, Alice would set the
switch either to full transmission or to full reflection, while for the
energy-basis she would set it at 50\%. This illustrates how considerations
initiated on photon pairs can yield advances on faint pulses systems.

\subsubsection{Quantum secret sharing}
In addition to QC using phase-time coding, we used the setup
depicted in Fig. \ref{fig5_6} for the first proof-of-principle
demonstration of quantum secret sharing -- the generalization of
quantum key distribution to more than two parties (Tittel {\it et
al.}, 2001). In this new application of quantum communication,
Alice distributes a secret key to two other users, Bob and
Charlie, in a way that neither Bob nor Charlie alone have any
information about the key, but that together they have full
information. Like with traditional QC, an eavesdropper trying to
get some information about the key creates errors in the
transmission data and thus reveals her presence. The motivation
behind quantum secret sharing is to guarantee that Bob and Charlie
cooperate -- one of them might be dishonest -- in order to obtain
a given piece of information. In contrast with previous proposals
using three-particle GHZ states (\.Zukowski {\it et al.},1998, and
Hillery {\it et al.}, 1999), pairs of entangled photons in
so-called energy-time Bell states were used to mimic the necessary
quantum correlation of three entangled qubits, albeit only two
photons exist at the same time. This is possible because of the
symmetry between the preparation device acting on the pump pulse
and the devices analyzing the down-converted photons. Therefore,
the emission of a pump pulse can be considered as the detection of
a photon with 100\% efficiency, and the scheme features a much
higher coincidence rate than that expected with the initially
proposed ``triple-photon'' schemes.

\newpage

\section{Eavesdropping}\label{Eve}
\subsection{Problems and Objectives}
After the qubit exchange and bases reconciliation, Alice and Bob each have a sifted key.
Ideally, these are identical. But in real life, there are always some errors and Alice and
Bob must apply some classical information processing protocols, like error correction and
privacy amplification, to their data (see paragraph \ref{ECPA}). The first protocol is
necessary to obtain identical keys, the second to obtain a secret key. Essentially, the
problem of eavesdropping is to find protocols which, given that Alice and Bob
can only measure the QBER, either provides Alice and Bob with a provenly secure key,
or stops the protocol and informs the users that the key distribution has failed.
This is a delicate question, really at the intersection
between quantum physics and information theory. Actually,
there is not one, but several eavesdropping problems, depending on the precise
protocol, on the degree of idealization one admits, on the
technological power one assumes Eve has and on the assumed
fidelity of Alice and Bob's equipment. Let us immediately
stress that the complete analysis of eavesdropping on quantum channel is
by far not yet finished. In this chapter we review some of the problems and solutions,
without any claim of mathematical rigor nor complete cover of the huge and fast evolving
literature.

The general objective of eavesdropping analysis is to find ultimate and practical
proofs of security for some quantum cryptosystems. Ultimate means that the security is
guaranteed against entire classes of eavesdropping attacks, even if Eve uses not only the best
of today's technology, but any conceivable technology of tomorrow. They take the
form of theorems, with clearly stated assumptions expressed in mathematical terms.
In contrast, practical proofs deal with some actual pieces of hardware and software.
There is thus a tension between ``ultimate'' and ``practical'' proofs. Indeed the first ones
favor general abstract assumptions, whereas the second ones concentrate on physical implementations
of the general concepts. Nevertheless, it is worth
aiming at finding such proofs. In addition to the security issue, they provide illuminating lessons for
our general understanding of quantum information.

In the ideal game Eve has perfect technology: she is only limited by
the laws of quantum mechanics, but not at all by today's technology \footnote{The question
whether QC would survive the discovery of the currently unknown validity limits of quantum
mechanics is interesting. Let us argue that it is likely that quantum mechanics will always
adequately describe photons at telecom and vsible wavelengths,
like classical mechanics always adequately describes
the fall of apples, whatever the future of physics might be.}. In particular,
Eve cannot clone the qubits, as this is incompatible with quantum dynamics (see paragraph \ref{NoQCM}),
but Eve is free to use any unitary interaction between one or several qubits and an auxiliary
system of her choice. Moreover, after the interaction,
Eve may keep her auxiliary system unperturbed, in particular
in complete isolation from the environment, for an arbitrarily long time. Finally, after
listening to all the public discussion between Alice and Bob, she can perform the
measurement of her choice on her system, being again limited only by the laws of quantum
mechanics.
Moreover, one assumes that all errors are due to Eve. It is tempting to assume
that some errors are due to Alice's and Bob's instruments and this probably makes sense in practice.
But there is the danger that Eve replaces them with higher quality instruments (see next
section)!

In the next section we elaborate on the most relevant differences between the above
ideal game (ideal especially from Eve's point of view!) and real
systems. Next, we return to the idealized situation and present several eavesdropping strategies,
starting from the simplest ones, where explicit formulas can be written down
and ending with a general abstract security proof. Finally, we discus practical eavesdropping
attacks and comment on the complexity of real system's security.

\subsection{Idealized versus real implementation}\label{IdealReal}
Alice and Bob use technology available today. This trivial remark has several implications.
First, all real components are imperfect, so that the qubits are prepared and
detected not exactly in the basis described by the theory. Moreover,
a real source always has a finite probability to produce more than one photon.
Depending on the details of the encoding device, all photons carry the same qubit (see
section \ref{Passivebasis}).
Hence, in principle, Eve could measure the photon number, without perturbing the qubit. This is
discussed in section \ref{QND}. Recall that
ideally, Alice should emit single qubit-photons, i.e. each logical qubit should be encoded
in a single degree of freedom of a single photon.

On Bob's side the situation is, first, that the efficiency of his detectors is quite limited and,
next, that the dark counts (spontaneous counts not produced by photons) are non negligible. The
limited efficiency is analogous to the losses in the quantum channel. The analysis
of the dark counts is more delicate and no complete solution is known. Conservatively,
L\"utkenhaus (2000) assumes in his analysis that all dark counts provide information to Eve. He also
advises that whenever two detectors fire simultaneously (generally due to a real photon and
a dark count), Bob should not disregard such events but choose a value at random. Note also
that the different contributions of dark count to the total QBER depend on whether Bob's choice
of basis is implemented using an active or a passive switch (see section \ref{SQBER}).

Next, one usually assumes that Alice and Bob have thoroughly
checked their equipments and that it is functioning according to the specifications.
This is not particular to quantum cryptography, but is quite a delicate
question, as Eve could be the actual manufacturer of the equipment!
Classical crypto-systems must also be carefully tested, like any commercial apparatuses.
Testing a crypto-system is however delicate, because
in cryptography the client buys confidence and security, two qualities difficult to quantify.
D. Mayers and A. Yao (1998)
proposed to use Bell inequality to test that the equipments really obey quantum
mechanics,
but even this is not entirely satisfactory. Indeed and interestingly, one of the most subtle
loopholes in all present day tests of Bell inequality, the detection loophole,
can be exploited to produce a purely
classical software mimicking all quantum correlation (Gisin and Gisin 1999).
This illustrates once again
how close practical issues in QC are to philosophical debates
about the foundations of quantum physics!

Finally, one has to assume that Alice and Bob are perfectly isolated from Eve.
Without such an assumption the entire
game would be meaningless: clearly, Eve is not allowed to look over Alice's shoulder!
But this elementary assumption is again a nontrivial one. What if Eve uses the quantum channel
connecting Alice to the outside world? Ideally, the channel should incorporate an isolator
\footnote{Optical isolators, based on the Faraday effect, let light pass through only in one
direction.} to keep Eve from shining light into Alice's output port to examine the
interior of her laboratory.
But all isolators operate only on a finite bandwidth, hence there should also be a filter. But
filters have only a finite efficiency. And so on. Except for section \ref{Trojan} where
this assumption is discussed, we henceforth assume that
Alice and Bob are isolated from Eve.

\subsection{Individual, joint and collective attacks}\label{EveAttacks}
In order to simplify the problem, several eavesdropping strategies of restricted
generalities have been defined (L\"utkenhaus 1996, Biham and Mor 1997a and 1997b) and analyzed.
Of particular interest is the assumption that Eve attaches independent probes to each qubit
and measures her probes one after the other. This class of attacks is called
{\it individual attacks}, also known as {\it incoherent
attacks}. This important class is analyzed in sections \ref{int-resend} and \ref{SymAttack}.
Two other classes of eavesdropping strategies let Eve process several qubits coherently, hence
the name of  {\it coherent attacks}.
The most general coherent attacks are called {\it joint attacks}, while an intermediate
class assumes that Eve attaches one probe per qubit, like in individual attacks, but can
measure several probes coherently, like in coherent attacks. This intermediate class is
called {\it collective attacks}. It is not known whether this class is less efficient than
the most general joint one. It is also not known whether it is more
efficient than the simpler individual attacks. Actually, it is not even known whether
joint attacks are more efficient than individual ones!

For joint and collective
attacks, the usual assumption is that Eve measures her probe only after Alice and Bob have completed all
their public discussion about bases reconciliation, error correction and privacy
amplification. But for the more realistic individual attacks, one assumes that Eve
waits only until the bases reconciliation phase of the public discussion\footnote{With
today's technology, it might even be fair to assume, in individual attacks, that Eve
must measure her probe before the basis reconciliation.}.
The motivation for this is that one hardly sees what Eve could gain waiting for the public
discussion on error correction and privacy amplification before measuring her probes, since
she is anyway going to measure them independently.

Individual attacks have the nice feature that the problem can be
entirely translated into a classical one: Alice, Bob and Eve all have classical
information in the form of random variables $\alpha$, $\beta$ an $\epsilon$,
respectively, and the laws of quantum mechanics imposes constraints on the joint probability
distribution $P(\alpha,\beta,\epsilon)$. Such classical scenarios have been widely studied by the classical
cryptology community and many results can thus be directly applied.

\subsection{Simple individual attacks: intercept-resend, measurement in the intermediate basis}\label{int-resend}
The simplest attack for Eve consists in intercepting all photons individually, to measure them in a basis
chosen randomly among the two bases used by Alice and to send new photons to Bob prepared
according to her result. As
 presented in paragraph \ref{IRstrat} and assuming that the BB84 protocol is used,
Eve gets thus 0.5 bit of information per bit in
the sifted key, for an induced QBER of 25\%.
Let us illustrate the general formalism on this
simple example. Eve's mean information gain on Alice's bit, $I(\alpha,\epsilon)$,
equals their relative entropy decrease:
\beq
I(\alpha,\epsilon)=H_{a~priori}-H_{a~posteriori}
\eeq
i.e. $I(\alpha,\beta)$ is the number of bits one can save writing $\alpha$
when knowing $\beta$.
Since the a priori probability for Alice's bit is uniform, $H_{a~priori}=1$. The a posteriori
entropy has to be averaged over all possible results $r$ that Eve might get:
\beq
H_{a~posteriori}=\sum_r P(r)H(i|r)
\eeq
\beq
H(i|r)=-\sum_i P(i|r)\log(P(i|r))
\eeq
where the a posteriori probability of bit $i$ given Eve's result $r$ is given by Bayes's
theorem:
\beq
P(i|r)=\frac{P(r|i)P(i)}{P(r)}
\eeq
with $P(r)=\sum_i P(r|i)P(i)$. In the case of intercept-resend, Eve gets one out of 4 possible
results: $r\in\{\up,\down,\lefta,\righta\}$. After the basis has been revealed, Alice's input
assumes one out of 2 values: $i\in\{\up,\down\}$ (assuming the $\up\down$ basis was used, the other
case is completely analogous). One gets $P(i=\up|r=\up)=1$, $P(i=\up|r=\righta)=\half$ and
$P(r)=\half$. Hence, $I(\alpha,\epsilon)=1-\half h(1)-\half h(\half)=1-\half=\half$
(with $h(p)=p\log_2(p)+(1-p)\log_2(1-p)$).

\begin{figure}
\infig{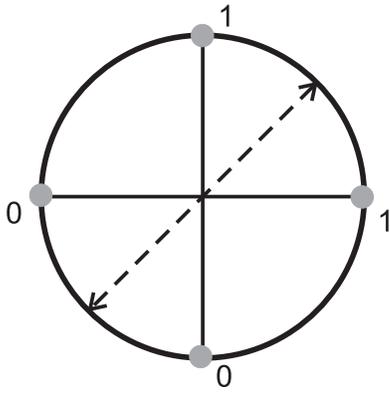}{0.6\columnwidth} \caption{Poincar\'{e}
representation of the BB84 states and the intermediate basis, also
known as the Breidbart basis, that can be used by Eve.}
\label{fig6_1}
\end{figure}

Another strategy for Eve, not more difficult to implement,
consists in measuring the photons in the intermediate basis (see
Fig. \ref{fig6_1}), also known as the Breidbart basis (Bennett et
al. 1992a). In this way the probability that Eve guesses the
correct bit value is
$p=\cos(\pi/8)^2=\half+\frac{\sqrt{2}}{4}\approx0.854$,
corresponding to a QBER=$2p(1-p)=25\%$ and Shannon information
gain per bit of
\beq
I=1-H(p)\approx 0.399.
\eeq
Consequently, this strategy is less advantageous for Eve than
the intercept-resend one. Note however, that with this strategy
Eve's probability to guess the correct bit value is 85.\%,
compared to only 75\% in the intercept-resend case. This is
possible because in the latter case Eve's information is
deterministic in half the cases, while in the first one Eve's information is always probabilistic (formally this results from the
convexity of the entropy function).

\subsection{Symmetric individual attacks}\label{SymAttack}
In this section we present in some details how Eve could get a maximum Shannon
information for a fixed QBER, assuming a perfect single qubit source and restricting
Eve to attacks on one qubit after the other (i.e. individual attacks).
The motivation is that this
idealized situation is rather easy to treat and nicely illustrates several of the
subtleties of the subject. Here we concentrate on the BB84 4-state protocol, for
related results on the 2-state and the 6-state protocols see Fuchs and Peres (1996) and
Bechmann-Pasquinucci and Gisin (1999), respectively.

\begin{figure}
\infig{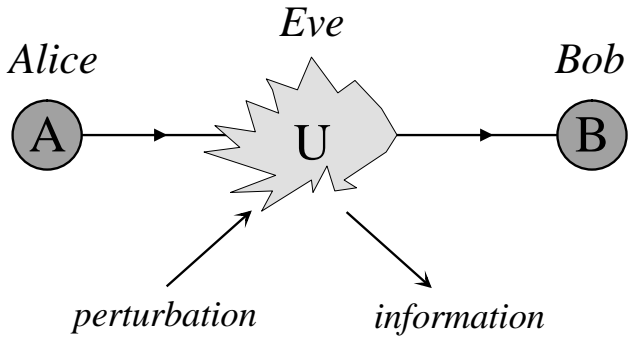}{0.95\columnwidth} \caption{Eavesdropping
on a quantum channel. Eve extracts information out of the quantum
channel between Alice and Bob at the cost of introducing noise
into that channel.} \label{fig6_2}
\end{figure}

The general idea of eavesdropping on a quantum channel goes as
follows. When a qubit propagates from Alice to Bob, Eve can let a
system of her choice, called a probe, interact with the qubit (see
Fig. \ref{fig6_2}). She can freely choose the probe and its
initial state, but it has to be a system satisfying the quantum
rules (i.e. described in some Hilbert space). Eve can also choose
the interaction, but it should be independent of the qubit state and she should follow the laws of quantum mechanics, i.e.
her interaction is described by a unitary operator.
After the interaction a qubit
has to go to Bob (in section \ref{QND} we consider lossy channels,
so that Bob does not always expect a qubit, a fact that Eve can
take advantage of). It makes no difference whether this qubit is
the original one (possibly in a modified state) or not. Actually
the question does not even make sense since a qubit is nothing but
a qubit! But in the formalism it is convenient to use the same
Hilbert space for the qubit sent by Alice and that received by Bob
(this is no loss of generality, since the swap operator -- defined by
$\psi\otimes\phi\rightarrow\phi\otimes\psi$ for all $\psi$,$\phi$ -- is unitary
and could be appended to Eve's interaction).

\begin{figure}
\infig{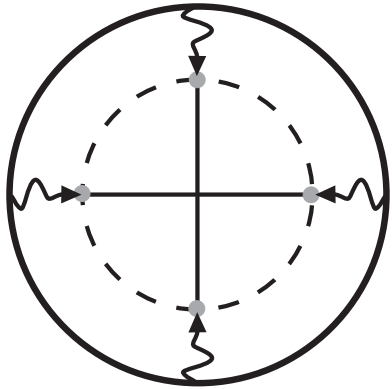}{0.6\columnwidth} \caption{Poincar\'{e}
representation of the BB84 states in the event of a symmetrical
attack. The state received by Bob after the interaction of Eve's
probe is related to the one sent by Alice by a simple shrinking
factor. When the unitary operator $U$ entangles the qubit and Eve's
probe, Bob's state (eq. \ref{SymmAtt}) is mixed and is represented by a point inside
the Poincar\'e sphere.} \label{fig6_3}
\end{figure}

Let $\H_{Eve}$ and $\C^2\otimes\H_{Eve}$ be the Hilbert spaces of
Eve's probe and of the total qubit+probe system, respectively. If
$|\vec m\>$, $|0\>$ and $U$ denote the qubit and the probe's
initial states and the unitary interaction, respectively, then the
state of the qubit received by Bob is given by the density matrix
obtained by tracing out Eve's probe: \beq \rho_{Bob}(\vec
m)=Tr_{\H_{Eve}}(U|\vec m,0\>\<\vec m,0|U^\dagger). \eeq The
symmetry of the BB84 protocol makes it very natural to assume that
Bob's state is related to Alice's $|\vec m\>$ by a simple
shrinking factor\footnote{Chris Fuchs and Asher Peres were the
first ones to derive the result presented in this section, using
numerical optimization. Almost simultaneously Robert Griffiths and
his student Chi-Sheng Niu derived it under very general conditions
and Nicolas Gisin using the symmetry argument used here. These 5
authors joined efforts in a common paper (Fuchs {\it et al.} 1997). The
result of this section is thus also valid without this symmetry
assumption.} $\eta\in[0,1]$ (see Fig. \ref{fig6_3}): \beq
\rho_{Bob}(\vec m)=\frac{\opone+\eta\vec m\vec\sigma}{2}.
\label{SymmAtt} \eeq Eavesdroppings that satisfy the above
condition are called symmetric attacks.

Since the qubit state space is 2-dimensional, the unitary operator is entirely determined
by its action on two states, for example the $|\up\>$ and $|\down\>$ states (in this section
we use spin $\half$ notations for the qubits). It is
convenient to write the states after the unitary interaction in the Schmidt form (Peres 1997):
\beq
U|\up,0\> = |\up\>\otimes\phi_\up + |\down\>\otimes\theta_\up
\eeq
\beq
U|\down,0\> = |\down\>\otimes\phi_\down + |\up\>\otimes\theta_\down
\eeq
where the 4 states $\phi_\up$, $\phi_\down$, $\theta_\up$ and $\theta_\down$ belong to
Eve's probe Hilbert space $\H_{Eve}$ and satisfy
$\phi_\up~\perp~\theta_\up$ and $\phi_\down~\perp~\theta_\down$. By symmetry
$|\phi_\up|^2=|\phi_\down|^2\equiv \F$ and $|\theta_\up|^2=|\theta_\down|^2\equiv \D$.
Unitarity imposes $\F+\D=1$ and
\beq
\<\phi_\up|\theta_\down\>+\<\theta_\up|\phi_\down\>=0.
\label{unitarity}
\eeq
The $\phi$'s correspond to Eve's state when Bob gets the qubit undisturbed, while the
$\theta$'s are Eve's state when the qubit is disturbed.

Let us emphasize that this is
the most general unitary interaction satisfying (\ref{SymmAtt}). One finds that the shrinking
factor is given by: $\eta=\F-\D$. Accordingly, if Alice sends $|\up\>$ and Bob measures in the
compatible basis, then $\<\up|\rho_{Bob}(\vec m)|\up\>=\F$ is the probability that Bob
gets the correct result. Hence $\F$ is the fidelity and $\D$ the QBER.

Note that only 4 states span Eve's relevant state space. Hence, Eve's effective Hilbert
space is at most of dimension 4, no matter how subtle she might be\footnote{Actually, Niu and Griffiths
(1999) showed that 2-dimensional probes suffice for Eve to get as much information as with the strategy
presented here, though in their case the attack is not symmetric (one basis is more disturbed
than the other).}! This greatly simplifies the
analysis.

The symmetry imposes that the attack on the other basis satisfies:
\beqa
U|\righta,0\>&=&U\frac{|\up,0\>+|\down,0\>}{\sqrt{2}} \\
&=&\frac{1}{\sqrt{2}}(|\up\>\otimes\phi_\up + |\down\>\otimes\theta_\up \\
&&\hspace{3mm}+~|\down\>\otimes\phi_\down + |\up\>\otimes\theta_\down) \\
&=&|\righta\>\otimes\phi_\righta + |\lefta\>\otimes\theta_\righta
\eeqa
where
\beqa
\phi_\righta=\half(\phi_\up+\theta_\up+\phi_\down+ \theta_\down) \\
\theta_\righta=\half(\phi_\up-\theta_\up-\phi_\down+ \theta_\down)
\eeqa
Similarly,
\beqa
\phi_\lefta=\half(\phi_\up-\theta_\up+\phi_\down- \theta_\down) \\
\theta_\lefta=\half(\phi_\up+\theta_\up-\phi_\down- \theta_\down)
\eeqa
Condition (\ref{SymmAtt}) for the $\{|\righta\>,|\lefta\>\}$ basis implies:
$\theta_\righta\perp\phi_\righta$ and $\theta_\lefta\perp\phi_\lefta$.
By proper choice of the phases, $\<\phi_\up|\theta_\down\>$ can be made real. By condition
(\ref{unitarity}) $\<\theta_\up|\phi_\down\>$ is then also real. Symmetry implies then
$\<\theta_\righta|\phi_\lefta\>\in\Re$.
A straightforward computation concludes that all scalar products among Eve's states are
real and that the $\phi$'s generate a subspace orthogonal to the $\theta$'s:
\beq
\<\phi_\up|\theta_\down\>=\<\phi_\down|\theta_\up\>=0.
\eeq
Finally, using $|\phi_\righta|^2=\F$, i.e. that the shrinking is the same for all states, one obtains
a relation between the probe states' overlaps and the fidelity:
\beq
\F = \frac{1+\<\hat\theta_\up|\hat\theta_\down\>}
{2-\<\hat\phi_\up|\hat\phi_\down\>+\<\hat\theta_\up|\hat\theta_\down\>}
\label{F}
\eeq
where the hats denote normalized states, e.g. $\hat\phi_\up=\frac{\phi_\up}{\sqrt{\D}}$.

Consequently, the entire class of symmetric individual attacks depends only on 2 real
parameters\footnote{Interestingly, when the symmetry is extended to a third maximally
conjugated basis, as natural in the 6-state protocol of paragraph \ref{6state}, then the number
of parameters reduces to one. This parameter measures the relative quality of Bob's and
Eve's ``copy'' of the qubit send by Alice. When both copies are of equal quality, one recovers
the optimal cloning presented in section \ref{Ampli} (Bechmann-Pasquinucci and Gisin 1999).}:
$\cos(x)\equiv\<\hat\phi_\up|\hat\phi_\down\>$ and
$\cos(y)\equiv\<\hat\theta_\up|\hat\theta_\down\>$!

Thanks to the symmetry, it suffices to analyze this scenario for the case that Alice sends
the $|\up\>$ state and Bob measures in
the $\{\up,\down\}$ basis (if not, Alice, Bob and Eve disregard the data). Since Eve
knows the basis, she knows that her probe is in one of the following two mixed states:
\beqa
\rho_{Eve}(\up)=\F P(\phi_\up) + \D P(\theta_\up)  \\
\rho_{Eve}(\down)=\F P(\phi_\down) + \D P(\theta_\down).
\eeqa
An optimum measurement strategy for Eve to distinguish between $\rho_{Eve}(\up)$ and
$\rho_{Eve}(\down)$ consists in first distinguishing whether her state is in the subspace
generated by $\phi_\up$ and $\phi_\down$ or the one generated by $\theta_\up$ and
$\theta_\down$. This is possible,
since the two subspaces are mutually orthogonal. Eve has then to distinguish between two
pure states, either with overlap $\cos(x)$, or with overlap $\cos(y)$.
The first alternative happens with probability $\F$, the second one with probability $\D$.
The optimal measurement distinguishing two states with overlap $\cos(x)$ is known
to provide Eve with the correct guess with probability $\frac{1+\sin(x)}{2}$
(Peres 1997). Eve's maximal Shannon
information, attained when she does the optimal measurements, is thus given by:
\beqa
I(\alpha,\epsilon)&=&\F\cdot\left(1-h(\frac{1+\sin(x)}{2})\right) \\
&+& \D\cdot\left(1-h(\frac{1+\sin(y)}{2})\right)
\eeqa
where $h(p)=-p\log_2(p)-(1-)\log_2(1-p)$.
For a given error rate $\D$, this information is maximal when $x=y$. Consequently,
for $\D=\frac{1-\cos(x)}{2}$, one has:
\beq
I^{\max}(\alpha,\epsilon)=1-h(\frac{1+\sin(x)}{2}).
\label{Ginfo}
\eeq
This provides the explicit and analytic optimum eavesdropping strategy. For $x=0$
the QBER (i.e. $\D$) and the information gain are zero. For $x=\pi/2$ the QBER is $\half$
and the information gain 1. For small QBERs, the information gain grows linearly:
\beq
I^{\max}(\alpha,\epsilon)= \frac{2}{\ln(2)}\D +O(\D)^2 \approx 2.9~\D
\eeq

\begin{figure}
\infig{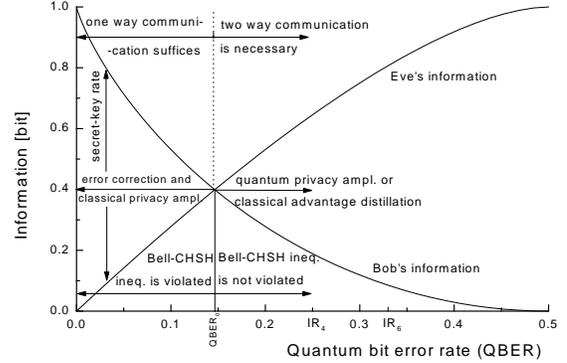}{0.95\columnwidth} \caption{Eve and Bob
information versus the QBER, here plotted for incoherent
eavesdropping on the 4-state protocol. For QBERs below QBER$_0$,
Bob has more information than Eve and secret-key agreement can be
achieved using classical error correction and privacy
amplification. These can, in principle, be implemented using only
1-way communication. The secret-key rate can be as large as the
information differences. For QBERs above QBER$_0$ ($\equiv\D_0$), Bob has a
disadvantage with respect to Eve. Nevertheless, Alice and Bob can
apply quantum privacy amplification up to the QBER corresponding
to the intercept-resend eavesdropping strategies, IR$_4$ and
IR$_6$ for the 4-state and 6-state protocols, respectively.
Alternatively, they can apply a classical protocol called
advantage distillation which is effective precisely up to the same
maximal QBER IR$_4$ and IR$_6$. Both the quantum and the classical
protocols require then 2-way communication. Note that for the
eavesdropping strategy optimal from Eve' Shannon point of view on
the 4-state protocol, QBER$_0$ correspond precisely to the noise
threshold above which a Bell inequality can no longer be violated.}
\label{fig6_4}
\end{figure}

Once Alice, Bob and Eve have measured their quantum systems, they are left with classical
random variables $\alpha$, $\beta$ and $\epsilon$, respectively. Secret key agreement between Alice and Bob
is then possible using only error correction and privacy amplification if and only if
the Alice-Bob mutual Shannon information $I(\alpha,\beta)$ is larger than the Alice-Eve
or the Bob-Eve
mutual information\footnote{Note, however, that if this condition is not satisfied, other
protocols might sometimes be used, see paragraph \ref{AdvDist}. These protocols are significantly
less efficient and are usually not considered as part of ``standard'' QC. Note also that in the
scenario analysed in this section $I(\beta,\epsilon)=I(\alpha,\epsilon)$.},
$I(\alpha,\beta)>I(\alpha,\epsilon)$
or $I(\alpha,\beta)>I(\beta,\epsilon)$.
It is thus interesting to compare Eve's maximal information (\ref{Ginfo}) with Bob's
Shannon information. The latter depends only on the error rate $\D$:
\beqa
I(\alpha,\beta)&=&1-h(\D) \label{Ginfo2}\\
&=&1+\D\log_2(\D)+(1-\D)\log_2(1-\D) \eeqa Bob's and Eve's
information are plotted on Fig. \ref{fig6_4}. As expected, for low
error rates $\D$, Bob's information is larger. But, more errors
provide Eve with more information, while Bob's information gets
lower. Hence, both information curves cross at a specific error
rate $\D_0$: \beq I(\alpha,\beta)=I^{\max}(\alpha,\epsilon)
\Longleftrightarrow \D=\D_0\equiv\frac{1-1/\sqrt{2}}{2}\approx
15\% \label{D0} \eeq Consequently, the security criteria against
individual attacks for the BB84 protocol reads: \beq BB84~secure
\Longleftrightarrow \D<\D_0\equiv\frac{1-1/\sqrt{2}}{2}
\label{BB84secure} \eeq

For QBERs larger than $\D_0$ no (one-way communication) error
correction and privacy amplification protocol can provide Alice
and Bob with a secret key immune against any individual attacks.

Let us mention that more general classical protocols, called advantage distillation
(paragraph \ref{AdvDist}), using two way communication, exist.
These can guarantee secrecy if and only if Eve's intervention does
not disentangle Alice and Bob's qubits (assuming they use the
Ekert version of the BB84 protocol) (Gisin and Wolf 2000). If Eve
optimizes her Shannon information, as discussed in this section,
this disentanglement-limit corresponds to a
QBER$=1-1/\sqrt{2}\approx30\%$ (Gisin and Wolf 1999). But, using
more brutal strategies, Eve can disentangled Alice and Bob already
for a QBER of 25\%, see Fig. \ref{fig6_4}. The latter is thus the
absolute upper limit, taking into account the most general
secret-key protocols. In practice, the limit (\ref{D0}) is
more realistic, since advantage distillation algorithms are much
less efficient than the classical privacy amplification ones.

\subsection{Connection to Bell inequality}\label{Bell}
There is an intriguing connection between the above tight bound (\ref{BB84secure})
and the CHSH form of Bell inequality (Bell 1964, Clauser {\it et al.} 1969, Clauser and Shimony 1978,
Zeilinger 1999):
\beq
S\equiv E(a,b)+E(a,b')+E(a',b)-E(a',b')\le2
\eeq
where $E(a,b)$ is the correlation between Alice and Bob's data when measuring $\sigma_a\otimes\opone$
and $\opone\otimes\sigma_b$, where $\sigma_a$ denotes an observable with eigenvalues $\pm1$
parameterized by the label $a$.
Recall that Bell inequalities are necessarily satisfied by all local models, but are
violated by quantum mechanics%
\footnote{Let us stress that the CHSH-Bell inequality is the strongest possible for two qubits.
Indeed, this inequality is violated if and only if the correlation can't be reproduced by a
local hidden variable model (Pitowski 1989).}.
To establish this connection, assume that the same quantum channel is used to
test Bell inequality. It is well-known that for error free channels, a maximal violation by a
factor $\sqrt{2}$ is achievable: $S_{max}=2\sqrt{2}>2$.
However, if the channel is imperfect, or equivalently if some
perturbator Eve acts on the channel, then the quantum correlation $E(a,b|\D)$ is
reduced,
\beqa
E(a,b|\D)&=&\F\cdot E(a,b) - \D\cdot E(a,b) \\
&=&(1-2\D)\cdot E(a,b)
\eeqa
where $E(a,b)$ denote the correlation for the unperturbed channel.
The achievable amount of violation is then reduced to $S_{max}(\D)=(1-2\D)2\sqrt{2}$ and for
large perturbations no violation at all can be achieved. Interestingly, the critical perturbation
$\D$ up to which a violation can be observed is precisely the same $\D_0$ as the limit derived
in the previous section for the security of the BB84 protocol:
\beq
S_{max}(\D)>2 \Longleftrightarrow \D<\D_0\equiv\frac{1-1/\sqrt{2}}{2}.
\label{Smax}
\eeq
This is a surprising and appealing connection between the security of QC
and tests of quantum nonlocality. One could argue that this
connection is quite natural, since, if Bell inequality were not violated, then quantum
mechanics would be incomplete and no secure communication could be based on such an
incomplete theory. In some sense, Eve's information is like probabilistic local hidden
variables. However, the connection between (\ref{BB84secure}) and (\ref{Smax})
has not been generalized to other protocols. A complete picture of these connections
is thus not yet available.

Let us emphasize that nonlocality plays no direct role in
QC. Indeed, generally, Alice is in the absolute past of Bob. Nevertheless, Bell inequality
can be violated as well by space like separated events as by time like separated events.
However, the independence assumption necessary to derive Bell inequality is justified by
locality considerations only for space-like separated events.

\subsection{Ultimate security proofs}\label{SecProofs}
The security proof of QC with perfect apparatuses and a noise-free channel is
straightforward. However, the fact that security can still be proven for imperfect
apparatuses and noisy channels is far from obvious. Clearly, something has to be
assumed about the apparatuses. In this section we simply make the hypothesis that they are perfect.
For the channel which is not under Alice and Bob's control, however, nothing
is assumed. The question is then: up to which QBER can Alice and Bob apply
error correction and privacy amplification to their classical bits? In the previous
sections we found that the threshold is close to a QBER of 15\%, assuming individual
attacks. But in principle Eve could manipulate several qubits coherently. How much help to Eve
this possibility provides is still unknown, though some bounds are known.
Already in 1996, Dominic Mayers (1996b) presented the main ideas on how to prove security%
\footnote{I (NG) vividly remember the 1996 ISI workshop in Torino, sponsored by Elsag-Bailey,
were I ended my talk stressing the importance of security proofs. Dominic Mayers stood up,
gave some explanation, and wrote a formula on a transparency, claiming that this was the
result of his proof. I think it is fair to say that no one in the audience understood Mayers'
explanation. But I kept the transparency and it contains the basic eq. (\ref{MayersBound}) (up
to a factor 2, which corresponds to an improvement of Mayers result obtained in 2000
by Shor and Preskill, using also ideas from Lo and Chau)!}.
In 1998, two major papers were made public on the Los Alamos archives (Mayers 1998,
and Lo and Chau 1999). Nowadays, these proofs are
generally considered as valid, thanks -- among others -- to the works of P. Shor and J. Preskill
(2000), H. Inamori {\it et al.} (2001) and of E. Biham {\it et al.} (1999).
But it is worth noting that during the first years after the first disclosure of these proofs,
essentially nobody in the community understood them!

Here we shall present the argument in a form quite different from the original proofs.
Our presentation aims at being transparent in the sense that it rests on two theorems.
The proofs of the theorems are hard and will be omitted. However, their claims are
easy to understand and rather intuitive.
Once one accepts the theorems, the security proof is rather straightforward.

The general idea is that at some point Alice, Bob and Eve perform measurements
on their quantum systems. The outcomes provide them with classical random
variables $\alpha$, $\beta$ and $\epsilon$, respectively,
with $P(\alpha,\beta,\epsilon)$ the joint probability distribution.
The first theorem, a standard of classical information based cryptography, states
necessary and sufficient condition on $P(\alpha,\beta,\epsilon)$ for the
possibility that Alice and Bob
extract a secret key from $P(\alpha,\beta,\epsilon)$ (Csisz\'{a}r and K\"orner 1978). The second
theorem is a clever version of
Heisenberg's uncertainty relation expressed in terms of available information
(Hall 1995):
it sets a bound on the sum of
the information available to Bob and to Eve on Alice's key.

{\bf Theorem 1.} For a given $P(\alpha,\beta,\epsilon)$, Alice and Bob can establish a secret key
(using only error correction and classical privacy amplification) if and only if
$I(\alpha,\beta)\ge I(\alpha,\epsilon)$ or $I(\alpha,\beta)\ge I(\beta,\epsilon)$, where
$I(\alpha,\beta)=H(\alpha)-H(\alpha|\beta)$ denotes the
mutual information, with $H$ the Shannon entropy.

{\bf Theorem 2.} Let E and B be two observables in an N dimensional Hilbert
space. Denote $\epsilon$, $\beta$, $|\epsilon\rangle$ and $|\beta\rangle$ the corresponding
eigenvalues and eigenvectors, respectively, and let
$c=\max_{\epsilon,\beta}\{|\langle\epsilon|\beta\rangle|\}$. Then
\beq
I(\alpha,\epsilon)+I(\alpha,\beta)\le 2\log_2(Nc),
\eeq
where $I(\alpha,\epsilon)=H(\alpha)-H(\alpha|\epsilon)$ and
$I(\alpha,\beta)=H(\alpha)-H(\alpha|\beta)$
are the entropy differences corresponding to the probability distribution of the
eigenvalues $\alpha$ prior to and deduced from any measurement by Eve and Bob,
respectively.

\begin{figure}
\infig{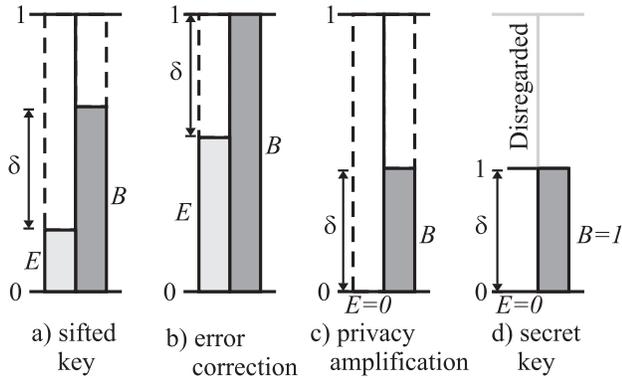}{0.95\columnwidth} \caption{Intuitive
illustration of theorem 1. The initial situation is depicted in
a). During the 1-way public discussion phase of the
protocol Eve receives as much information as Bob,
the initial information difference $\delta$ thus remains. After
error correction, Bob's information equals 1, as illustrated on
b). After privacy amplification Eve's information is zero. In c)
Bob has replaced all bits to be disregarded by random bits. Hence
the key has still the original length, but his information has
decreased. Finally, removing the random bits, the key is shortened
to the initial information difference, see d). Bob has full
information on this final key, while Eve has none.} \label{fig6_5}
\end{figure}

The first theorem states that Bob must have more information on
Alice's bits than Eve (see Fig. \ref{fig6_5}). Since error
correction and privacy amplification can be implemented using only
1-way communication, theorem 1 can be understood intuitively as
follows. The initial situation is depicted in a). During the
public phase of the protocol, because of the 1-way communication,
Eve receives as much information as Bob, the initial
information difference $\delta$ thus remains. After error correction, Bob's
information equals 1, as illustrated on b). After privacy
amplification Eve's information is zero. In c) Bob has replaced
all bits to be disregarded by random bits. Hence the key has still
the original length, but his information has decreased. Finally,
removing the random bits, the key is shortened to the initial
information difference, see d). Bob has full information on this
final key, while Eve has none.

The second theorem states that if Eve performs a measurement providing her with some
information $I(\alpha,\epsilon)$, then, because of the perturbation, Bob's information is necessarily
limited. Using these two theorems, the argument now runs as follows. Suppose Alice sends
out a large number of qubits and that n where received by Bob in the correct basis.
The relevant Hilbert space's dimension is thus $N=2^n$. Let us re-label the bases used
for each of the n qubits such that Alice used n times the x-basis. Hence, Bob's observable
is the n-time tensor product $\sigma_x\otimes...\otimes\sigma_x$. By symmetry, Eve's
optimal information on the correct bases is precisely the
same as her optimal information on the incorrect ones (Mayers 1998). Hence one can bound her information
assuming she measures $\sigma_z\otimes...\otimes\sigma_z$. Accordingly, $c=2^{-n/2}$ and
theorem 2 implies:
\beq
I(\alpha,\epsilon)+I(\alpha,\beta)\le 2\log_2(2^n2^{-n/2})=n
\label{Imax}
\eeq
That is, the sum of Eve's and Bob's information per qubit is smaller or equal to 1.
This is quite an intuitive result: together, Eve and Bob cannot get more information
than sent out by Alice! Next, combining the bound (\ref{Imax}) with theorem 1, one
deduces that a secret key is achievable whenever $I(\alpha,\beta)\ge n/2$.
Using $I(\alpha,\beta)=n\left(1-\D\log_2(\D)-(1-\D)\log_2(1-\D)\right)$ one obtains the
sufficient condition on the error rate $\D$ (i.e. the QBER):
\beq
\D\log_2(\D)+(1-\D)\log_2(1-\D) \le \half
\label{MayersBound}
\eeq
i.e. $\D\le11\%$.

This bound, QBER$\le$11\%, is precisely that obtained in Mayers
proof (after improvement by P. Shor and J. Preskill (2000)).
The above proof is, strickly speaking, only valid if the key is
much longer than the number of qubits that Eve attacks coherently,
so that the Shannon informations we used represent averages over
many independent realisations of classical random variables. In
other words, assuming that Eve can attack coherently a large but
finite number $n_0$ of qubits, Alice and Bob can use the above
proof to secure keys much longer than $n_0$ bits. If one assumes
that Eve has an unlimited power, able to attack coherently any
number of qubits, then the above proof does not apply, but Mayer's
proof can still be used and provides precisely the same bound.

This 11\% bound for coherent attacks is clearly compatible with
the 15\% bound found for individual attacks. The 15\% bound is
also a necessary one, since an explicit eavesdropping strategy
reaching this bound is presented in section \ref{SymAttack}. It is
not known what happens in the intermediate range $11\% < QBER <
15\%$, but the following is plausible. If Eve is limited to
coherent attacks on a finite number of qubits, then in the limit
of arbitrarily long keys, she has a negligibly small probability
that the bits combined by Alice and Bob during the error
correction and privacy amplification protocols originate from
qubits attacked coherently. Consequently, the 15\% bound would
still be valid (partial results in favor of this conjecture can be
found in Cirac and Gisin 1997, and in Bechmann-Pasquinucci and
Gisin 1999). However, if Eve has unlimited power, in particular,
if she can coherently attack an unlimited number of qubits, then
the 11\% bound might be required.

To conclude this section, let us stress that the above security
proof equally applies to the 6-state protocol (paragraph
\ref{6state}). It also extends straightforwardly to protocols
using larger alphabets (Bechmann-Pasquinucci and Tittel 2000,
Bechmann-Pasquinucci and Peres 2000, Bourennane {\it et al.} 2001a, Bourennane
{\it et al.} 2001b).

\subsection{Photon number measurements, lossless channels}\label{QND}
In section \ref{PhotonSources} we saw that all real photon sources have a finite probability
to emit more than 1 photon. If all emitted photons encode the same qubit,
Eve can take advantage of this. In principle, she can first measure the number of photons in
each pulse, without disturbing the degree of freedom encoding the qubits\footnote{For
polarization coding, this is quite clear. But for phase coding one may think (incorrectly) that phase
and photon number are incompatible!
However, the phase used for encoding is a relative phase between two modes.
Whether these modes are polarization modes or correspond to different times (determined e.g. by the
relative length of interferometers), does not matter.}. Such measurements are sometimes called
Quantum Non Demolition (QND) measurements, because they do not perturb the qubit, in
particular they do not destroy the photons. This is possible because Eve knows in advance that
Alice sends a mixture of states with well defined photon numbers\footnote{Recall that a mixture
of coherent states $|e^{i\phi}\alpha\>$ with a random phase $\phi$, as produced by lasers
when no phase reference in available, is equal to a mixture of photon number states $|n\>$
with Poisson statistics: $\int_0^{2\pi}|e^{i\phi}\alpha\>\<e^{i\phi}\alpha|\frac{d\phi}{2\pi}
=\sum_{n\ge0}\frac{\mu ^{n}}{n!}e^{-^{\mu }}|n\>\<n|$, where $\mu=|\alpha|^2$.}, (see section \ref{Ampli}).
Next, if Eve finds more than one photon, she keeps one and sends the other(s) to Bob. In order
to prevent that Bob detects a lower qubit rate, Eve must use a channel with
lower losses. Using an ideally lossless quantum channel, Eve can even, under certain conditions, keep one photon and
increase the probability that pulses with more than one photon get to Bob!
Thirdly, when Eve finds one photon, she may
destroy it with a certain probability, such that she does not affect the total number of qubits
received by Bob. Consequently, if the probability that a non-empty pulse has more than
one photon (on Alice's side) is larger than the probability that a non-empty pulse
is detected by Bob, then Eve can get full information without introducing any perturbation!
This is possible only when the QC protocol is not perfectly implemented, but this is a
realistic situation (Huttner {\it et al.} 1995, Yuen 1997).

The QND atacks have recently received a lot of attention (L\"utkenhaus 2000, Brassard {\it et al.} 2000).
The debate is not yet settled. We would
like to argue that it might be unrealistic, or even unphysical, to assume that Eve can
perform ideal QND attacks. Indeed, first she needs the capacity to perform QND photon number
measurements. Although impossible with today's technology, this is a reasonable
assumption (Nogues {\it et al.} 1999). Next, she should be able to keep her photon until Alice and Bob reveal the
basis. In principle this could be achieved using a lossless channel in a loop. We discuss this
eventuality below. Another possibility would be that Eve maps her photon to a quantum
memory. This does not exist today, but might well exist in the future. Note that the quantum
memory should have essentially unlimited time, since Alice and Bob could easily wait for minutes before
revealing the bases\footnote{The quantum part of the protocol could run continuously, storing
large ammount of raw classical data. But the classical part of the protocol, processing
these raw data, could take place just seconds before the key is used.}.
Finally, Eve must access a lossless
channel, or at least a channel with losses lower than that used by Alice and Bob. This might
be the most tricky point. Indeed, besides using a shorter channel, what can Eve do? The
telecom fibers are already at the physical limits of what can be achieved (Thomas {\it et al.} 2000).
The loss is almost
entirely due to the Rayleigh scattering which is unavoidable: solve the Schr\"odinger
equation in a medium with inhomogeneities and you get scattering. And when the inhomogeneities
are due to the molecular stucture of the medium, it is difficult to imagine lossless fibers!
The 0.18 dB/km attenuation in silica fibers at 1550 nm is a lower bound which is based on
physics, not on technology\footnote{Photonics crystal fibers have the potential to overcome the Rayleigh
scaterring limit. Actually, there are two kinds of such fibers. The
first kind guides light by total internal reflection, like in ordinary fibers. In
these most of the light also propagates in silica, and thus the loss limit is
similar. In the second kind, most of the light propagates in air, thus the
theoretical loss limit is lower. However, today the losses are extremely
high, in the range of hundreds of dB/km. The best reported result that we
are aware of is 11 dB/km and it was obtained with a fiber of the first kind
(Canning {\it et al.} 2000).}.
Note that using the air is not a viable solution, since the attenuation at the telecom
wavelengths is rather high. Vacuum, the only way to avoid Rayleigh scattering, has also
limitations, due to diffraction, again an unavoidable physical phenomenon. In the end, it seems
that Eve has only two possibilities left. Either she uses teleportation (with extremely high success
probability and fidelity) or she converts the
photons to another wavelength (without perturbing the qubit). Both of these ``solutions'' are
seemingly unrealistic in any foreseeable future.

Consequently, when considering the type of attacks discussed in this section, it is
essential to distinguish the ultimate proofs from the practical ones discussed in the
first part of this chapter. Indeed, the assumptions about the defects of Alice and
Bob's apparatuses must be very specific and might thus be of limited interest. While for
practical considerations, these assumptions must be very general and might thus be excessive.

\subsection{A realistic beamsplitter attack}\label{beamsplitter}
\begin{figure}
\infig{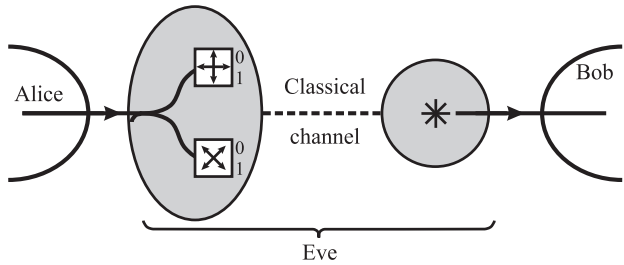}{0.95\columnwidth} \caption{Realistic
beamsplitter attack. Eve stops all pulses. The two photon pulses
have a 50\% probability to be analyzed by the same analyzer. If
this analyzer is compatible with the state prepared by Alice, then
both photon are detected at the same outcome; if not there is a
50\% chance that they are detected at the same outcome. Hence,
there is a probability of 3/8 that Eve detects both photons at the
same outcome. In such a case, and only in such a case, she resends
a photon to Bob. In 2/3 of these cases she introduces no errors
since she identified the correct state and gets full information; in
the remaining cases she has a probability 1/2 to introduce an
error and gains no information. The total QBER is thus 1/6 and
Eve's information gain 2/3.} \label{fig6_6}
\end{figure}

The attack presented in the previous section takes advantage of
the pulses containing more than one photon. However, as discussed,
it uses unrealistic assumptions. In this section, following N.
L\"utkenhaus (2000) and M. Dusek et al (2000), we briefly comment
on a realistic attack, also exploiting the multiphoton pulses
(for details, see Felix {\it et al.} 2001, where this and another examples are presented).
Assume that Eve splits all pulses in two, analysing each half in
one of the two bases, using photon counting devices able to
distinguish pulses with 0, 1 and 2 photons (see Fig. \ref{fig6_6}).
In practice this could be realized using many single photon
counters in parallel. This requires nearly perfect detectors, but at
least one does not need to assume technology completely out of
today's realm. Whenever Eve detects two photons in the same
output, she sends a photon in the corresponding state into Bob's
apparatus. Since Eve's information is classical, she can overcome
all the losses of the quantum channel. In all other cases, Eve
sends nothing to Bob. In this way, Eve sends a fraction 3/8 of the
pulses containing at least 2 photons to Bob. On these, she
introduces a QBER=1/6 and gets an information $I(A,E)=2/3=4\cdot
QBER$. Bob doesn't see any reduction in the number of detected
photons, provided the transmission coefficient of the quantum
channel $t$ satisfies: \beq t\le
\frac{3}{8}Prob(n\ge2|n\ge1)\approx\frac{3\mu}{16} \eeq where
the last expression assumes Poissonian photon distribution.
Accordingly, for a fixed QBER, this attacks provides Eve with
twice the information she would get using the intercept resend
strategy. To counter such an attack, Alice should use a mean
photon number $\mu$ such that Eve can only use this attack on a
fraction of the pulses. For example, Alice could use pulses weak
enough that Eve's mean information gain is identical to the one
she would obtain with the simple intercept resend strategy (see
paragraph \ref{IRstrat}). For 10, 14 and 20 dB attenuation, this
corresponds to $\mu=0.25,\ 0.1$ and 0.025, respectively.

\subsection{Multi-photon pulses and passive choice of states}\label{Passivebasis}
Multi-photon pulses do not necessarily constitute a threat for the key security, but limit the key creation rate
because they imply that more bits must be discarded during key distillation.
This fact is based on the assumption that all photons in a pulse carry the same
qubit, so that Eve does not need to copy the qubit going to Bob, but merely keeps the copy
that Alice inadvertently provides. When using weak pulses, it seems unavoidable that all
the photons in a pulse carry the same qubit. However, in 2-photon implementations,
each photon on Alice's side chooses
independently a state (in the experiments of Ribordy {\it et al.} 2001 and Tittel {\it et al.} 2000, each photon chooses randomly both its basis and its bit value; in the experiments of Naik {\it et al.} 2000 and Jennewein {\it et al.} 2000b, the bit value choice only is random).
Hence, when two
photon pairs are simultaneously produced, by accident, the two twins carry independent qubits.
Consequently, Eve can't take advantage of such multi-photon twin-pulses. This might be one of
the main advantages of the 2-photon schemes compared to the much simpler weak-pulse schemes.
But the multi-photon problem is then on Bob's side who gets a noisy signal, consisting partly in
photons not in  Alice's state!

\subsection{Trojan Horse Attacks}\label{Trojan}
All eavesdropping strategies discussed up to now consisted
of Eve's attempt to get a maximum information out of the qubits exchanged by
Alice and Bob. But Eve can also follow a completely different strategy:
she can herself send signals that enter Alice and Bob's offices
through the quantum channel. This kind of strategies are called Trojan horse
attacks. For example, Eve can send
light pulses into the fiber entering Alice or Bob apparatuses and analyze the backreflected light.
In this way, it is in principle possible to detect which laser just flashed,
or which detector just fired, or the settings of phase and polarization modulators.
This cannot be simply prevented by using a shutter, since Alice and Bob
must leave the ``door open'' for the photons to go out and in, respectively.

In most QC-setups the amount of backreflected light can be made very small and
sensing the apparatuses with light pulses through the quantum channel is
difficult. Nevertheless, this attack is especially threatening in the
{\it plug-\&-play} scheme on Alice's side (section \ref{PandP1}),
since a mirror is used to send the light pulses back to Bob. So in
principle, Eve can send strong light pulses to Alice and sense the applied
phase shift. However, by applying
the phase shift only during a short time $\Delta t_{phase}$(a few
nanoseconds), Alice can oblige Eve to send the spying pulse at the same time
as Bob. Remember that in the {\it plug-\&-play} scheme pulse coming from Bob are
macroscopic and an attenuator at Alice reduces them to the below one photon
level, say 0.1 photons per pulse. Hence, if Eve wants to get, say 1 photon
per pulse, she has to send 10 times Bob's pulse energy. Since Alice is
detecting Bob's pulses for triggering her apparatus, she must be able
to detect an increase of energy of these pulses in order to reveal the
presence of a spying pulse. This is a relatively easy task, provided that Eve's pulses
look the same as Bob's. But, Eve could of course use another wavelength or
ultrashort pulses (or very long pulses with low intensity, hence the
importance of $\Delta t_{phase}$), therefore Alice must introduce an optical
bandpass filter with a transmission spectrum corresponding to the
sensitivity spectrum of her detector, and choose a $\Delta t_{phase}$ that
fits to the bandwidth of her detector.

There is no doubt that Trojan horse attacks can be prevented by technical
measures. However, the fact that this class of attacks
exist illustrates that the security of QC can never be guaranteed only by the
principles of quantum mechanics, but necessarily relies also on technical measures
that are subject to discussions
\footnote{Another technological loophole, recently pointed out by Kurtsiefer {\it et al.}, is the possible information leakage
caused by light emitted by APDs during their breakdown (2001).}.

\subsection{Real security: technology, cost and complexity}\label{RealSecurity}
Despite the elegant and generality of security proofs, the dream of a QC system whose
security relies entirely on quantum principles is unrealistic. The technological
implementation of the abstract principles will always be questionable. It is
likely that they will remain the weakest point in all systems. Moreover, one should
remember the obvious equation:
\beqa
Infinite\ security\ &\Rightarrow& \ Infinite\ cost\\
&\Rightarrow& \ Zero\ practical\ interest \nonumber
\eeqa

On the other hand, however, one should not underestimate the following two advantages of QC.
First, it is much easier to forecast progress in technology than in mathematics: the danger that
QC breaks down overnight is negligible, contrary to public-key cryptosystems. Next, the
security of QC depends on the technological level of the adversary {\it at the time of
the key exchange}, contrary to complexity based systems whose coded message can be registered
and broken thanks to future progress. The latter point is relevant for secrets whose value
last many years.

One often points at the low bit rate as one of the current limitations of QC.
However, it is important to stress that QC must not necessarily be used in conjunction with one-time pad encryption.
It can also be used to provide a key for a symmetrical cipher -- such as AES --
whose security is greatly enhanced by frequent key changes.

To conclude this chapter, let us briefly elaborate on the differences and similarities
between technological
and mathematical complexity and on their possible connections and implications.
Mathematical complexity means that the number of steps needed to run
complex algorithms explodes exponentially when the size of the input data grows
linearly. Similarly, one can define technological complexity of a quantum computer by
an exploding difficulty to process coherently all the qubits necessary to run a
(non-complex)
algorithm on a linearly growing number of input data.
It might be interesting to consider the possibility that the relation between these two
concepts of
complexity is deeper. It could be that the solution of a problem requires either a
complex classical algorithm or a quantum one which itself requires a complex
quantum computer\footnote{Penrose (1994) pushes these speculations even further, suggesting that
spontaneous collapses stop quantum computers whenever they try to compute beyond a certain
complexity.}.

\section{Conclusion}\label{Concl}
Quantum cryptography is a fascinating illustration of the dialog between
basic and applied physics. It is based on a beautiful combinations of concepts from
quantum physics and information theory and made possible thanks to the tremendous
progress in quantum optics and in the technology of optical fibers and of
free space optical communication. Its security
principle relies on deep theorems in classical
information theory and on a profound understanding of the Heisenberg's uncertainty
principle, as illustrated by theorems 1 and 2 in section \ref{SecProofs} (the only mathematically involved theorems
in this review!). Let us also emphasize the important contributions of QC to classical
cryptography: privacy amplification and classical bound information (paragraphs \ref{ECPA}
and \ref{AdvDist}) are examples of concepts in classical information
whose discovery were much inspired by QC.
Moreover, the fascinating tension between quantum physics and
relativity, as illustrated by Bell's inequality, is not far away, as discussed
in section \ref{Bell}. Now, despite the huge progress over the
recent years, many open questions and technological challenges remain.

One technological challenge at present concerns improved detectors compatible
with telecom fibers. Two other issues concern free space QC and quantum repeaters. The
first is presently the only way to realize QC over thousands of kilometers using near
future technology (see section
\ref{Satellite}). The idea of quantum repeaters (section \ref{Qrepeaters})
is to encode the qubits in such a way that if
the error rate is low, then errors can be detected and corrected entirely in the quantum
domain. The hope is that such techniques could extend the range of quantum communication to
essentially unlimited distances. Indeed, Hans Briegel, then at Innsbruck University (1998),
and coworkers, showed that the number of additional qubits needed for quantum repeaters can be
made smaller than the numbers of qubits needed to improved the fidelity of the quantum
channel (Dur {\it et al.} 1999).
One could thus overcome the decoherence problem. However, the main practical limitation
is not decoherence but loss (most photons never get to Bob, but those which get there, exhibit high
fidelity).

On the open questions side, let us emphasize three main concerns. First,
complete and realistic analyses of the security issues are still missing. Next, figures of
merit to compare QC schemes based on different quantum systems (with different
dimensions for example) are still awaited. Finally, the delicate question of how to
test the apparatuses did not yet receive enough attention. Indeed, a potential
customer of quantum cryptography buys confidence and secrecy, two qualities hard
to quantify. Interestingly, both of these issues have a connection with Bell
inequality (see sections \ref{Bell} and \ref{IdealReal}). But, clearly, this
connection can not be phrased  in the old context of local hidden variables, but rather in
the context of the security of tomorrows communications. Here, like in all the field of
quantum information, old concepts
are renewed by looking at them from a fresh perspective: let's
exploit the quantum weirdness!

QC could well be the first application of quantum mechanics at the single quanta level.
Experiments have demonstrated that keys can be exchanged over distances of a few tens of
kilometers at rates at least of the order of a thousand bits per second.
There is no doubt that the technology can be mastered and the question is not whether QC
will find commercial applications, but when. Indeed, presently QC is still very limited in
distance and in secret-bit rate. Moreover, public key systems occupy the market and,
being pure software, are tremendously easier to manage. Every so often, the news is that
some classical ciphersystem has been broken. This would be impossible with properly implemented
QC. But this apparent strength of QC might turn out to be its weak point: the security agencies
would equally be unable to break quantum cryptograms!

\small
\section*{Acknowledgments}
Work supported by the Swiss FNRS and the European projects EQCSPOT
and QUCOMM financed by the Swiss OFES. The authors would also like
to thank Richard Hughes for providing Fig. \ref{fig3_4}, and
acknowledge both referees, Charles H. Bennett and Paul G. Kwiat,  for their very careful reading of the
manuscript and their helpful remarks.

\newpage
\section*{References}
Ardehali, M., H. F. Chau and H.-K. Lo, 1998, ``Efficient Quantum Key Distribution'',
quant-ph/9803007.

Aspect, A., J. Dalibard, and G. Roger, 1982, ``Experimental Test of Bell's
Inequalities Using Time-Varying Analyzers'', Phys. Rev. Lett. {\bf49}, 1804-1807.

Bechmann-Pasquinucci, H., and N. Gisin, 1999, ``Incoherent and Coherent
Eavesdropping in the 6-state Protocol of Quantum Cryptography'',
Phys. Rev. A {\bf59}, 4238-4248.

Bechmann-Pasquinucci, H., and A. Peres, 2000, ``Quantum cryptography with 3-state
systems'', Phys. Rev. Lett. {\bf85}, 3313-3316.

Bechmann-Pasquinucci, H., and W. Tittel, 2000, ``Quantum cryptography using larger
alphabets'', Phys. Rev. A {\bf61}, 062308-1.

Bell, J.S., 1964, ``On the problem of hidden variables in quantummechanics'',
Review of Modern Phys. {\bf38}, 447-452; reprinted in ``Speakable and unspeakable
in quantum mechanics'', Cambridge University Press, New-York 1987.

Bennett, Ch.H., 1992, ``Quantum cryptography using any two nonorthogonal states'',
Phys. Rev. Lett. {\bf68}, 3121-3124.

Bennett, Ch.H. and G. Brassard, 1984, ``Quantum cryptography: public key
distribution and coin tossing'', Int. conf. Computers, Systems \& Signal
Processing, Bangalore, India, December 10-12, 175-179.

Bennett, Ch.H. and G. Brassard, 1985, ``Quantum public key distribution system'',
IBM Technical Disclosure Bulletin, 28, 3153-3163.

Bennett, Ch.H., G. Brassard and J.-M. Robert, 1988, ``Privacy amplification by
public discussion'' SIAM J. Comp. {\bf 17}, 210-229.

Bennett, Ch.H., F. Bessette, G. Brassard, L. Salvail, and J. Smolin, 1992a,
``Experimental Quantum Cryptography'', J. Cryptology {\bf5}, 3-28.

Bennett, Ch.H., G. Brassard and Mermin N.D., 1992b, ``Quantum cryptography
without Bell's theorem'', Phys. Rev. Lett. {\bf68}, 557-559.

Bennett, Ch.H., G. Brassard and A. Ekert, 1992c, ``Quantum cryptography'', Scientific Am.
{\bf267}, 26-33 (int. ed.).

Bennett, Ch.H., G. Brassard, C. Cr\'epeau, R. Jozsa, A. Peres and W.K. Wootters, 1993,
``Teleporting an unknown quantum state via dual classical and Einstein-Podolsky-Rosen channels'',
Phys. Rev. Lett. {\bf70}, 1895-1899.

Bennett, Ch.H., G. Brassard, C. Cr\'epeau, and U.M. Maurer, 1995, ``Generalized
privacy amplification'', IEEE Trans. Information th., 41, 1915-1923.

Berry, M.V., 1984, ``Quantal phase factors accompanying adiabatic changes'',
Proc. Roy. Soc. Lond. A {\bf392}, 45-57.

Bethune, D., and W. Risk, 2000, ``An Autocompensating Fiber-Optic Quantum
Cryptography System Based on Polarization Splitting of Light'', IEEE J.
Quantum Electron., {\bf36}, 340-347.

Biham, E. and T. Mor, 1997a, ``Security of quantum cryptograophy
against collective attacks'', Phys. Rev. Lett. {\bf78}, 2256-1159.

Biham, E. and T. Mor, 1997b, ``Bounds on Information and the Security of Quantum Cryptography'',
Phys. Rev. Lett. {\bf79}, 4034-4037.

Biham, E., M. Boyer, P.O. Boykin, T. Mor and V. Roychowdhury, 1999, ``A proof of the
security of quantum key distribution'', quant-ph/9912053.

Bourennane, M., F. Gibson, A. Karlsson, A. Hening, P. Jonsson, T. Tsegaye, D.
Ljunggren, and E. Sundberg, 1999, ``Experiments on long wavelength (1550nm)
'plug and play' quantum cryptography systems', Opt. Express {\bf4},383-387

Bourennane, M., D. Ljunggren, A. Karlsson, P. Jonsson, A. Hening,
and J.P. Ciscar, 2000, ``Experimental long wavelength quantum
cryptography: from single photon transmission to key extraction
protocols'', J. Mod. Optics {\bf47}, 563-579.

Bourennane, M., A. Karlsson and G. Bj\"orn, 2001a, ``Quantum Key Distribution using
multilevel encoding'', Phys. Rev A {\bf64}, 012306.

Bourennane, M., A. Karlsson, G. Bj\"orn, N. Gisin and N. Cerf,
2001b, ``Quantum Key distribution using multilevel encoding :
security analysis'', quant-ph/0106049.

Braginsky, V.B. and F.Ya. Khalili, 1992, ``Quantum Measurements'', Cambridge University Press.

Brassard, G., 1988, ``Modern cryptology'', Springer-Verlag, Lecture Notes in Computer Science,
vol. 325.

Brassard, G. and L. Salvail, 1993, ``Secrete-key reconciliation by public discussion''
In {\em Advances in Cryptology, Eurocrypt '93 Proceedings}.

Brassard, G., C. Cr\'epeau, D. Mayers and L. Salvail, 1998, ``The Security of quantum bit commitment schemes'',
Proceedings of Randomized Algorithms, Satellite Workshop of 23rd International Symposium on
Mathematical Foundations of Computer Science, Brno, Czech Republic, 13-15.

Brassard, G., N. L\"utkenhaus, T. Mor, and B.C. Sanders, 2000, ``Limitations
on Practical Quantum Cryptography'', Phys. Rev. Lett. {\bf85}, 1330-1333.

Breguet, J., A. Muller and N. Gisin, 1994, ``Quantum cryptography with polarized photons
in optical fibers: experimental and practical limits'', J. Modern optics {\bf41}, 2405-2412.

Breguet, J. and N. Gisin, 1995, ``New interferometer using a 3x3 coupler and Faraday
mirrors'', Optics Lett. {\bf20}, 1447-1449.

Brendel, J., W. Dultz and W. Martienssen, 1995, ``Geometric phase in 2-photon
interference experiments'', Phys. rev. A {\bf52}, 2551-2556.

Brendel, J., N. Gisin, W. Tittel, and H. Zbinden, 1999, ``Pulsed Energy-Time
Entangled Twin-Photon Source for Quantum Communication'', Phys. Rev. Lett.
{\bf82} (12), 2594-2597.

Briegel, H.-J., Dur W., J.I. Cirac, and P. Zoller, 1998, ``Quantum Repeaters: The Role of Imperfect
Local Operations in Quantum Communication'', Phys. Rev. Lett. {\bf81}, 5932-5935.

Brouri, R., A. Beveratios, J.-P. Poizat, P. Grangier, 2000, ``Photon antibunching in the
fluorescence of individual colored centers in diamond'', Opt. Lett. {\bf 25}, 1294-1296.

Brown, R.G.W. and M. Daniels, 1989,
``Characterization of silicon avalanche photodiodes for photon correlation measurements.
3: Sub-Geiger operation'', Applied Optics {\bf28}, 4616-4621.

Brown, R.G.W., K. D. Ridley, and J. G. Rarity, 1986,
``Characterization of silicon avalanche photodiodes for photon correlation measurements.
1: Passive quenching'', Applied Optics {\bf25}, 4122-4126.

Brown, R.G.W., R. Jones, J. G. Rarity, and Kevin D. Ridley, 1987,
``Characterization of silicon avalanche photodiodes for photon correlation measurements.
2: Active quenching'', Applied Optics {\bf26}, 2383-2389.

Brunel, Ch., B. Lounis, Ph. Tamarat, and M. Orrit, 1999, ``Triggered Source of Single
Photons based on Controlled Single Molecule Fluorescence'', Phys. Rev. Lett. {\bf83}, 2722-2725.

Bruss, D., 1998, ``Optimal eavesdropping in quantum cryptography with six
states'', Phys. Rev. Lett. 81, 3018-3021.

Bruss, D., A. Ekert and C. Macchiavello, 1998, ``Optimal universal quantum cloning and state estimation'',
Phys. Rev. Lett. {\bf81}, 2598-2601.

Buttler, W.T., R.J. Hughes, P.G. Kwiat, S. K. Lamoreaux, G.G. Luther, G.L. Morgan, J.E. Nordholt,
C.G. Peterson, and C. Simmons, 1998, ``Practical free-space quantum key distribution over
1 km'', Phys. Rev. Lett. {\bf81}, 3283-3286.

Buttler, W.T., R.J. Hughes, S.K. Lamoreaux, G.L. Morgan, J.E. Nordholt, and C.G.
Peterson, 2000, ``Daylight Quantum key distribution over 1.6 km'', Phys. Rev.
Lett, 84, pp. 5652-5655.

Bu\v{z}ek, V. and M. Hillery, 1996, ``Quantum copying: Beyond the no-cloning theorem'',
Phys. Rev. A {\bf 54}, 1844-1852.

Cancellieri, G., 1993, ``Single-mode optical fiber measurement: characterization and
sensing'', Artech House, Boston \& London.

Canning, J., M. A. van Eijkelenborg, T. Ryan, M. Kristensen and K. Lyytikainen,
2000, ``Complex mode coupling within air-silica structured optical fibers
and applications'', Optics Commun. {\bf185}, 321-324

Cirac, J.I., and N. Gisin, 1997, ``Coherent eavesdropping strategies for the 4-
state quantum cryptography protocol'', Phys. Lett. A {\bf229}, 1-7.

Clarke, M., R.B., A. Chefles, S.M. Barnett and E. Riis, 2000, ``Experimental Demonstration of
Optimal Unambiguous State Discrimination'', Phys. Rev. A {\bf63}, 040305.

Clauser, J.F., M.A.~Horne, A.~Shimony and R.A. Holt, 1969, ``Proposed experiment to test local
hidden variable theories'', Phys. Rev. Lett. {\bf23}, 880-884.

Clauser, J.F. and A.~Shimony, 1978, ``Bell's theorem: experimental tests and implications'', Rep.
Prog. Phys. {\bf41}, 1881-1927.

Cova, S., A. Lacaita, M. Ghioni, and G. Ripamonti, 1989, ``High-accuracy
picosecond characterization of gain-switched laser diodes'', Optics Letters {\bf14}, 1341-1343.

Cova, S., M. Ghioni, A. Lacaita, C. Samori, and F. Zappa, 1996,
``Avalanche photodiodes and quenching circuits for single-photon
detection'', Applied Optics {\bf35}(129), 1956-1976.

Csisz\'{a}r, I. and K\"orner, J., 1978, ``Broadcast channels with
confidential messages'', IEEE Transactions on Information Theory,
Vol.~IT-24, 339-348.

De Martini, F., V. Mussi and F. Bovino, 2000, ``Schroedinger cat states and optimum universal
Quantum cloning by entangled parametric amplification'', Optics Commun. {\bf179}, 581-589.

Desurvire, E., 1994, ``The golden age of optical fiber amplifiers'', Phys.
Today, Jan. 94, 20-27.

Deutsch, D., ``Quantum theory, the Church-Turing principle and the universal quantum computer'',
1985, Proc. Royal Soc. London, Ser. A {\bf400}, 97-105.

Deutsch, D., A. Ekert, R. Jozsa, C. Macchiavello, S. Popescu, and A. Sanpera,
1996, ``Quantum privacy amplification and the security of quantum cryptography
over noisy channels'', Phys. Rev. Lett. {\bf77}, 2818-2821; Erratum-ibid. 80,
(1998), 2022.

Dieks, D., 1982, ``Communication by EPR devices'', Phys. Lett. A {\bf 92}, 271-272.

Diffie, W. and Hellman M.E., 1976, ``New directions in cryptography'',
IEEE Trans. on Information Theory {\bf IT-22}, pp 644-654.

Dur, W., H.-J. Briegel, J.I. Cirac, and P. Zoller, 1999, ``Quantum repeaters
based on entanglement purification'', Phys. Rev. A {\bf59}, 169-181 (see also ibid
{\bf60}, 725-725).

Dusek, M., M. Jahma, and N. L\"utkenhaus, 2000, ``Unambiguous state discrimination
in quantum cryptography with weak coherent states'', Phys. Rev. A {\bf62}, 022306.

Einstein, A., B. Podolsky, and N. Rosen, 1935, ``Can quantum-mechanical description of
physical reality be considered complete?'', Phys. Rev. {\bf47}, 777-780.

Ekert, A.K., 1991, ``Quantum cryptography based on Bell's theorem'',
Phys. Rev. Lett. {\bf67}, 661-663.

Ekert, A.K., J.G. Rarity, P.R. Tapster, and G.M. Palma, 1992, ``Practical quantum cryptography
based on two-photon interferometry'', Phys. Rev. Lett. {\bf69}, 1293-1296.

Ekert, A.K., B. Huttner, 1994, ``Eavesdropping Techniques in Quantum Cryptosystems'',
J. Modern Optics {\bf41}, 2455-2466.

Ekert, A.K., 2000, ``Coded secrets cracked open'', Physics World {\bf13}, 39-40.

Elamari, A., H. Zbinden, B. Perny and Ch. Zimmer, 1998, ``Statistical prediction and
experimental verification of concatenations of fibre optic components with polarization
dependent loss'', J. Lightwave Techn. {\bf16}, 332-339.

Enzer, D., P. Hadley, R. Hughes, G. Peterson, and P. Kwiat, 2001, private communication.

Felix, S., A. Stefanov, H. Zbinden and N. Gisin, 2001, ``Faint laser
quantum key distribution: Eavesdropping exploiting multiphoton pulses'',
quant-ph/0102062.

Fleury, L., J.-M. Segura, G. Zumofen, B. Hecht, and U.P. Wild, 2000, ``Nonclassical Photon
Statistics in Single-Molecule Fluorescence at Room Temperature'', Phys. Rev. Lett. {\bf84}, 1148-1151.

Franson J.D., 1989, ``Bell Inequality for Position and Time'', Phys. Rev.
Lett. {\bf62}, 2205-2208.

Franson, J.D., 1992, ``Nonlocal cancellation of
dispersion'', Phys. Rev. A {\bf45}, 3126-3132.

Franson, J.D., and B.C. Jacobs, 1995, ``Operational system for Quantum
cryptography'', Elect. Lett. {\bf 31}, 232-234.

Freedmann, S.J. and J.F. Clauser, 1972, ``Experimental test of local hidden variable
theories'', Phys. rev. Lett. {\bf28}, 938-941.

Fry, E.S. and R.C. Thompson, 1976, ``Experimental test of local hidden variable
theories'', Phys. rev. Lett. {\bf37}, 465-468.

Fuchs, C.A., and A. Peres, 1996, ``Quantum State Disturbance vs. Information
Gain: Uncertainty Relations for Quantum Information'', Phys. Rev. A {\bf53}, 2038-2045.

Fuchs, C.A., N. Gisin, R.B. Griffiths, C.-S. Niu, and A. Peres, 1997, ``Optimal
Eavesdropping in Quantum Cryptography. I'', Phys. Rev. A {\bf56}, 1163-172.

G\'erard, J.-M., B. Sermage, B. Gayral, B. Legrand, E. Costard, and V. Thierry-Mieg, 1998, ``Enhanced Spontaneous Emission by
Quantum Boxes in a Monolithic Optical Microcavity'', Phys. Rev. Lett., {\bf 81}, 1110-1113.

G\'erard, J.-M., and B. Gayral, 1999, ``Strong Purcell Effect for InAs Qantum Boxes
in Three-Dimensional Solid-State Microcavities'', J. Lightwave Technology {\bf17}, 2089-2095.

Gilbert, G., and M. Hamrick, 2000, ``Practical Quantum Cryptography: A Comprehensive Analysis (Part One)'',
MITRE Technical Report (MITRE, McLean USA), quant-ph/0009027.

Gisin, N., 1998, ``Quantum cloning without signaling'', Phys. Lett. A {\bf242}, 1-3.

Gisin, N. et al., 1995, ``Definition of Polarization Mode Dispersion and First Results of the COST 241
Round-Robin Measurements, with the members of the COST 241 group'',
JEOS Pure \& Applied Optics {\bf4}, 511-522.

Gisin, N. and S. Massar, 1997, ``Optimal quantum cloning machines'', Phys. Rev. Lett. {\bf79}, 2153-2156.

Gisin, B. and N. Gisin, 1999, ``A local hidden variable model of quantum correlation
exploiting the detection loophole'', Phys. Lett. A {\bf260}, 323-327.

Gisin, N., and S. Wolf, 1999, ``Quantum cryptography on noisy channels: quantum
versus classical key-agreement protocols'', Phys. Rev. Lett. 83, 4200-4203.

Gisin, N., and H. Zbinden, 1999, ``Bell inequality and the locality loophole: Active
versus passive switches'', Phys. Lett. A {\bf264}, 103-107.

Gisin, N., and S. Wolf, 2000a, ``Linking Classical and Quantum Key Agreement: Is There
``Bound  Information''?, Advances in cryptology - Proceedings of Crypto 2000,
Lecture Notes in Computer Science, Vol. 1880, 482-500.

Gisin, N., R. Renner and S. Wolf, 2000b, ``Bound information : the classical analog to
bound quantum entanglement, Proceedingsof the Third European Congress of Mathematics,
Barcelona, July 2000.

Goldenberg, L., and L. Vaidman, 1995, ``Quantum Cryptography Based on Orthogonal
States'', Phys. Rev. Lett. 75, 1239-1243.

Gorman, P.M., P.R. Tapster and J.G. Rarity, 2000, ``Secure Free-space Key Exchange Over
a 1.2 km Range Using Quantum Cryptography'' (DERA Malvern, United Kingdom).

Haecker, W., O. Groezinger, and M.H. Pilkuhn, 1971, ``Infrared photon counting by
Ge avalanche diodes'', Appl. Phys. Lett. {\bf19}, 113-115.

Hall, M.J.W., 1995, ``Information excusion principle for complementary observables'',
Phys. Rev. Lett. {\bf74}, 3307-3310.

Hariharan, P., M. Roy, P.A. Robinson and O'Byrne J.W., 1993, ``The geometric phase
observation at the single photon level'', J. Modern optics {\bf40}, 871-877.

Hart, A.C., R.G. Huff and K.L. Walker, 1994, ``Method of making a fiber having low polarization
mode dispersion due to a permanent spin'', U.S. Patent 5,298,047.

Hildebrand, E., 2001, Ph. D. thesis (Johann-Wolfgang
Goethe-Universit\"at, Frankfurt).

Hillery, M., V. Buzek, and A. Berthiaume, 1999, ``Quantum secret sharing'',
Phys. Rev. A {\bf59}, 1829-1834.

Hiskett, P. A., G. S. Buller, A. Y. Loudon, J. M. Smith, I. Gontijo,
A. C. Walker, P. D. Townsend, and M. J. Robertson, 2000, ``Performance
and Design of InGaAs/InP Photodiodes for Single-Photon Counting
at 1.55 $\mu m$'', Appl. Opt. {\bf 39}, 6818-6829.

Hong, C.K. and L. Mandel, 1985, ``Theory of parametric frequency down conversion of light'',
Phys. Rev. A {\bf31}, 2409-2418.

Hong, C.K. and L. Mandel, 1986, ``Experimental realization of a localized one-photon state'',
Phys. Rev. Lett. {\bf56}, 58-60.

Horodecki, M., R. Horodecki and P. Horodecki, 1996, ``Separability of Mixed States: Necessary and
Sufficient Conditions'', Phys. Lett. A {\bf223}, 1-8.

Hughes, R., G.G. Luther, G.L. Morgan and C. Simmons, 1996, ``Quantum Cryptography over Underground Optical
Fibers'', Lecture Notes in Computer Science {\bf 1109}, 329-342.

Hughes, R., W. Buttler, P. Kwiat, S. Lamoreaux, G. Morgan, J.
Nordhold, G. Peterson, 2000a, ``Free-space quantum key
distribution in daylight'', J. Modern Opt. {\bf47}, 549-562.

Hughes, R., G. Morgan, C. Peterson, 2000b, ``Quantum key distribution over a
48km optical fibre network'', J. Modern Opt. {\bf47}, 533-547.

Huttner, B., N. Imoto, N. Gisin, and T. Mor, 1995, ``Quantum Cryptography with
Coherent States'', Phys. rev. A {\bf51}, 1863-1869.

Huttner, B., J.D. Gautier, A. Muller H. Zbinden, and N. Gisin, 1996a, ``Unambiguous quantum
measurement of non-orthogonal states'', Phys. Rev. A {\bf54}, 3783-3789.

Huttner, B., N. Imoto, and S.M. Barnett, 1996b, ``Short distance applications of
Quantum cryptography'', J. Nonlinear Opt. Phys. \& Materials, {\bf5}, 823-832.

Imamoglu, A., and Y. Yamamoto, 1994, ``Turnstile Device for Heralded Single Photons :
Coulomb Blockade of Electron and Hole Tunneling in Quantum Confined p-i-n
Heterojunctions'', Phys. Rev. Lett. {\bf72}, 210-213.

Inamori, H., L. Rallan, and V. Vedral, 2000, ``Security of
EPR-based Quantum Cryptography against Incoherent Symmetric
Attacks'', quant-ph/0103058.

Ingerson, T.E., R.J. Kearney, and R.L. Coulter, 1983, ``Photon counting with photodiodes'',
Applied Optics {\bf22}, 2013-2018.

Ivanovic, I.D., 1987, ``How to differentiate between non-orthogonal states'', Phys. Lett. A {\bf123}, 257-259.

Jacobs, B., and J. Franson, 1996, ``Quantum cryptography in free space'',
Optics Letters {\bf 21}, 1854-1856.

Jennewein, T., U. Achleitner, G. Weihs, H. Weinfurter and A. Zeilinger, 2000a
``A fast and compact quantum random number generator'', Rev. Sci. Inst. {\bf71}, 1675-1680
and quantph/9912118.

Jennewein, T., C. Simon, G. Weihs, H. Weinfurter, and A. Zeilinger, 2000b
``Quantum Cryptography with Entangled Photons'', Phys. Rev. Lett. {\bf84}, 4729-4732

Karlsson, A., M. Bourennane, G. Ribordy, H. Zbinden, J. Brendel, J. Rarity,
and P. Tapster, 1999, ``A single-photon counter for long-haul telecom'', IEEE
Circuits \& Devices {\bf 15}, 34-40.

Kempe, J., Simon Ch., G. Weihs and A. Zeilinger, 2000, ``Optimal photon cloning'',
Phys. Rev. A 62, 032302.

Kim, J., O. Benson, H. Kan, and Y. Yamamoto, 1999, ``A single-photon turnstile device'', Nature, {\bf 397}, 500-503.

Kimble, H. J., M. Dagenais, and L. Mandel, 1977, ``Photon antibunching in resonance fluorescence'',
Phys. Rev. Lett. {\bf 39}, 691-694.

Kitson, S.C., P. Jonsson, J.G. Rarity, and P.R. Tapster, 1998, ``Intensity fluctuation
spectroscopy of small numbers of dye molecules in a microcavity'', Phys. Rev. A {\bf58},
620-6627.

Kolmogorow, A.N., 1956, ``Foundations of the theory of probabilities'', Chelsa Pub., New-York.

Kurtsiefer, Ch., S. Mayer, P. Zarda, and H. Weinfurter, 2000, ``Stable Solid-State Source of Single Photons'',
Phys. Rev. Lett., {\bf 85}, 290-293.

Kurtsiefer, Ch., P. Zarda, S. Mayer, and H. Weinfurter, 2001, ``The breakdown flash of Silicon
Avalanche Photodiodes -- backdoor for eavesdropper attacks?'', quant-ph/0104103.

Kwiat, P.G., A.M. Steinberg, R.Y. Chiao, P.H. Eberhard, M.D. Petroff, 1993,
``High-efficiency single-photon detectors'', Phys. Rev.A, {\bf48}, R867-R870.

Kwiat, P.G., E. Waks, A.G. White, I. Appelbaum, and
P.H. Eberhard, 1999, ``Ultrabright source of polarization-entangled
photons'', Phys. Rev. A, {\bf60}, R773-776.

Lacaita, A., P.A. Francese, F. Zappa, and S. Cova, 1994,
``Single-photon detection beyond 1 $\mu m$: performance of
comercially available germanium photodiodes'', Applied Optics
{\bf33}, 6902-6918.

Lacaita, A., F. Zappa, S. Cova, and P. Lovati, 1996,
``Single-photon detection beyond 1 $\mu m$: performance of
commercially available InGaAs/InP detectors. Appl. Optics
{\bf35}(16), 2986-2996.

Larchuk, T.S., M.V. Teich and B.E.A. Saleh, 1995, ``Nonlocal cancellation of dispersive
broadening in Mach-Zehnder interferometers'', Phys. Rev. A {\bf 52}, 4145-4154.

Levine, B.F., C.G. Bethea, and J.C. Campbell, 1985, ``Room-temperature 1.3-$\mu m$ optical time
domain reflectometer using a photon counting InGaAs/InP avalanche
detector'', Appl. Phys. Lettt. {\bf45}(4), 333-335.

Li, M.J., and D.A. Nolan, 1998, ``Fiber spin-profile designs for producing fibers with low PMD'',
Optics Lett. {\bf23}, 1659-1661.

Lo, H.-K., and H.F. Chau, 1998, ``Why Quantum Bit Commitment And Ideal Quantum
Coin Tossing Are Impossible'', Physica D {\bf120}, 177-187.

Lo, H.-K. and H.F. Chau, 1999, ``Unconditional security of quantum key distribution over
arbitrary long distances'' Science {\bf283}, 2050-2056; also quant-ph/9803006.

L\"utkenhaus, N., 1996, ``Security against eavesdropping in Quantum cryptography'',
Phys. Rev. A, {\bf54}, 97-111.

L\"utkenhaus, N., 2000, ``Security against individual attacks for realistic quantum key
distribution'', Phys. Rev. A, {\bf61}, 052304.

Marand, C., and P.D. Townsend, 1995, ``Quantum key distribution over distances as
long as 30 km'', Optics Letters {\bf 20}, 1695-1697.

Martinelli, M., 1992, ``Time reversal for the polarization state in optical systems'',
J. Modern Opt. {\bf39}, 451-455.

Martinelli, M., 1989, ``A universal compensator for polarization changes
induced by birefringence on a retracing beam'', Opt. Commun. {\bf72}, 341-344.

Maurer, U.M., 1993, ``Secret key agreement by public discussion from common
information'', IEEE Transacions on Information Theory {\bf39}, 733-742.

Maurer, U.M., and S. Wolf, 1999, ``Unconditionnally secure key agreement and
intrinsic information'', IEEE Transactions on Information Theory, {\bf45}, 499-514.

Mayers, D., 1996a, ``The Trouble with Quantum Bit Commitment'', quant-ph/9603015.

Mayers, D., 1996b, ``Quantum key distribution and string oblivious transfer in noisy
channels'', {\em Advances in Cryptology \,---\, Proceedings of Crypto '96},
Springer\,-\, Verlag, 343-357.

Mayers, D., 1997, ``Unconditionally secure Q bit commitment is impossible'',
Phys. Rev. Lett. {\bf78}, 3414-3417.

Mayers, D., 1998, ``Unconditional security in quantum cryptography'',
Journal for the Association of Computing Machinery (to be published);
also in quant-ph/9802025.

Mayers, D., and A. Yao, 1998, ``Quantum Cryptography with Imperfect Apparatus'',
Proceedings of the 39th IEEE Conference on Foundations of Computer Science.

Mazurenko, Y., R. Giust, and J.P. Goedgebuer, 1997, ``Spectral coding for secure
optical communications using refractive index dispersion'', Optics Commun. {\bf133}, 87-92.

M\'erolla, J-M., Y. Mazurenko, J.P. Goedgebuer, and W.T. Rhodes, 1999, ``Single-
photon interference in sidebands of phase-modulated light for Quantum
cryptography'', Phys. Rev. Lett, {\bf82}, 1656-1659.

Michler, P., A. Kiraz, C. Becher, W. V. Schoenfeld, P. M. Petroff, L. Zhang, E. Hu, and A. Imamoglu,
2000, ``A quantum dot single photon turnstile device'', Science (in press).

Milonni, P.W. and Hardies, M.L., 1982, ``Photons cannot always be replicated'',
Phys. Lett. A {\bf92}, 321-322.

Molotkov, S.N., 1998, ``Quantum crypto using photon frequency states (example of
a possible relaization)'', J. Exp. \& Theor. Physics {\bf87}, 288-293.

Muller, A., J. Breguet and N. Gisin, 1993, ``Experimental demonstration of quantum
cryptography using polarized photons in optical fiber over more than 1 km'',
Europhysics Lett. {\bf23}, 383-388.

Muller, A., H. Zbinden and N. Gisin, 1995, ``Underwater quantum coding'',
Nature {\bf378}, 449-449.

Muller, A., H. Zbinden and N. Gisin, 1996, ``Quantum cryptography over 23 km
in installed under-lake telecom fibre'', Europhysics Lett. {\bf33}, 335-339

Muller, A., T. Herzog, B. Huttner, W. Tittel, H. Zbinden, and N. Gisin, 1997,
``\,`Plug and play' systems for quantum cryptography'', Applied Phys. Lett. {\bf70}, 793-795.

Naik, D., C. Peterson, A. White, A. Berglund, and P. Kwiat, 2000, ``Entangled
State Quantum Cryptography: Eavesdropping on the Ekert Protocol'', Phys.
Rev. Lett. {\bf84}, 4733-4736

Neumann, E.-G., 1988, ``Single-mode fibers: fundamentals'', Springer series in Optical Sciences,
vol. 57.

Niu, C. S. and R. B. Griffiths, 1999, ``Two-qubit copying machine for
economical quantum eavesdropping'' Phys. Rev. A {\bf60}, 2764-2776.

Nogues, G., A. Rauschenbeutel, S. Osnaghi, M. Brune, J.M. Raimond and S. Haroche, 1999,
``Seeing a single photon without destroying it'',  Nature {\bf400}, 239-242.

Owens, P.C.M., J.G. Rarity, P.R. Tapster, D. Knight,
and P.D. Townsend, 1994, ``Photon counting with passively quenched
germanium avalanche'', Applied Optics {\bf33}, 6895-6901.

Penrose, R., 1994, ``Shadows of the mind'', Oxford University Press.

Peres, A., 1988, ``How to differentiate between two non-orthogonal states'',
Phys. Lett. A {\bf128}, 19.

Peres, A., 1996, ``Separability criteria for density matrices'', Phys. Rev. Lett.  {\bf76},
1413-1415.

Peres, A., 1997, {\it Quantum Theory: Concepts and Methods\/}, Kluwer, Dordrecht.

Phoenix, S.J.D., S.M. Barnett, P.D. Townsend, and K.J. Blow, 1995, ``Multi-user
Quantum cryptography on optical networks'', J. Modern optics, 6, 1155-1163.

Piron, C., 1990, ``M\'ecanique quantique'', Presses Polytechniques et Universitaires
Romandes, Lausanne, Switzerland, pp 66-67.

Pitowsky, I., 1989, ``Quantum probability, quantum logic'', Lecture Notes in
Physics 321, Heidelberg, Springer.

Rarity, J. G. and P.R. Tapster, 1988, ``Nonclassical effects in parametric downconversion'',
in  ``Photons \& Quantum Fluctuations'', eds Pike \& Walther, Adam Hilger.

Rarity, J. G., P.C.M. Owens and P.R. Tapster, 1994, ``Quantum random-number generation
and key sharing'', Journal of Modern Optics {\bf41}, 2435-2444.

Rarity, J. G., T. E. Wall, K. D. Ridley, P. C. M. Owens, and P. R. Tapster, 2000, ``Single-Photon
Counting for the 1300-1600-nm Range by Use of Peltier-Cooled and Passively Quenched InGaAs Avalanche Photodiodes'',
Appl. Opt. {\bf 39}, 6746-6753.

Ribordy, G., J. Brendel, J.D. Gautier, N. Gisin, and H. Zbinden, 2001, ``Long distance
entanglement based quantum key distribution'', Phys. Rev. A {\bf63}, 012309.

Ribordy, G., J.-D. Gautier, N. Gisin, O. Guinnard, H. Zbinden, 2000, ``Fast
and user-friendly quantum key distribution'', J. Modern Opt., {\bf47}, 517-531

Ribordy, G., J.D. Gautier, H. Zbinden and N. Gisin, 1998, ``Performance of InGaAsInP
avalanche photodiodes as gated-mode photon counters'', Applied Optics {\bf37}, 2272-2277.

Rivest, R.L., Shamir A. and Adleman L.M., 1978, ``A Method
of Obtaining Digital Signatures and Public-Key Cryptosystems'' {\em
Communications of the ACM} {\bf 21}, 120-126.

Santori, C., M. Pelton, G. Solomon, Y. Dale, and Y. Yamamoto, 2000, ``Triggered single photons from
a quantum dot'' (Stanford University, Palo Alto, California).

Shannon, C.E., 1949, ``Communication theory of secrecy systems'', Bell System
Technical Journal {\bf 28}, 656-715.

Shih, Y.H. and C.O. Alley, 1988, ``New type of Einstein-Podolsky-Rosen-Bohm
Experiment Using Pairs of Light Quanta Produced by Optical Parametric Down Conversion'',
Phys. Rev. Lett. {\bf61}, 2921-2924.

Shor, P.W., 1994, ``Algoritms for quantum computation: discrete
logarithms and factoring'', {\it{Proceedings of the 35th Symposium
on Foundations of Computer Science}}, Los Alamitos, edited by
Shafi Goldwasser (IEEE Computer Society Press), 124-134.

Shor, P.W., and J. Preskill, 2000, ``Simple proof of security of the BB84 Quantum
key distribution protocol'', Phys. Rev. Lett. {\bf85}, 441-444.

Simon, C., G. Weihs, and A. Zeilinger, 1999, ``Quantum Cloning and Signaling'', Acta Phys.
Slov. {\bf49}, 755-760.

Simon, C., G. Weihs, A. Zeilinger, 2000, ``Optimal Quantum Cloning via Stimulated
Emission'', Phys. Rev. Lett. {\bf84}, 2993-2996.

Singh, S., 1999, ``The code book: The Science of Secrecy from Ancient Egypt
to Quantum Cryptography'' (Fourh Estate, London), see Ekert 2000 for a review.

Snyder, A.W., 1983, ``Optical waveguide theory'', Chapman \& Hall, London.

Spinelli, A., L.M. Davis, H. Dauted, 1996, ``Actively quenched single-photon
avalanche diode for high repetition rate time-gated photon counting'',
Rev. Sci. Instrum {\bf67}, 55-61.

Stallings, W., 1999, ``Cryptography and network security: principles and practices'',
(Prentice Hall, Upper Saddle River, New Jersey, United States).

Stefanov, A., O. Guinnard, L. Guinnard, H. Zbinden and N. Gisin, 2000,
``Optical Quantum Random Number Generator'', J. Modern Optics {\bf47}, 595-598.

Steinberg, A.M., P. Kwiat and R.Y. Chiao, 1992a, ``Dispersion cancellation and
high-resolution time measurements in a fourth-order optical
interferometer'', Phys. Rev. A {\bf 45}, 6659-6665.

Steinberg, A.M., P. Kwiat and R.Y. Chiao,
1992b, ``Dispersion Cancellation in a Measurement of the
Single-Photon Propagation Velocity in Glass'', Phys. Rev. Lett. {\bf68}, 2421-2424.

Stucki, D., G. Ribordy, A. Stefanov, H. Zbinden, J. Rarity and T. Wall, 2001,
``Photon counting for quantum key distribution with Peltier cooled InGaAs/InP APD's'',
preprint, University of Geneva, Geneva.

Sun, P.C., Y. Mazurenko, and Y. Fainman, 1995, ``Long-distance frequency-division
interferometer for communication and quantum cryptography'', Opt. Lett. {\bf20}, 1062-1063.

Tanzilli, S., H. De Riedmatten, W. Tittel, H.
Zbinden, P. Baldi, M. De Micheli, D.B. Ostrowsky, and N. Gisin,
2001, ``Highly efficient photon-pair source using a Periodically
Poled Lithium Niobate waveguide'', Electr. Lett. {\bf37}, 26-28.

Tapster, P.R., J.G. Rarity, and P.C.M. Owens, 1994, ``Violation of Bell's
Inequality over 4 km of Optical Fiber'', Phys. Rev. Lett. {\bf73}, 1923-1926.

Thomas, G.A., B.I. Shraiman, P.F. Glodis and M.J. Stephen, 2000, ``Towards the
clarity limit in optical fiber'', Nature {\bf404}, 262-264.

Tittel, W., J. Brendel, H. Zbinden, and N. Gisin, 1998, ``Violation of Bell
inequalities by photons more than 10 km apart'', Phys. Rev. Lett. {\bf81}, 3563-3566.

Tittel, W., J. Brendel, H. Zbinden and N. Gisin, 1999, ``Long-distance Bell-type
tests using energy-time entangled photons'', Phys. Rev. A {\bf59}, 4150-4163.

Tittel, W., J. Brendel, H. Zbinden, and N. Gisin,\ 2000, ``Quantum
Cryptography Using Entangled Photons in Energy-Time Bell States'', Phys.
Rev. Lett. {\bf84}, 4737-4740

Tittel, W., H. Zbinden, and N. Gisin, 2001, ``Experimental demonstration of quantum
secret sharing'', Phys. Rev. A {\bf63}, 042301.

Tomita, A. and R. Y. Chiao, 1986, ``Observation of Berry's topological phase by use of
an optical fiber'', Phys. Rev. Lett. {\bf 57}, 937-940.

Townsend, P., 1994, ``Secure key distribution system based on Quantum
cryptography'', Elect. Lett. {\bf30}, 809-811.

Townsend, P., 1997a, ``Simultaneous Quantum cryptographic key distribution and
conventional data transmission over installed fibre using WDM'', Elect. Lett. {\bf33},
188-190.

Townsend, P., 1997b, ``Quantum cryptography on multiuser optical fiber
networks'', Nature {\bf385}, 47-49.

Townsend, P., 1998a, ``Experimental Investigation of the Performance Limits
for First Telecommunications-Window Quantum Cryptography Systems'', IEEE
Photonics Tech. Lett. {\bf10}, 1048-1050.

Townsend, P., 1998b, ``Quantum Cryptography on Optical Fiber Networks'', Opt.
Fiber Tech. {\bf4}, 345-370.

Townsend, P., J.G. Rarity, and P.R. Tapster, 1993a, ``Single photon interference
in a 10 km long optical fiber interferometer'', Electron. Lett. {\bf29}, 634-639.

Townsend, P., J. Rarity, and P. Tapster, 1993b, ``Enhanced single photon
fringe visibility in a 10km-long prototype quantum cryptography channel'',
Electron. Lett. {\bf29}, 1291-1293.

Townsend, P.D., S.J.D. Phoenix, K.J. Blow, and S.M. Barnett, 1994, ``Design of QC
systems for passive optical networks'', Elect. Lett, 30, pp. 1875-1876.

Vernam, G., 1926, ``Cipher printing telegraph systems for secret wire and radio
telegraphic communications'', J. Am. Institute of Electrical
Engineers Vol. XLV, 109-115.

Vinegoni, C., M. Wegmuller and N. Gisin, 2000a, ``Determination of nonlinear coefficient
n2/Aeff using self-aligned interferometer and Faraday mirror'', Electron. Lett. {\bf36}, 886-888.

Vinegoni, C., M. Wegmuller, B. Huttner and N. Gisin, 2000b, ``Measurement of nonlinear
polarization rotation in a highly birefringent optical fiber using a Faraday mirror'',
J. of Optics A {\bf2}, 314-318.

Walls, D.F. and G.J. Milburn, 1995, ``Quantum optics'', Springer-verlag.

Weihs, G., T. Jennewein, C. Simon, H. Weinfurter, and A. Zeilinger, 1998,
''Violation of Bell's Inequality under Strict Einstein Locality
Conditions'', Phys. Rev. Lett. {\bf81}, 5039-5043.

Wiesner, S., 1983, ``Conjugate coding'', Sigact news, 15:1, 78-88.

Wigner, E.P., 1961, ``The probability of the existence of a self-reproducing unit'',
in ``The logic of personal knowledge''
Essays presented to Michael Polanyi in his Seventieth birthday, 11 March 1961
Routledge \& Kegan Paul, London, pp 231-238.

Wooters, W. K. and Zurek, W.H., 1982, ``A single quanta cannot be cloned'', Nature {\bf 299}, 802-803.

Yuen, H.P., 1997, ``Quantum amplifiers, Quantum duplicators and Quantum
cryptography'', Quantum \& Semiclassical optics, 8, p. 939.

Zappa, F., A. Lacaita, S. Cova, and P. Webb, 1994, ``Nanosecond
single-photon timing with InGaAs/InP photodiodes'', Opt. Lett. {\bf19}, 846-848.

Zbinden, H., J.-D. Gautier, N. Gisin, B. Huttner, A. Muller, and W. Tittel,
1997, ``Interferometry with Faraday mirrors for quantum cryptography'',
Electron. Lett. {\bf33}, 586-588.

Zeilinger, A., 1999, ``Experiment and the foundations of quantum physics'',
Rev. Mod. Phys. {\bf71}, S288-S297.

Zissis, G., and A. Larocca, 1978, ``Optical Radiators and Sources'',
Handbook of Optics, edited by W. G. Driscoll (McGraw-Hill, New York), Sec. 3.

\.Zukowski, M., A. Zeilinger, M.A. Horne and A. Ekert, 1993, ``\,`Event-ready-detectors' Bell experiment via entanglement swapping'', Phys. Rev.
Lett. {\bf71}, 4287-4290.

\.Zukowski, M., A. Zeilinger, M. Horne, and H. Weinfurter, 1998, ``Quest for
GHZ states'', Acta Phys. Pol. A {\bf93}, 187-195.

\newpage
\normalsize
\section*{Figures}

\newpage

\bigskip

\end{document}